\documentclass{article}

\newcommand{\secref}[1]{Subsection~\ref{sec.#1}}
\newcommand{\seclabel}[1]{\label{sec.#1}}

\newcommand{\appref}[1]{Appendix~\ref{app.#1}}
\newcommand{\applabel}[1]{\label{app.#1}}

\newcommand{\tabnum}[1]{\ref{tab.#1}}
\newcommand{\tabref}[1]{Table~\tabnum{#1}}
\newcommand{\tablabel}[1]{\label{tab.#1}}

\newcommand{\eqnlabel}[1]{\label{Equations.#1}}

\newcommand{\figref}[1]{Figure~\ref{Figures.#1}}
\newcommand{\figlabel}[1]{\label{Figures.#1}}

\newcommand{\easyfig}[4]{
\begin{figure}
\includegraphics[#2]{#1}
\caption{ \figlabel{#3} #4}
\end{figure}}

\newcommand{\pngfig}[2]{\easyfig{#1.png}{}{#1}{#2}}
\newcommand{\pdffig}[2]{\easyfig{#1-fig.pdf}{}{#1}{#2}}

\newcommand{\widepngfig}[2]{\easyfig{#1.png}{width=\textwidth}{#1}{#2}}

\newcommand{\widepdffig}[2]{\easyfig{#1-fig.pdf}{width=\textwidth}{#1}{#2}}

\newcommand{\sidewayspngfig}[2]{
\begin{sidewaysfigure}
\includegraphics[width=\textwidth]{#1.png}
\caption{\figlabel{#1} #2}
\end{sidewaysfigure}
}

\usepackage{graphicx}
\usepackage{url}
\usepackage{amsmath}
\usepackage{amssymb}
\usepackage{color} 
\usepackage{ifthen}
\usepackage{rotating}
\newcommand{\hl}[1]{#1}

\newcommand{\profileSizeLimit}{p}

\def\pterror#1{\errmessage{Parsetree ERROR: #1}}

\newdimen\pthgap\def\pthorgap#1{\pthgap=#1}
\newdimen\ptvgap\def\ptvergap#1{\ptvgap=#1}

\newbox\ptnodestrutbox\def\ptnodestrut{\unhcopy\ptnodestrutbox}
\newbox\ptleafstrutbox\def\ptleafstrut{\unhcopy\ptleafstrutbox}

\def\ptnodefont#1#2#3{\def\ptnodefn{#1}
  \setbox\ptnodestrutbox=\hbox{\vrule height#2 width0pt depth#3}}
\def\ptleaffont#1#2#3{\def\ptleaffn{#1}
  \setbox\ptleafstrutbox=\hbox{\vrule height#2 width0pt depth#3}}

\pthorgap{12pt}                         
\ptvergap{12pt}                         
\ptnodefont{\normalsize\rm}{11pt}{3pt}  
\ptleaffont{\normalsize\it}{11pt}{3pt}  

\newcount\ptdepth           
\newcount\ptn               
\newbox\ptm \newdimen\ptmx  
\newbox\pta \newdimen\ptax  
\newbox\ptb \newdimen\ptbx  
\newbox\ptc \newdimen\ptcx  
\newbox\ptx \newdimen\ptxx  
\newif\ifpttri              

\def\ptnext{\advance\ptn by 1 \ifcase\ptn
  \or \setbox\ptm=\box\ptx \ptmx=\ptxx \or \setbox\pta=\box\ptx \ptax=\ptxx
  \or \setbox\ptb=\box\ptx \ptbx=\ptxx \or \setbox\ptc=\box\ptx \ptcx=\ptxx
  \else \pterror{More than 3 daughters in (sub)tree}\fi}

\def\ptbegtree{\ptdepth=0}
\def\ptendtree
  {\ifnum\ptdepth>0 \pterror{Mismatched bracketing: too few ')'s!}\fi}

\def\ptbeg{\ifnum\ptdepth=0 \leavevmode\fi\begingroup
  \advance\ptdepth1 \ptn=0\pttrifalse}
\def\ptend{\ifnum\ptdepth=0 \pterror{Mismatched bracketing: too many ')'s!}
  \else\ptcons\endgroup\ifnum\ptdepth=0 \box\ptx\else\ptnext\fi\fi}

\def\ptnodeaux#1{\setbox\ptx=\hbox{#1}\ptxx=0.5\wd\ptx\ptnext}
\def\ptnode#1{\ptnodeaux{\ptnodefn\ptnodestrut #1}}
\def\ptleaf#1{\ptnodeaux{\ptleaffn\ptleafstrut #1}}

\def\pthoradjust#1{\ifcase\ptn
  \or \pthadjbox{\ptm}{#1} \or \pthadjbox{\pta}{#1}
  \or \pthadjbox{\ptb}{#1} \or \pthadjbox{\ptc}{#1}
  \else \pterror{More than 3 daughters in (sub)tree}\fi}
\def\pthadjbox#1#2{\setbox#1=\hbox{\box#1\kern#2}}

\def\ptcons
 {\ifnum\ptn=0 \ptconsz 
  \else
    \ifpttri\ptconstri\else
      \ifcase\ptn \or\ptconsm\or\ptconsma\or\ptconsmab\or\ptconsmabc \fi \fi
    \ptax=\ptxx \advance\ptax-\ptmx \ptbx=0pt
    \ifdim\ptax<0pt \ptbx=-\ptax\ptax=0pt\ptxx=\ptmx \fi
    \setbox\ptx=\vtop{\hbox{\kern\ptax\box\ptm}\nointerlineskip
                      \hbox{\kern\ptbx\box\ptx}}\fi
  \global\ptxx=\ptxx\global\setbox\ptx=\box\ptx}

\def\ptavg#1#2#3{#1=#2\advance#1#3#1=0.5#1}     
\def\ptadv#1#2{\advance#1#2\advance#1\pthgap}   

\def\ptconsz{\ptxx=0pt \setbox\ptx=\vtop{}}     

\def\ptconsm{\ptxx=0pt 
  \setbox\ptx=\hbox{\ptedge{1}{0}{}{}}}         

\def\ptconsma                                   
 {\ptxx=\ptax\setbox\ptx=\vtop{
    \hbox{\ptedge{1}\ptax{}{}}\nointerlineskip
    \hbox{\box\pta}}}

\def\ptconsmab                                  
 {\ptadv\ptbx{\wd\pta}\ptavg\ptxx\ptax\ptbx
  \setbox\ptx=\vtop{
    \hbox{\ptedge{2}\ptax\ptbx{}}\nointerlineskip
    \hbox{\box\pta\kern\pthgap\box\ptb}}}

\def\ptconsmabc                                 
 {\ptadv\ptbx{\wd\pta}\ptadv\ptcx{\wd\pta}%
  \ptadv\ptcx{\wd\ptb}\ptavg\ptxx\ptax\ptcx
  \setbox\ptx=\vtop{
    \hbox{\ptedge{3}\ptax\ptbx\ptcx}\nointerlineskip
    \hbox{\box\pta\kern\pthgap\box\ptb\kern\pthgap\box\ptc}}}

\def\ptconstri                                  
 {\ifcase\ptn\or\setbox\pta\hbox{\kern2\pthgap}\or
  \or\setbox\pta=\hbox{\box\pta\kern\pthgap\box\ptb}
  \or\setbox\pta=\hbox{\box\pta\kern\pthgap\box\ptb\kern\pthgap\box\ptc}\fi
  \ptxx=0.5\wd\pta
  \setbox\ptx=\vtop{
    \hbox{\ptedge{0}{0}{\wd\pta}{}}\nointerlineskip
    \box\pta}}

\newcount\pted          
\newcount\ptedm         
\newcount\pteda         
\newcount\ptedb         
\newcount\ptedc         
\newcount\ptedl         
\newcount\ptedh         
\newcount\ptedhs        
\newcount\ptedvs        
\newcount\ptedtemp      

\def\ptedge#1#2#3#4{\pted=#1%
  \pteda=#2\ifcase\pted\ptedb=#3\or\or\ptedb=#3\or\ptedb=#3\ptedc=#4\fi
  \ptedm=\pteda\advance\ptedm\ifcase\pted\ptedb\or\pteda\or\ptedb\or\ptedc\fi
  \divide\ptedm by 2
  \ptedh=\ptvgap\ptedtemp=\ptedm\advance\ptedtemp-\pteda\divide\ptedtemp by 6
  \ifnum\ptedh<\ptedtemp\ptedh=\ptedtemp\fi
  \unitlength=1sp%
  \begin{picture}(0,\ptedh)
    \ifnum\pted=3 \ptedput\ptedc\fi
    \ifnum\pted=1 \else\ptedput\ptedb\fi
    \ptedput\pteda
    \ifnum\pted=0 \ptedbot\fi 
  \end{picture}}

\def\ptedput#1{\ptedl=#1\advance\ptedl-\ptedm
  \ifnum\ptedl>0 \ptedslope\else
    \ptedl=-\ptedl\ptedslope\ptedhs=-\ptedhs\fi
  \ifnum\ptedhs=0 \ptedl=\ptedh\fi
  \put(\ptedm,\ptedh){\line(\ptedhs,-\ptedvs){\ptedl}}}

\def\ptedbot
 {\ptedtemp=\ptedl\multiply\ptedtemp by \ptedvs\divide\ptedtemp by \ptedhs
  \ifnum\ptedtemp>0\ptedtemp=-\ptedtemp\fi \advance\ptedtemp-1
  \advance\ptedtemp\ptedh\multiply\ptedl by 2
  \put(\pteda,\ptedtemp){\line(1,0){\ptedl}}}

\def\ptedslope
 {\ifnum\ptedl>\ptedh\ptedhs=\ptedl\ptedvs=\ptedh
  \else              \ptedvs=\ptedl\ptedhs=\ptedh \fi
  \divide\ptedhs by 60 \divide\ptedvs by \ptedhs
  \ifnum \ptedvs <  5 \ptedvs=0 \ptedhs=1 \else
  \ifnum \ptedvs < 11 \ptedvs=1 \ptedhs=6 \else
  \ifnum \ptedvs < 13 \ptedvs=1 \ptedhs=5 \else
  \ifnum \ptedvs < 17 \ptedvs=1 \ptedhs=4 \else
  \ifnum \ptedvs < 22 \ptedvs=1 \ptedhs=3 \else
  \ifnum \ptedvs < 27 \ptedvs=2 \ptedhs=5 \else
  \ifnum \ptedvs < 33 \ptedvs=1 \ptedhs=2 \else
  \ifnum \ptedvs < 38 \ptedvs=3 \ptedhs=5 \else
  \ifnum \ptedvs < 42 \ptedvs=2 \ptedhs=3 \else
  \ifnum \ptedvs < 46 \ptedvs=3 \ptedhs=4 \else
  \ifnum \ptedvs < 49 \ptedvs=4 \ptedhs=5 \else
  \ifnum \ptedvs < 55 \ptedvs=5 \ptedhs=6 \else
                      \ptedvs=1 \ptedhs=1 
    \fi \fi \fi \fi \fi \fi \fi \fi \fi \fi \fi \fi
  \ifnum \ptedl < \ptedh
    \ptedtemp=\ptedhs \ptedhs=\ptedvs \ptedvs=\ptedtemp \fi}

\newenvironment{parsetree}{\ptactivechardefs\ptbegtree}{\ptendtree}

\def\ptcatcodes
 {\catcode`(=\active \catcode`)=\active
  \catcode`.=\active \catcode``=\active
  \catcode`~=\active}

{\ptcatcodes
\gdef\ptactivechardefs
 {\ptcatcodes
  \def({\ptbeg\ignorespaces}
  \def){\ptend\ignorespaces}
  \def.##1.{\ptnode{##1}\ignorespaces} 
  \def`##1'{\ptleaf{##1}\ignorespaces}
  \def~{\pttritrue\ignorespaces}}}

\usepackage[boxed,noend]{algorithm2e}

\newcommand\wikipediaTransducerUrl{http://en.wikipedia.org/w/index.php?title=Finite\_state\_transducer\&oldid=486381386}

\newcommand\nullmodel{{\cal N}}
\newcommand\substitute{{\cal S}}
\newcommand\tkf{{\cal B}}
\newcommand\tkfroot{{\cal R}}
\newcommand\profile{{\cal P}}

\newcommand\formaldefs{{\bf Formal definitions: }}

\newcommand\charat[2]{\mbox{symbol}(#1,#2)}

\newcommand\gappedalphabet[1]{(\Omega_{#1} \cup \{\epsilon\})}
\newcommand\gapsquared{\gappedalphabet{}^2}
\newcommand\gappedpair[2]{\gappedalphabet{#1} \times \gappedalphabet{#2}}

\newcommand\wtrans[4]{#1(#2 : [#3] : #4)}
\newcommand\transequiv{\equiv}

\newcommand\compose{}
\newcommand\identity{{\cal I}}

\newcommand\fork{\circ}
\newcommand\idfork{\Upsilon}
\newcommand\forkn[1]{\idfork(#1)}
\newcommand\forkfun[2]{\forkn{#1, #2}}

\newcommand\generate{\Delta}
\newcommand\recognize{\nabla}

\newcommand\States{\Phi}
\newcommand\statesof[1]{\States_{#1}}

\newcommand\Transitions{\tau}
\newcommand\transitionsof[1]{\Transitions_{#1}}

\newcommand\startstate{\phi_S}
\newcommand\laststate{\phi_E}
\newcommand\startstateof[1]{\phi_{S;#1}}
\newcommand\laststateof[1]{\phi_{E;#1}}

\newcommand\weight{{\cal W}}
\newcommand\weightfunof[1]{\weight_{#1}}
\newcommand\transweightfun[1]{\weightfunof{#1}^{\mbox{\small trans}}}
\newcommand\emitweightfun[1]{\weightfunof{#1}^{\mbox{\small emit}}}

\newcommand\sumoverpaths[1]{\transweightfun{#1}(\{\pi_{#1}\})}

\newcommand\transviawait[1]{\weightfunof{#1}^{\mbox{\small via-wait}}}
\newcommand\transtowait[1]{\weightfunof{#1}^{\mbox{\small to-wait}}}

\newcommand\numberofstates[1]{|\statesof{#1}|}
\newcommand\numberoftransitions[1]{|\transitionsof{#1}|}

\newcommand\statesoftype[1]{\States_{#1}}
\newcommand\statetype{\mbox{type}}

\newcommand\dup[1]{\left( \begin{array}{l} #1 \\ #1 \end{array} \right)}

\newcommand\numberofleaves{\kappa}
\newcommand\numberofinternalnodes{\numberofleaves - 1}
\newcommand\numberofnodes{2\numberofleaves - 1}
\newcounter{LeafIndex}
\newcommand\leafnode[1]{\numberofleaves \ifthenelse{\equal{#1}{1}}{}{\setcounter{LeafIndex}{#1} \addtocounter{LeafIndex}{-1} +\arabic{LeafIndex}}}
\newcommand\leaves{{\cal L}}

\newcommand\seqlen[1]{\mbox{len}(#1)}

\newcommand\outputs{{\cal D}}
\newcommand\outputn[1]{{\cal S}_{#1}}
\newcommand\outseqlen[1]{\seqlen{\outputn{#1}}}

\newcommand\order[1]{{\cal O}(#1)}

\newcommand\typeset[1]{{\cal T}_{\mbox{\small #1}}}
\newcommand\stateset[1]{\statesof{\mbox{\small #1}}}

\newcommand\hstate{(\upsilon,b_l,e_l,b_r,e_r)}
\newcommand\hstatedest{(\upsilon',b'_l,e'_l,b'_r,e'_r)}

\newcommand\externalsuffix{ext}
\newcommand\internalsuffix{int}
\newcommand\leftsuffix{left-int}
\newcommand\rightsuffix{right-int}
\newcommand\waitsuffix{wait}

\newcommand\externalcascades{\stateset{\externalsuffix}}
\newcommand\internalcascades{\stateset{\internalsuffix}}
\newcommand\leftcascades{\stateset{\leftsuffix}}
\newcommand\rightcascades{\stateset{\rightsuffix}}
\newcommand\waitstates{\stateset{\waitsuffix}}

\newcommand\externaltypes{\typeset{\externalsuffix}}
\newcommand\internaltypes{\typeset{\internalsuffix}}
\newcommand\lefttypes{\typeset{\leftsuffix}}
\newcommand\righttypes{\typeset{\rightsuffix}}
\newcommand\waittypes{\typeset{\waitsuffix}}

\newcommand\mstate{(\rho,\upsilon,b_l,e_l,b_r,e_r)}
\newcommand\mstatedest{(\rho',\upsilon',b'_l,e'_l,b'_r,e'_r)}

\newcommand\qstate{(\rho,\upsilon,b_l,b_r)}
\newcommand\qstatedest{(\rho',\upsilon',b'_l,b'_r)}

\newcommand\matchsuffix{match}
\newcommand\nullsuffix{null}
\newcommand\leftinsertsuffix{left-ins}
\newcommand\rightinsertsuffix{right-ins}
\newcommand\leftdeletesuffix{left-del}
\newcommand\rightdeletesuffix{right-del}
\newcommand\leftemitsuffix{left-emit}
\newcommand\rightemitsuffix{right-emit}
\newcommand\qwaitsuffix{wait}

\newcommand\matchstates{\stateset{\matchsuffix}}
\newcommand\nullstates{\stateset{\nullsuffix}}
\newcommand\leftinsertstates{\stateset{\leftinsertsuffix}}
\newcommand\rightinsertstates{\stateset{\rightinsertsuffix}}
\newcommand\leftdeletestates{\stateset{\leftdeletesuffix}}
\newcommand\rightdeletestates{\stateset{\rightdeletesuffix}}
\newcommand\leftemitstates{\stateset{\leftemitsuffix}}
\newcommand\rightemitstates{\stateset{\rightemitsuffix}}
\newcommand\qwaitstates{\stateset{\qwaitsuffix}}

\newcommand\matchtypes{\typeset{\matchsuffix}}
\newcommand\nulltypes{\typeset{\nullsuffix}}
\newcommand\leftinserttypes{\typeset{\leftinsertsuffix}}
\newcommand\rightinserttypes{\typeset{\rightinsertsuffix}}
\newcommand\leftdeletetypes{\typeset{\leftdeletesuffix}}
\newcommand\rightdeletetypes{\typeset{\rightdeletesuffix}}
\newcommand\leftemittypes{\typeset{\leftemitsuffix}}
\newcommand\rightemittypes{\typeset{\rightemitsuffix}}
\newcommand\qwaittypes{\typeset{\qwaitsuffix}}

\newcommand\envelope[2]{\mbox{is\_in\_envelope}(#1,#2)}

\newcommand\newTransName[1]{t_{#1}}

\newcommand\numStates[1]{N_{#1}}

\newcommand\leftFromI{i_l}
\newcommand\rightFromI{i_r}

\newcommand\rightToI{i'_r}
\newcommand\leftToI{i'_l}

\newcommand\rightProfTo{e'_r}
\newcommand\leftProfTo{e'_l}

\newcommand\rightProfFrom{e_r}
\newcommand\leftProfFrom{e_l}

\newcommand\mTo{m'}
\newcommand\mFrom{m}

\newcommand\qTo{q'}
\newcommand\qFrom{q}

\newcommand\emitProb{\mathcal{E}}
\newcommand\getprofiletype{\mbox{get\_state\_type}}
\newcommand\incomingLeftProfile[1]{\mbox{incoming\_left\_profile\_indices}(#1)}
\newcommand\incomingRightProfile[1]{\mbox{incoming\_right\_profile\_indices}(#1)}
\newcommand\incomingM[1]{\mbox{incoming\_match\_states}(#1)}
\newcommand\incomingL[1]{\mbox{incoming\_left\_emit\_states}(#1)}
\newcommand\incomingR[1]{\mbox{incoming\_right\_emit\_states}(#1)}

\newcommand\addToDPFunction{sum\_paths\_to}   
\newcommand\addToDP[3]{\addToDPFunction(#1,#2,#3)} 

\newcommand\profTrans[1]{t_{#1}}
\newcommand\profiledelete[1]{\phi_D^{(#1)}}
\newcommand\profileunknown[1]{\phi_{\tau}^{(#1)}}
\newcommand\profilewait[1]{\phi_W^{(#1)}}
\newcommand\profileterminate{\profilewait{\mbox{\small end}}}

\newcommand\currentstate{m'}
\newcommand\newstate{m}
\newcommand\instates{\mbox{states\_in}}
\newcommand\inweights{\mbox{weights\_in}}
\newcommand\sample{\mbox{sample}}

\newcommand\emptyarray{ ( )}

\begin{document}

\newcommand\authorstring{
Oscar Westesson$^{1}$, 
Gerton Lunter$^{2}$, 
Benedict Paten$^{3}$, 
Ian Holmes$^{1,\ast}$
\\
\textbf{1} Department of Bioengineering, University of California, Berkeley, CA, USA; \\
\textbf{2} Wellcome Trust Center for Human Genetics, Oxford, Oxford, UK;\\
\textbf{3} Baskin School of Engineering, UC Santa Cruz, Santa Cruz, CA, USA
\\
$\ast$ E-mail: ihh@berkeley.edu
}

\newcommand\titlestring{Phylogenetic automata, pruning, and multiple alignment}
\newcommand\shorttitlestring{Phylogenetic automata, pruning, and alignment}
\markboth{\shorttitlestring}{\shorttitlestring}

\begin{flushleft}
  {\Large
    \textbf{\titlestring}
  }
\\
\authorstring
\end{flushleft}

\pagebreak

\section*{Abstract}

We present an extension of Felsenstein's algorithm to indel models defined on entire sequences,
without the need to condition on one multiple alignment.
The algorithm makes use of a generalization from probabilistic substitution matrices to weighted finite-state transducers.
Our approach may equivalently be viewed as a probabilistic formulation of progressive multiple sequence alignment,
using partial-order graphs to represent ensemble profiles of ancestral sequences.
We present a hierarchical stochastic approximation technique which makes this algorithm tractable for alignment analyses of reasonable size.  


\paragraph{Keywords:} Multiple alignment, Felsenstein pruning, transducers, insertion deletion, ancestral sequence reconstruction. 

\tableofcontents

\section{Background}

Felsenstein's pruning algorithm is routinely used throughout bioinformatics and molecular evolution \cite{Felsenstein81}.
A few common applications include estimation of substitution rates \cite{Yang94b};
reconstruction of phylogenetic trees \cite{RannalaYang96};
identification of conserved (slow-evolving) or recently-adapted (fast-evolving) elements in proteins and DNA \cite{SiepelHaussler04b};
detection of different substitution matrix ``signatures''
(e.g. purifying vs diversifying selection at synonymous codon positions \cite{YangEtAl2000},
hydrophobic vs hydrophilic amino acid signatures \cite{ThorneEtAl96},
CpG methylation in genomes \cite{SiepelHaussler04},
or basepair covariation in RNA structures \cite{KnudsenHein99});
annotation of structures in genomes \cite{SiepelHaussler04c,PedersenEtAl2006};
and placement of metagenomic reads on phylogenetic trees \cite{MatsenEtAl2010}.

The pruning algorithm computes the likelihood of observing a single column of a multiple sequence alignment,
 given knowledge of an underlying phylogenetic tree (including a map from leaf-nodes of the tree to rows in the alignment)
 and a substitution probability matrix associated with each branch of the tree.
Crucially, the algorithm sums over all unobserved substitution histories on internal branches of the tree.
For a tree containing $N$ taxa, the algorithm achieves ${\cal O}(N)$ time and memory complexity by computing and tabulating intermediate probability functions of the form $G_n(x) = P(Y_n|x_n=x)$,
where $x_n$ represents the individual residue state of ancestral node $n$,
and $Y_n=\{y_m\}$ represents all the data at leaves $\{m\}$ descended from node $n$ in the tree (i.e. the observed residues at all leaf nodes $m$ whose ancestors include node $n$).

The pruning recursion visits all nodes in postorder.
Each $G_n$ function is computed in terms of the functions $G_l$ and $G_r$ of its immediate left and right children (assuming a binary tree):
\begin{eqnarray*}
G_n(x) & = & P(Y_n|x_n = x) \\
& = & \left\{
\begin{array}{ll}
\left( \sum_{x_l} B^{(l)}_{x,\ x_l} G_l(x_l) \right) \left( \sum_{x_r} B^{(r)}_{x,\ x_r} G_r(x_r) \right) & \mbox{if $n$ is not a leaf}
 \\
\delta(x=y_n) & \mbox{if $n$ is a leaf}
\end{array}
\right.
\end{eqnarray*}
where $B^{(n)}_{ab} = P(x_n=b|x_m=a)$ is the probability that node $n$ has state $b$, given that its parent node $m$ has state $a$;
and $\delta(x=y_n)$ is a Kronecker delta function terminating the recursion at the leaf nodes of the tree.

The ``states'' in the above description may represent individual residues (nucleotides, amino acids), base-pairs (in RNA secondary structures) or base-triples (codons).
Sometimes, the state space is augmented to include gap characters, or latent variables.
In the machine learning literature, the $G_n$ functions are often described as ``messages'' propagated from the leaves to the root of the tree \cite{KschischangEtAl98},
and corresponding to a summary of the information in the subtree rooted at $n$.

The usual method for extending this approach from individual residues to full-length sequences assumes both that one knows the alignment of the sequences, 
and that the columns of this alignment are each independent realizations of single-residue evolution.  
One uses pruning to compute the above likelihood for a single alignment column,
then multiplies together the probabilities across every column in the alignment.
For an alignment of length $L$, the time complexity is ${\cal O}(LN)$ and the memory complexity ${\cal O}(N)$.
This approach works well for marginalizing substitution histories consistent with a single alignment,
but does not readily generalize to summation over indel histories or alignments.

The purpose of this manuscript is to introduce another way of extending Felsenstein's recursion from single residues (or small groups of residues)
to entire, full-length sequences, without needing to condition on a single alignment.
With no constraints on the algorithm, \hl{using a branch transducer with $c$ states,}
the time and memory complexities are ${\cal O}((cL)^N)$,
with close similarity to the algorithms of Sankoff \cite{SankoffCedergren83} and Hein \cite{Hein2001}.
\hl{For a user-specified maximum internal profile size  $\profileSizeLimit \geq L$, }
the worst-case time complexity drops to \hl{ ${\cal O}(c^2\profileSizeLimit^4 N)$ 
(typical case is ${\cal O}((c\profileSizeLimit)^2 N)$}) when
 a stochastic lower-bound approximation is used; 
memory complexity is ${\cal O}((c\profileSizeLimit)^2 + pN)$. 
In this form,
the algorithm is similar to the partial order graph for multiple sequence alignment \cite{LeeGrassoSharlow2002}.
\hl{If an``alignment envelope'' is provided as a clue to the algorithm,
the typical-case time complexity drops further to
 ${\cal O}(c^2\profileSizeLimit^\alpha N)$,
with empirical tests indicating that $\alpha = 1.55$. 
That is, the alignment envelope brings the complexity to sub-quadratic with respect to 
sequence length. }
The  alignment envelope is not a hard constraint,
and may be controllably relaxed, or dispensed with altogether.

The new algorithm is, essentially, algebraically equivalent to Felsenstein's algorithm,
if the concept of a ``substitution matrix'' over a particular alphabet is extended to the countably-infinite set of all sequences over that alphabet.
Our chosen class of ``infinite substitution matrix'' is one that has a finite representation:
namely, the {\em finite-state transducer}, a probabilistic automaton that transforms an input sequence to an output sequence,
a familiar tool of statistical linguistics \cite{MohriPereiraRiley2000}.

In vector form, Felsenstein's pruning recursion is
\[
G_n = \left\{
\begin{array}{ll}
\left( B^{(l)} G_l \right) \fork \left( B^{(r)} G_r \right) & \mbox{if $n$ is not a leaf} \\
\recognize(y_n) & \mbox{if $n$ is a leaf}
\end{array}
\right.
\]
where $A \fork B$ is the pointwise (Hadamard) product
and $\recognize(x)$ is the unit column vector in dimension $x$.
By generalizing a few key algebraic ideas from matrices to transducers
(matrix multiplication, the pointwise product, row vectors and column vectors),
we are able to interpret this vector-space form of Felsenstein's algorithm
as the specification of a composite phylogenetic transducer
that spans all possible alignments (see \secref{EvidenceExpandedModel}).

The transducer approach offers a natural generalization of Felsenstein's pruning recursion to indels,
since it can be used to calculate
\[
P(S|T,\theta) = \sum_A P(S,A|T,\theta)
\]
i.e. the likelihood of sequences $S$ given tree $T$ and parameters $\theta$, summed over all alignments $A$.
Previous attempts to address indels phylogenetically have mostly returned $P(S|\hat{A},T,\theta)$ where $\hat{A}$ represents a single alignment
(typically estimated by a separate alignment program, which may introduce undetermined biases).
The exceptions to this rule are the ``statistical alignment'' methods \cite{HeinEtal2000,Hein2001,HolmesBruno2001,Metzler2003,SuchardRedelings2006}
which also marginalize alignments in an unbiased way---albeit more slowly, since they use Markov Chain Monte Carlo methods (MCMC).
In this sense, the new algorithm may be thought of as a fast, non-MCMC approximation to statistical alignment.

The purpose of this manuscript is a clean theoretical presentation of the algorithm.
In separate work \cite{WestessonEtAl2012} we find that the algorithm appears to recover more accurate reconstructions of simulated phylogenetic indel histories,
as indicated by proxy statistics such as the estimated indel rate.

The use of transducers in bioinformatics has been reported before \cite{Searls95,Holmes2003,BradleyHolmes2007,SatijaEtAl2008,PatenEtAl2008}
including an application to genome reconstruction that is conceptually similar to what we do here for proteins \cite{PatenEtAl2008}.
In particular, to maximize accessibility, we have chosen to use a formulation of finite-state transducers
that closely mirrors the formulation available on Wikipedia at the time of writing ({\tt \wikipediaTransducerUrl}).
This presentation is consistent with others described in the computer science literature \cite{MohriPereiraRiley2000}.

\subsection{Document structure}  
We will begin with a narrative, ``tutorial'' overview that introduces the main
 theoretical concepts using a small worked example.
Following this we will present general, precise, technical definitions.
The informal overview makes extensive use of illustrative figures,
and is intended to be easily read, even by someone not familiar with transducer theory. 
The technical description is intended primarily for those wishing to integrate the 
internals of this method into their own algorithms.  
Either section may be read in isolation.

Each example presented in this informal section
 (e.g. single transducers, composite transducers, sampled paths) 
correspond to rigorously-defined mathematical constructs defined in the technical section. 
Whenever possible, we provide references between the examples and their technical definitions.

The final section of the tutorial, \secref{FormalTutorialRelationship}, gives a detailed account of the connections between the tutorial and formal sections.

\section{Informal tutorial on transducer composition}

In this section we introduce (via verbal descriptions and graphical representations)
the various machines and manners of combining them necessary for the
task of modeling evolution on the tree shown in  \figref{cs-mf-liv-tree}.  
(The arrangement of machines required for this tree is shown in \figref{cs-mf-liv-machines}.)
While the conceptual underpinnings of our algorithm are not unusually complex, 
a complete mathematical description demands a significant amount of technical notation (which we provide in \secref{Formal}). 
For this reason, we aim to minimize notation in this section,
instead focusing on a selection of illustrative example machines ranging from simple to complex. 

We first describe the sorts of state machines used, beginning
 with simple linear machines
 (which appear at the leaves of the tree in \figref{cs-mf-liv-machines}) 
and moving on to the various possibilities of the branch model. 
Then we describe (and provide examples for) the techniques which allow us to 
co-ordinate these machines on a phylogeny: composition and intersection. 

Finally, we outline how combinations of these machines allow a straightforward definition
of Felsenstein's pruning algorithm for models allowing insertion/deletion events,
and a stochastic approximation technique which will allow inference on datasets
of common practical size.  

\subsection{Transducers as input-output machines}

We begin with a brief definition of transducers from Wikipedia.  
These ideas are defined with greater mathematical precision in \secref{Transducer}.

{\em A finite state transducer is a finite state machine with two tapes: an input tape and an output tape. ... An automaton can be said to } recognize {\em a string if we view the content of its tape as input. In other words, the automaton computes a function that maps strings into the set $\{0,1\}$. $(\dagger\dagger\dagger)$ Alternatively, we can say that an automaton } generates {\em strings, which means viewing its tape as an output tape. On this view, the automaton generates a formal language, which is a set of strings. The two views of automata are equivalent: the function that the automaton computes is precisely the indicator function of the set of strings it generates... Finite State Transducers can be weighted, where each transition is labeled with a weight in addition to the input and output labels. }
{\tt \wikipediaTransducerUrl}
\\
$(\dagger\dagger\dagger)$ For a weighted transducer this mapping is,
more generally, to the nonnegative real axis $[0,\infty)$
rather than just the binary set $\{0,1\}$.

In this tutorial section we are going to work through a small examples of using transducers on a tree
for three tiny protein sequences (MF, CS, LIV).
Specifically, we will compute the likelihood of the tree shown in \figref{cs-mf-liv-tree},
explaining the common descent of these three sequences
under the so-called TKF91 model (\figref{tkf91}),
as well as a simpler model that only allows point substitutions.
To do this we will construct (progressively, from the bottom up) the ensemble of transducer machines shown in \figref{cs-mf-liv-machines}.
We will see that the full state space of \figref{cs-mf-liv-machines} is equivalent to Hein's ${\cal O}(L^N)$ alignment algorithm for the TKF91 model \cite{Hein2001};
${\cal O}(NL^2)$ progressive alignment corresponds to a greedy Viterbi/maximum likelihood-traceback approximation,
and partial-order graph alignment corresponds to a Forward/stochastic-traceback approximation.

\pdffig{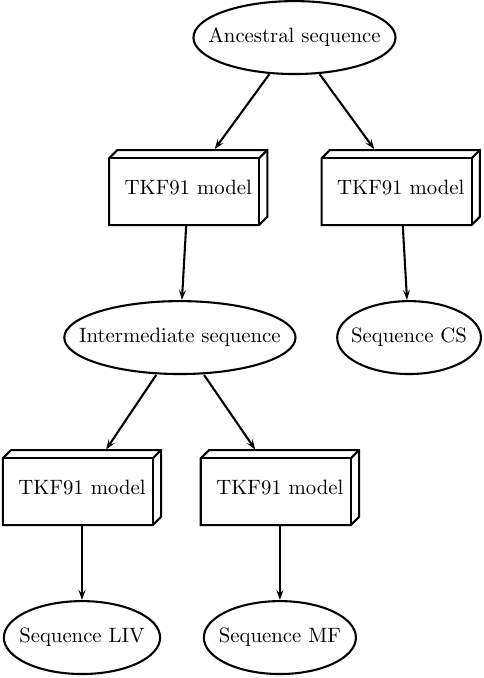}{Example tree used in this tutorial.  
The TKF91 model is used as the branch transducer model, but our approach 
is applicable to a wider range of string transducer models.  }

\pdffig{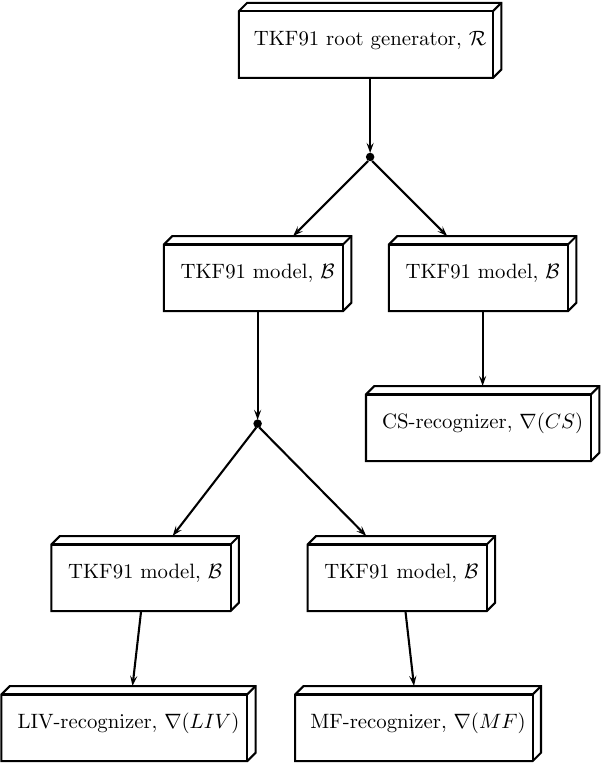}{An ensemble of transducers modeling the likelihood of the tree shown in \figref{cs-mf-liv-tree}.
We write this as $\tkfroot \cdot (\tkf \cdot (\tkf \cdot \recognize(LIV))\fork(\tkf \cdot \recognize(MF)))\fork(\tkf \cdot \recognize(CS))$.
The terms in this expression represent individual component transducers:
$\tkfroot$ is shown in \figref{tkf91-root},
$\tkf$ is shown in \figref{tkf91-labeled},
$\recognize(LIV)$ is in \figref{liv-labeled},
$\recognize(MF)$ in \figref{mf-labeled},
 and
$\recognize(CS)$ in \figref{cs-labeled}.
(The general notation $\recognize(\ldots)$ is
introduced in \secref{moore-genrec} and
formalized in \secref{ExactMatch}.)
The operations for combining these transducers, denoted ``$\cdot$'' and ``$\fork$'',
are---respectively---{\em transducer composition}
(introduced in \secref{Tutorial.Composition}, formalized in \secref{Composition})
and {\em transducer intersection}
(introduced in \secref{Tutorial.Intersection}, formalized in \secref{Fork}).
The full state graph of this transducer is not shown in this manuscript:
even for such a small tree and short sequences, it is too complex to visualize easily
(the closest thing is \figref{fork3-tkf91liv-tkf91mf-tkf91cs},
which represents this transducer configuration minus the root generator, $\tkfroot$).
 }

\subsubsection{Generators and recognizers}

As noted in the Wikipedia quote, transducers can be thought of as generalizations of the related
concepts of {\em generators} (state machines that emit output sequences, such as HMMs)
and parsers or {\em recognizers} (state machines that match/parse input sequences, such as the UNIX `lex' program).
Both generators and recognizers are separate special cases of transducers.
Of particular use in our treatment are generators/recognizers that generate/recognize a single unique sequence.
Generators and recognizers are defined with greater precision in \secref{GenSinglet} and \secref{ExactMatch}.

\figref{mf-generator} is an example of a generator that uniquely generates (inserts) the protein sequence MF.
\figref{liv-small} is an example of a recognizer that uniquely recognizes (and deletes) the protein sequence LIV.

\widepngfig{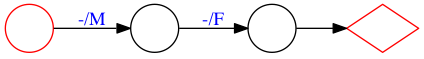}{Generator for protein sequence MF.  
This is a trivial state machine which emits (generates) the sequence MF with weight (probability) 1.
The red circle indicates the Start state, and the red diamond the End state.
}

\widepngfig{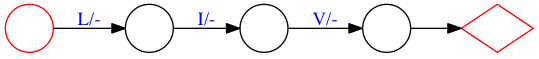}{Recognizer for protein sequence LIV.
This is a trivial state machine which absorbs (recognizes) the sequence LIV with weight (probability) 1, and all other sequences with weight 0.
The red circle indicates the Start state, and the red diamond the End state.
}

These Figures illustrate the visual notation we use throughout the illustrative Figures of this tutorial.
States and transitions are shown as a graph.
Transitions can be labeled with absorption/emission pairs,
written $x/y$ where $x$ is the absorbed character and $y$ the emitted character.
Either $x$ or $y$ is allowed to be the empty string (shown in these diagrams as the gap character, a hyphen).
In a Figure that shows absorption/emission pairs,
if there is no absorption/emission labeled on a transition, then it can be assumed to be $-/-$
(i.e. no character is absorbed or emitted) and the transition is said to be a ``null'' transition.

Some transitions are also labeled with weights.
If no transition label is present, the weight is usually 1
(some more complicated diagrams omit all the weights, to avoid clutter).
The weight of a path is defined to be the  product of transition weights occuring on the path. 

The weight of an input-output sequence-pair is the sum over all path weights that generate the
specified  input and output sequences.  
The weight of this sequence-pair can be interpreted as the joint probability of both
(a) successfully parsing the input sequence and
(b) emitting the specified output sequence.

Note sometimes this weight is zero---e.g. in \figref{liv-small} the weight is zero
except in the unique case that the input tape is LIV, when the weight is one---this
in fact makes \figref{liv-small} a special kind of recognizer:
one that only recognizes a single string
(and recognizes that string with weight one).
We call this an {\em exact-match} recognizer.

More generally, suppose that $G$, $R$ and $T$ are all probabilistically weighted finite-state transducers:
$G$ is a generator (output only), $R$ is a recognizer (input only) and $T$ is a general transducer (input {\em and} output).
Then, conventionally, $G$ defines a probability distribution $P(Y|G)$ over the emitted output sequence $Y$;
$R$ defines a probability $P(\mbox{accept}|X,R)$ of accepting a given input sequence $X$;
and $T$ defines a joint probability $P(\mbox{accept},Y|X,T)$
that input $X$ will be accepted and output $Y$ emitted.
According to this convention, it is reasonable to expect these weights to obey the following
(here $\Omega^\ast$ denotes the set of all sequences):
\begin{eqnarray*}
\sum_{Y \in \Omega^\ast} P(Y|G) & = & 1 \\
P(\mbox{accept}|X,R) & \leq & 1 \quad \forall X \in \Omega^\ast \\
\sum_{Y \in \Omega^\ast} P(\mbox{accept},Y|X,T) & \leq & 1 \quad \forall X \in \Omega^\ast
\end{eqnarray*}
{\bf It is important to state that these are just conventional interpretations of the computed weights:}
in principle the weights can mean anything we want,
but it is common to interpret them as probabilities in this way.

Thus, as noted in the Wikipedia quote, generators and recognizers are in some sense equivalent,
although the probabilistic interpretations of the weights are slightly different.
In particular, just as we can have a {\em generative profile}
that generates some sequences with higher probability than others (e.g. a profile HMM)
we can also have a {\em recognition profile}: a transducer
that recognizes some sequences with higher probability than others.
The exact-match transducer of \figref{liv-small} is a (trivial and deterministic) example of such a recognizer;
later we will see that the stored probability vectors in the Felsenstein pruning recursion can also
be thought of as recognition profiles.

\subsection{Moore machines}

In our mathematical descriptions, we will treat transducers as
{\em Mealy machines}, meaning that absorptions and emissions are associated with transitions between states.  
In the {\em Moore machine} view, absorptions and emissions are associated with states.  
In our case, the distinction between these two is primarily a semantic one, since the structure 
of the machines and the I/O functions of the states is intimately tied.

The latter view (Moore) can be more useful in bioinformatics,
where rapid point substitution means that all combinations of input and output characters 
are possible.  
In such situations, the Mealy type of machine can suffer from an excess of transitions, complicating the presentation.  
For example, consider the Mealy-machine-like view of  \figref{fanned-emission},
and compare it with the more compact Moore-machine-like view of \figref{condensed-emission}.  

The majority of this paper deals with Moore machines.
However, \appref{Mealy} reformulates the key algorithms of transducer composition (\secref{Composition})
and transducer intersection (\secref{Fork}) as Mealy machines.

\sidewayspngfig{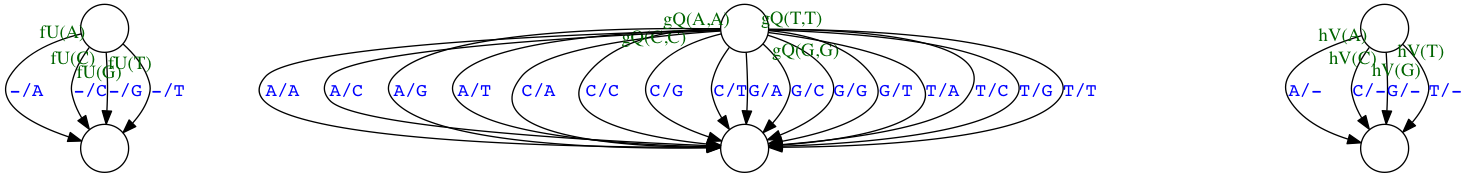}{All combinations of input and output characters are frequently observed.  
The transitions in this diagram include all possible deletion, insertion, and substitution transitions between a pair of transducer states.
Each transition is labeled with I/O characters (blue) and selected transitions are labeled with the transition weights (green).
Insertions (I/O label ``{\tt -/y}'') have weight $fU(y)$,
deletions (I/O label ``{\tt x/-}'') have weight $hV(x)$,
and substitutions (I/O label ``{\tt x/y}'') have weight $gQ(x,y)$.
The large number of transitions complicates the visualization of such ``Mealy-machine'' transducers.
We therefore use a ``Moore-machine'' representation, where all transitions of each type between a pair of states
are collapsed into a single transition, and I/O weights are associated with states (\figref{condensed-emission}).
}

\widepngfig{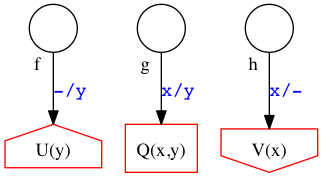}
{In a condensed Moore-machine-like representation, possible combinations of input and output characters are encoded in the distributions contained within each state,
simplifying the display.  
In this diagram, all four insertion transitions from \figref{fanned-emission} ({\tt -/A}, {\tt -/C}, etc.) are collapsed into a single {\tt -/y} transition;
similarly, all four deletion transitions from \figref{fanned-emission} ({\tt A/-}, etc.) are collapsed into one {\tt x/-},
and all sixteen substitution transitions ({\tt A/A}, {\tt A/C}, \ldots {\tt G/G}) are collapsed to one {\tt x/y}.
To allow this, the transition weights for all the collapsed transitions must factorize into independent I/O- and transition-associated components.
In this example (corresponding to \figref{fanned-emission}), the I/O weights are $U(y)$ for insertions, $Q(x,y)$ for substitutions and $V(x)$ for deletions;
while the transition weights are $f$ for insertions, $g$ for substitutions, and $h$ for deletions.
{\bf Visual conventions.}
The destination state node is bordered in red, to indicate that transitions into it have been collapsed.
Instead of just a plain circle, the node shape is an upward house (insertions), a downward house (deletions), or a rectangle (substitutions).
Instead of specific I/O character labels ({\tt A/G} for a substitution of A to G, {\tt -/C} for an insertion of C, etc.)
we now have generic labels like {\tt x/y} representing the set of all substitutions; the actual characters (and their weights) are encoded by the I/O functions.
For most Figures in the remainder of this manuscript, we will omit these blue generic I/O labels,
as they are implied by the node shape of the destination state.}

\figref{legend} shows the visual notation we use in this tutorial for Moore-form transducer state types.
There are seven {\em state types}: Start, Match, Insert, Delete, End, Wait, and Null, frequently abbreviated to $S,M,I,D,E,W,N$.
State types are defined precisely in \secref{StateTypes}.
Note that in the TKF91 model (\figref{tkf91}, the example we use for most of this tutorial)
there are exactly one of each of these types, but this is not a requirement.
For instance,
the transducer in \figref{protpal-mix2} has two Insert, two Delete and three Wait states,
while \figref{substituter} has no Insert or Delete states at all;
\figref{moore-mf-generator} has no Match states; and so on.

\widepdffig{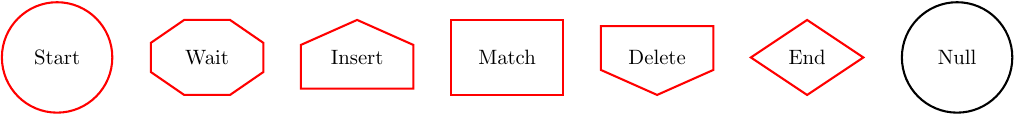}{In a Moore machine, each state falls into one of several {\em types} (a transducer may contain more than one state of each type).  
A state's type determines its I/O capabilities:
 an Insert state emits (writes) an output character,
 a Delete state absorbs (reads) an input character,
and
 a Match state both absorbs an input character and emits an output character.
Mnemonics: Insert states point upwards, Delete states point Downwards, Wait states are octagonal like U.S. Stop signs.
}

Some other features of this view include the following:
\begin{itemize}
\item The shape and color of states indicates their type.
The six Moore normal form states are all red.
Insert states point upwards; Delete states point downwards; Match states are rectangles; Wait states are octagons (like U.S. Stop signs);
Start is a circle and End is a diamond.
There is a seventh state type, Null, which is written as a black circle to distinguish it from the six Moore-form state types.
(Null states have no associated inputs or outputs; they arise as a side-effect of algorithmically-constructed transducers in \secref{Tutorial.Composition} and \secref{Tutorial.Intersection}.
In practice, they are nuisance states that must be eliminated by marginalization, or otherwise dealt with somehow.)
This visual shorthand will be used throughout.
\item We impose certain constraints on states that involve I/O:
 they must be typed  as Insert, Delete, or Match, and
 their type determines what kinds of I/O happens on transitions into those states (e.g. a Match state always involves an absorption and an emission).
\item We impose certain constraints on transitions into I/O states:
 their weights must be factorizable into transition and I/O components.
 Suppose $j$ is a Match state and $i$ is a state that precedes $j$;
 then all transitions $i \to j$ must both absorb a non-gap input character $x$
 and emit a non-gap output character $y$,
 so the transition can be written $i \stackrel{x/y}{\longrightarrow} j$
 and the transition weight must take the form $t_{ij} \times e_j(x,y)$
 where $t_{ij}$ is a component that can depend on the source and destination state
  (but not the I/O characters)
 and $e_j(x,y)$ is a component that can depend on the I/O characters and the destination state
  (but not the source state).
\item We can then associate the ``{\em I/O weight function}'' $e_j$ with Match state $j$
 and the ``{\em transition weight}'' $t_{ij}$ with a single conceptual transition $i \to j$
 that summarizes all the transitions $i \stackrel{x/y}{\to} j$
 (compare \figref{fanned-emission} and \figref{condensed-emission}).
\item The function $e_j$ can be thought of as a conditional-probability substitution matrix
 (for Match states, c.f. $Q$ in \figref{condensed-emission}),
a row vector representing a probability distribution
 (for Insert states, c.f. $U$ in \figref{condensed-emission}),
or a column vector of conditional probabilities
 (for Delete states, c.f. $V$ in \figref{condensed-emission}).
\item Note that we call $e_j$ an ``I/O function'' rather than an ``emit function''.
The latter term is more common in bioinformatics HMM theory;
however, $e_j$ also describes probabilistic weights of {\em absorptions} as well as {\em emissions},
and we seek to avoid ambiguity.
\end{itemize}

\pngfig{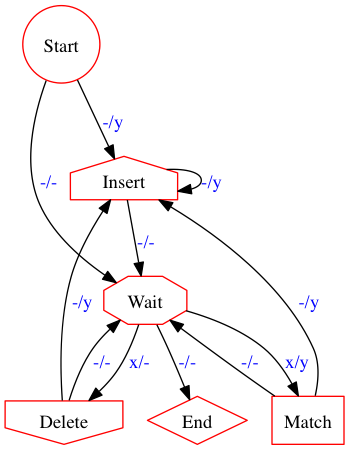}
{Allowed transitions between types of transducer states, along with their I/O requirements.
In particular, note the Wait state(s) which must precede all 
absorbing (Match and Delete) states---the primary departure from the familiar pair HMM
structure.  
Wait states are useful in co-ordinating multiple connected trandsucers, since they indicate that
the transducer is ``waiting'' for an upstream transducer to emit a character before entering an
absorbing state. 
Also note that this graph is not intended to depict a {\em particular} state machine, 
but rather it shows the transitions which are permitted between the {\em types}
of states of arbitrary machines under our formalism.  
Since the TKF91 model (\figref{tkf91-labeled}) contains exactly one state of each type,
 its structure is similar to this graph 
(but other transducers may have more or less than one state of each type).  }

\figref{transitions} shows the allowed types of transition in Moore-normal form transducers.
In our ``Moore-normal form'' for transducers, we require that all input states (Match, Delete)
are immediately preceded in the transition graph by a Wait state.
This is useful for co-ordinating multiple transducers connected together as described in later sections,
since it requires
the transducer to ``wait'' for a parent transducer to emit a character before entering an
absorbing state. 
The downside is that the state graph sometimes contains a few  Wait states which appear redundant
(for example, compare \figref{moore-mf-generator} with \figref{mf-generator},
or \figref{liv-labeled} with \figref{liv-small}).
For most Figures in the remainder of this manuscript, we will leave out the blue ``x/y'' labels on transitions,
as they are implied by the state type of the destination state.

Note also that this graph depicts the transitions between {\em types} of states  allowed under our formalism,
rather than a {\em particular} state machine.  
It happens that the TKF91 model (\figref{tkf91-labeled}) contains exactly one state of each type,
so its state graph appears similar to \figref{transitions},
but this is not true of all transducers.

\subsubsection{Generators and recognizers in Moore normal form}
\seclabel{moore-genrec}
We provide here several examples of small transducers in Moore normal form,
including versions of the transducers in \figref{mf-generator}
and \figref{liv-small}.

We introduce a notation for generators ($\generate$) and recognizers ($\recognize$);
a useful mnemonic for this notation (and for the state types in \figref{legend})
is ``insertions point up, deletions point down''.

\begin{itemize}

\item \figref{moore-mf-generator} uses our Moore-machine visual representation
to depict the generator in \figref{mf-generator}.
We write this transducer as $\generate(MF)$.

\item \figref{liv-labeled} is a Moore-form recognizer for sequence LIV.
We write this transducer as $\recognize(LIV)$.
The state labeled $\delta_Z$  (for $Z \in \{L,I,V\}$) has I/O function $\delta(x=Z)$,
defined to be 1 if $x=Z$, 0 otherwise.
The machine recognizes sequence LIV with weight 1, and all other sequences with weight 0. 

\item \figref{mf-labeled} is the Moore-machine recognizer for MF,
the same sequence whose generator is shown in \figref{moore-mf-generator}.
We write this transducer as $\recognize(MF)$.

\item \figref{cs-labeled} is the Moore-machine recognizer for sequence CS.
We write this transducer as $\recognize(CS)$.

\item \figref{null-model} is a ``null model'' generator that emits a single IID sequence
(with residue frequency distribution $\pi$)
of geometrically-distributed length (with geometric parameter $p$).

\end{itemize}

\widepdffig{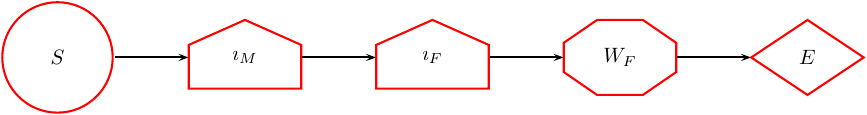}{Transducer $\generate(MF)$, the Moore-normal form generator for protein sequence MF.
The states are labeled $S$ (Start), $E$ (End),
$\imath_M$ and $\imath_F$ (Insert states that emit the respective amino acid symbols),
and $W_F$ (a Wait state that pauses after emitting the final amino acid;
this is a requirement imposed by our Moore normal form).
The state labeled $\imath_Z$ (for $Z \in \{M,F\}$) has I/O function $\delta(y=Z)$.}

\widepdffig{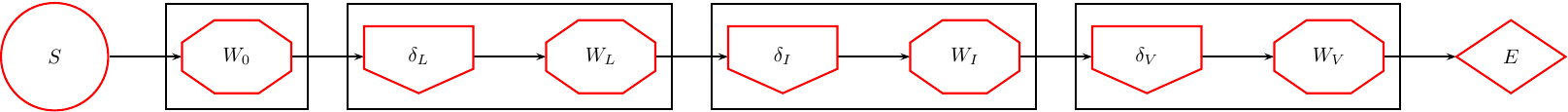}{Transducer $\recognize(LIV)$, the Moore-normal form recognizer for protein sequence LIV.
The states are labeled $S$ (Start), $E$ (End),
$\delta_L$, $\delta_I$ and $\delta_V$ (Delete states that recognize the respective amino acid symbols),
$W_L$, $W_I$ and $W_V$ (Wait states that pause after recognizing each amino acid;
these are requirements imposed by our Moore normal form).
The states have been grouped (enclosed by a rectangle) to show four clusters:
 states that are visited before any of the sequence has been recognized,
 states that are visited after ``L'' has been recognized,
 states that are visited after ``I'' has been recognized,
and
 states that are visited after ``V'' has been recognized.
The I/O function associated with each Delete state $\delta_Z$ is $\delta(x=Z)$.}
\widepdffig{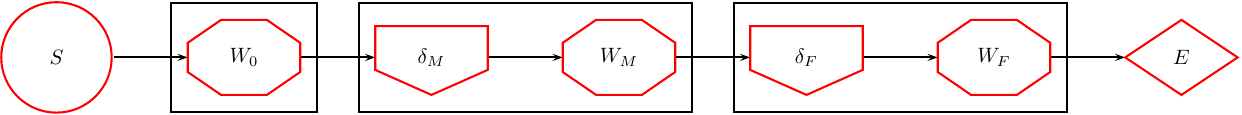}{Transducer $\recognize(MF)$, the Moore-normal form recognizer for protein sequence MF.
The states are labeled $S$ (Start), $E$ (End),
$\delta_M$ and $\delta_F$ (Delete states that recognize the respective amino acid symbols),
$W_M$ and $W_F$ (Wait states that pause after recognizing each amino acid;
these are requirements imposed by our Moore normal form).
The states have been grouped (enclosed by a rectangle) to show four clusters:
 states that are visited before any of the sequence has been recognized,
 states that are visited after ``M'' has been recognized,
and
 states that are visited after ``F'' has been recognized.
The I/O function associated with each Delete state $\delta_Z$ is $\delta(x=Z)$.}
\widepdffig{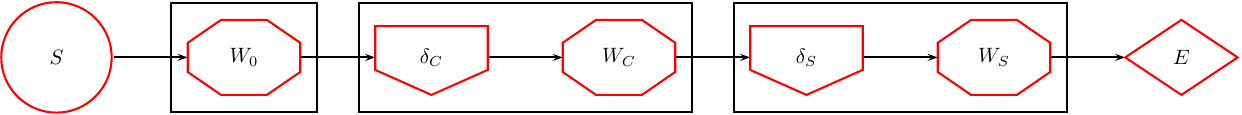}{Transducer $\recognize(CS)$, the Moore-normal form recognizer for protein sequence CS.
The states are labeled $S$ (Start), $E$ (End),
$\delta_C$ and $\delta_S$ (Delete states that recognize the respective amino acid symbols),
$W_C$ and $W_S$ (Wait states that pause after recognizing each amino acid;
these are requirements imposed by our Moore normal form).
The states have been grouped (enclosed by a rectangle) to show four clusters:
 states that are visited before any of the sequence has been recognized,
 states that are visited after ``C'' has been recognized,
and
 states that are visited after ``S'' has been recognized.
The I/O function associated with each Delete state $\delta_Z$ is $\delta(x=Z)$.}
\pdffig{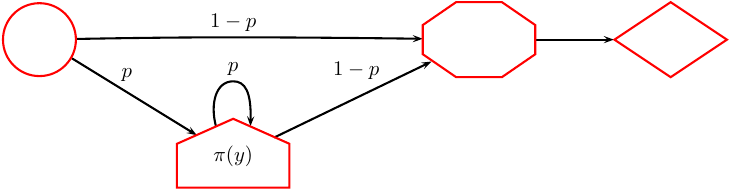}{Transducer $\nullmodel$, a simple null-model generator with geometric length parameter $p$ and residue frequency distribution $\pi$.}

\subsubsection{Substitution and identity}
\figref{substituter} shows how the Moore-normal notation can be  used to represent a substitution matrix.  
The machine pauses in the Wait state before absorbing each residue $x$ and emitting a residue $y$
according to the distribution $Q_{xy}$.  
Since there are no states of type Insert or Delete, the output sequence will necessarily
be the same length as the input.  

This is something of a trivial example, since it is certainly not necessary to use transducer machinery
to model point substitution processes. Our aim is to show explicitly how a familiar simple case
(the Felsenstein algorithm for point substitution) is represented using the more elaborate transducer notation.

\pdffig{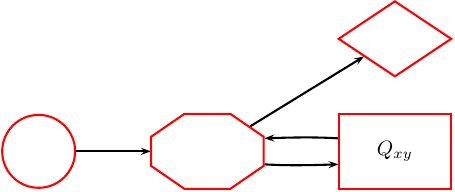}{Transducer $\substitute(Q)$ (``the substituter'') introduces substitutions (according to substitution matrix $Q$) but no indels.  
Whenever the machine makes a transition into the rectangular Match state, a character $x$ is read from the input and a character $y$ emitted to the output,
with the output character sampled from the conditional distribution $P(y|x) = Q_{xy}$.}

\figref{identity} shows the identity transducer, $\identity$, a special case of the substituter: $\identity = \substitute(\delta)$, where

\[
\delta_{xy} = \left\{
\begin{array}{ll}
1 & x=y \\
0 & x \neq y
\end{array}
\right.
\]
The identity essentially copies its input to its output directly.
It is defined formally in \secref{Identity}. 

\pdffig{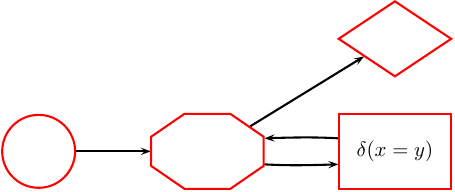}{Transducer $\identity$, the identity transducer, 
simply copies the input tape to the output tape.  This can be thought of as a special case of the substituter transducer in \figref{substituter}
where the substitution matrix is the identity matrix.
\formaldefs
Transducer $\identity$ is defined in \secref{Identity}.
}

\subsubsection{The TKF91 model as a transducer}
\seclabel{tkf91}

We use the TKF91 model of indels \cite{ThorneEtal91} as an example, not because it is the best model of indels
 (it has deficiencies, most notably the linear gap penalty);
rather, because it is canonical, widely-known, and illustrative of the general properties of transducers.  

The TKF91 transducer with I/O functions is shown in \figref{tkf91}.  
The underlying continuous-time indel model has insertion rate $\lambda$ and deletion rate $\mu$.
The transition weights of the transducer,
modeling the stochastic transformation of a sequence over a finite time interval $t$,
are
$a = \exp(-\mu t)$ is the ``match probability'' (equivalently, $1-a$ is the deletion probability),
$b = \lambda\frac{1 - a \exp(\lambda t)}{\mu - a \lambda \exp(\lambda t)}$ is the probability of an insertion at the start of the sequence or following a match,
and $c = \frac{\mu b}{\lambda(1-a)}$ is the probability of an insertion following a deletion. 
These may be derived from analysis of the underlying birth-death process \cite{ThorneEtal91}.

The three I/O functions for Match, Delete and Insert states are defined as follows:
conditional on absorbing character $x$, the Match state emits  $y$ with probability
$exp(Rt)_{xy}$, where $R$ is the rate matrix governing character substitutions and 
$t$ is the time separating the input and output sequences. 
Characters are all deleted with the same weight: once a Delete state is entered the 
absorbed character is deleted with weight 1 regardless of its value.  
Inserted characters $y$ are emitted according to the equilibrium distribution $\pi_y$.

In the language of \cite{ThorneEtal91}, every state path contains a Start$\to$Wait segment that corresponds to the ``immortal link'',
and every Wait$\to$Wait tour corresponds to an ``mortal link''.

\widepdffig{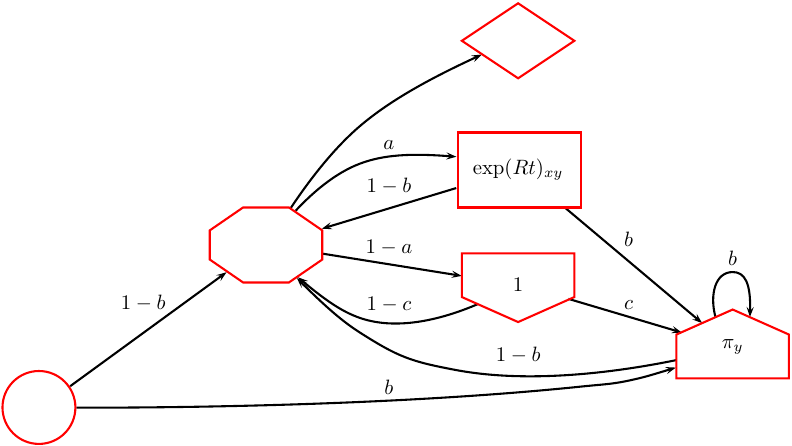}{The TKF91 transducer, labeled with transition weights and I/O functions.  
This transducer models the sequence transformations of the TKF91 model \cite{ThorneEtal91}.
The transition weights $a,b,c$ are defined in \secref{tkf91}.
\figref{tkf91-labeled} shows the same transducer with state type labels rather than I/O functions.}


\widepdffig{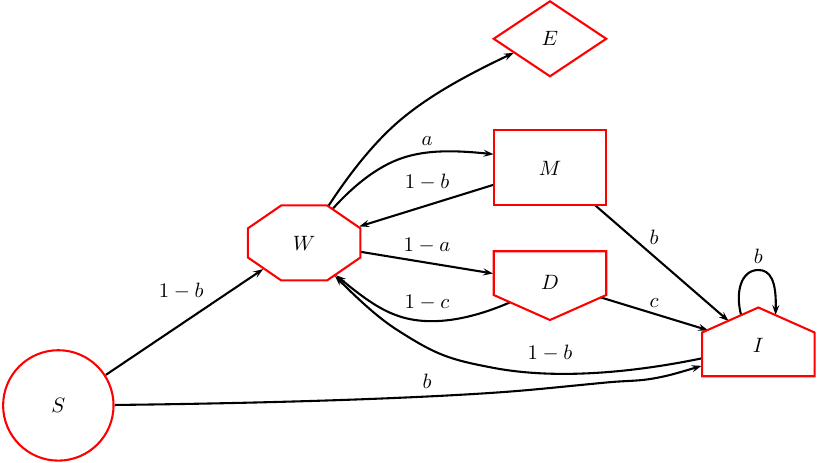}{
Transducer $\tkf$, the TKF91 model on a branch of length $t$.
This transducer models the sequence transformations of the TKF91 model \cite{ThorneEtal91}.
The machine shown here is identical to that of \figref{tkf91} in nearly all respects.
The only difference is this: in order that we can later refer to the states by name,
rather than writing the I/O weight functions directly on each state
we have instead written a state type label
 $S,M,I,D,E,W$ (Start, Match, Insert, Delete, End, Wait).
It so happens that the TKF91 transducer has one of each of these kinds of state.
Identically to \figref{tkf91}, the transition weights $a,b,c$ are defined in \secref{tkf91}.
For each of the I/O states ($I$, $D$ and $M$) we must, of course, still specify an I/O weight function.
These are also identical to \figref{tkf91}, and are reproduced here for reference:
 $\exp(Rt)$ is the substitution matrix for the $M$ (Match) state,
 $\pi$ is the vector of weights
  corresponding to the probability distribution of inserted characters for the $I$ (Insert) state,
 and
 $(1,1,\ldots,1)$
 is the vector of weights corresponding to
 the conditional probabilities that any given character will be deleted by the $D$ (Delete) state
 (in the TKF91 model, deletion rate is independent of the actual character being deleted,
 which is why these delete weights are all 1).
}

\figref{tkf91-labeled} is a version of \figref{tkf91}
where, rather than writing the I/O weight functions directly on each state (as in \figref{tkf91}),
we have instead written a state label (as in \figref{liv-labeled}, \figref{mf-labeled} and \figref{cs-labeled}).
The state labels are $S,M,I,D,E,W$ (interpretation: Start, Match, Insert, Delete, End, Wait).

\widepdffig{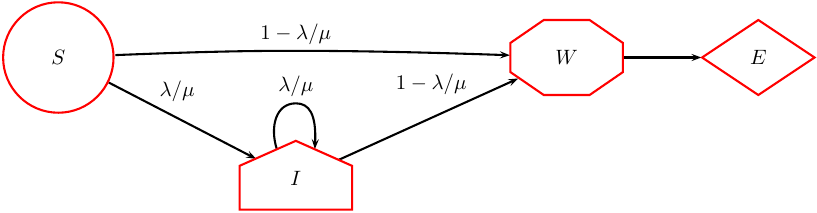}{Transducer $\tkfroot$, the equilibrium TKF91 generator.
The equilibrium distribution for the TKF91 model is essentially the same generator as \figref{null-model} with $p=\lambda/\mu$.
The I/O function for the $I$ (Insert) state is $\pi_y$.
Note that this is the limit of \figref{tkf91} on a long branch with empty ancestor:
$\tkfroot = \lim_{t\to\infty} \left( \generate(\epsilon) \cdot \tkf(t) \right)$.
}

It has been shown that affine-gap versions of the TKF91 model can also be represented using state machines \cite{MiklosLunterHolmes2004}.
An approximate affine-gap transducer, based on the work of Knudsen \cite{KnudsenMiyamoto2003}, Rivas \cite{Rivas05} {\em et al}, is shown in \figref{protpal};
a version that approximates a two-component mixture-of-geometric indel length distributions is shown in \figref{protpal-mix2},
while a transducer that only inserts/deletes in multiples of 3 is in \figref{triplet}.

\widepdffig{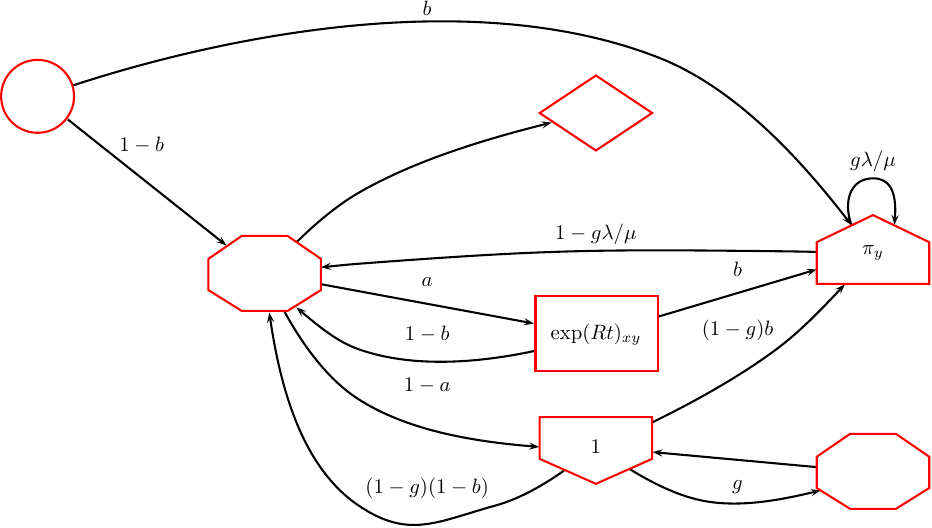}{An approximate affine-gap transducer.
Note that in contrast to \figref{tkf91}, this machine requires {\em two} Wait states,
with the extra Wait state serving to preserve the ``memory'' of being inside a deletion.
The parameters of this model are insertion rate $\lambda$, deletion rate $\mu$, gap extension probability $g$,
substitution rate matrix $R$, inserted residue frequencies $\pi$ (typically the equilibrium distribution, so $\pi R = 0$), and branch length $t$.
The transition weights use the quantities
$a = \exp(-\mu t)$ and $b = 1 - \exp(-\lambda t)$.
The transducer in \figref{tkf91-root} serves as a suitable root generator.
}

\widepdffig{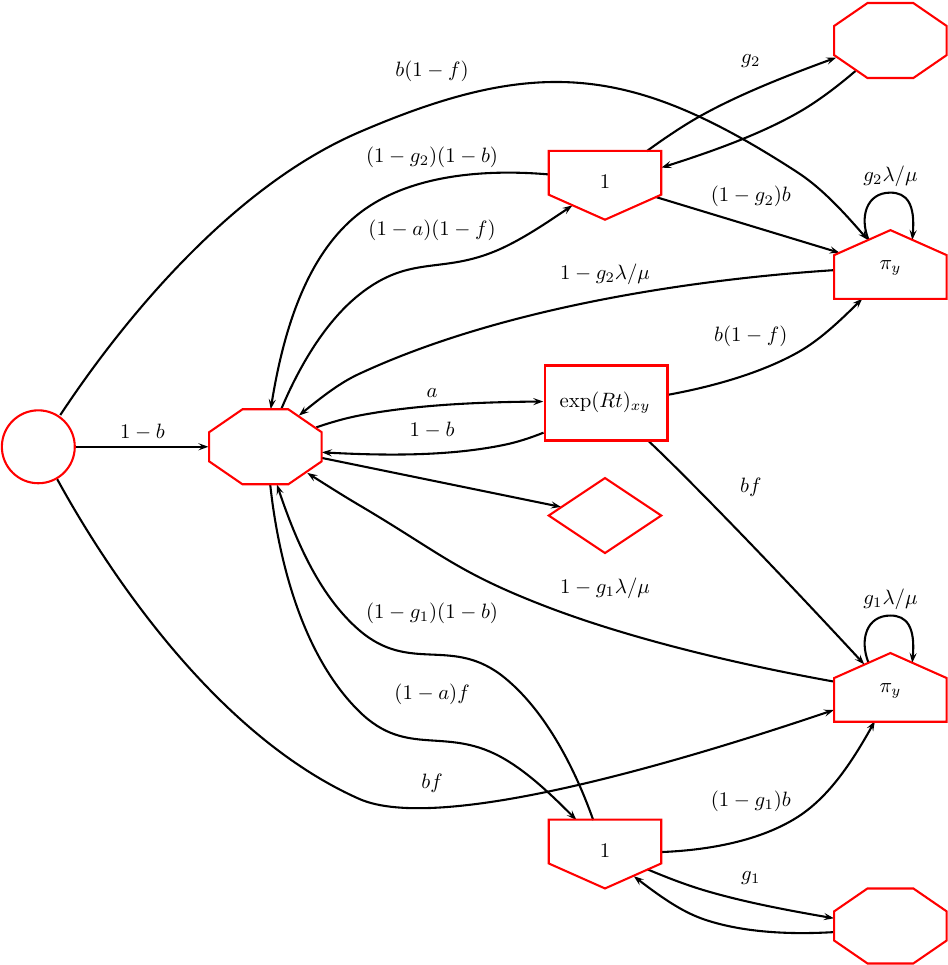}{A transducer that models indel length distributions as a 2-component mixture of geometric distributions.
Note that this machine requires
two separate copies of the Insert and Delete states, along with {\em three} Wait states,
with two of the Wait states serving to preserve the ``memory'' of being inside a deletion from the respective Delete states.
The parameters of this model are insertion rate $\lambda$, deletion rate $\mu$, gap extension probabilities $g_1$ and $g_2$ for the two mixture components,
first component mixture strength $f$, substitution rate matrix $R$, inserted residue frequencies $\pi$ (typically the equilibrium distribution, so $\pi R = 0$), and branch length $t$.
The transition weights use the quantities
$a = \exp(-\mu t)$ and $b = 1 - \exp(-\lambda t)$.
The transducer in \figref{tkf91-root} serves as a suitable root generator.
}

\widepdffig{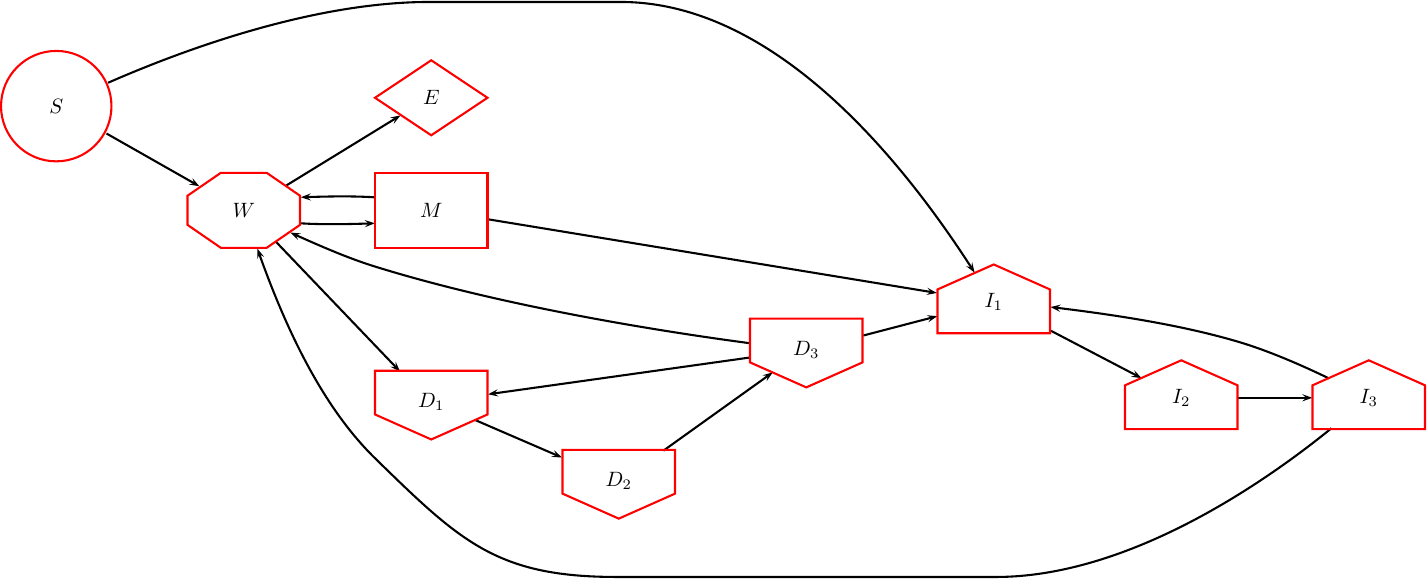}{A transducer whose indels are a multiple of 3 in length.
(This is not in strict normal form, but could be made so by adding a Wait state after every Delete state.)}

\subsubsection{A generator for the equilibrium distribution of the TKF91 model}
TKF91 is an input/output transducer that
operates on individual branches.  In order to arrange this model on a phylogenetic tree,
we need a generating transducer at the root.  
Shown in \figref{tkf91-root} is the equilibrium TKF91 generator, conceptually the same as
\figref{null-model}, with insertion weight $\lambda/\mu$
(which is the parameter of the geometric length distribution of sequences at equilibrium under the TKF91 model).

We have now defined all the component transducers we need to model evolution along a phylogeny. 
Conceptually, we can imagine the equilibrium machine generating a sequence at the root of the
tree which is then repeatedly passed down along the branches, each of which mutates the sequence
via a TKF91 machine (potentially introducing substitutions and indels). 
The result is a set of sequences (one  for each node),
  whose evolutionary history is recorded via  
the actions of each TKF91  branch machine.  
What remains is to detail the various ways of connecting transducers by
 constraining some of their tapes to be the same---namely {\em composition} and {\em intersection}.

\subsection{Composition of transducers}
\seclabel{Tutorial.Composition}
The first way of connecting transducers that we will consider is  ``composition'':
feeding the output of one transducer, $T$, into the input of another, $U$.
The two transducers are now connected, in a sense, since the output of $T$ must be
 synchronized with inputs from $U$: when $T$ emits a character, $U$ must be ready to absorb a character.  

We can equivalently  consider the two connected transducers as
a single, {\em composite} machine that represents
 the connected ensemble of $T$ followed by $U$.  
From two transducers $T$ and $U$,
we make a new transducer, written $TU$ (or $T \cdot U$),
wherein every state corresponds to a pair $(t,u)$ of $T$- and $U$-states.  
 
In computing the weight of an input/output sequence pair $(X,Y)$, where the input sequence $X$ is fed to $T$ and
output $Y$ is read from $U$, we must sum over all state paths through the composite machine $TU$ that are consistent with $(X,Y)$.
In doing so, we are effectively summing over the intermediate sequence
(the output of $T$, which is the input of $U$),
just as we sum over the intermediate index when doing matrix multiplication.
In fact, matrix multiplication of I/O functions is an explicit part of the algorithm that we use to automatically construct
composite transducers in \secref{Composition}.
The notation ($TU$ or $T \cdot U$) reflects the fact that transducer composition is  directly 
analogous to  matrix multiplication.  

Properties of composition include that
if $T$ is a generator then $TU$ is also a generator, while
if $U$ is a recognizer then $TU$ is also a recognizer.
A formal definition of transducer composition,
together with an algorithm for constructing the composite transducer $TU$,
is presented in \secref{Composition}. 
This algorithmic construction of $TU$
(which serves both as a proof of the existence of $TU$, and an upper bound on its complexity)
is essential as a formal means of verifying our later results.

\subsubsection{Multiplying two substitution models}

\pdffig{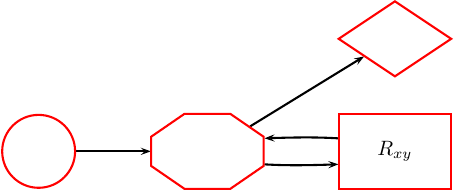}{Transducer $\substitute(R)$ introduces substitutions (rate matrix $R$) but no indels. Compare to \figref{substituter}, which is the same model but with substitution matrix $Q$ instead of $R$.}

\pdffig{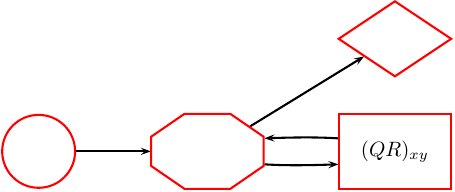}{Transducer $\substitute(Q) \cdot \substitute(R)$, the composition of \figref{substituter} and \figref{substituter2}.
Note that this is just $\substitute(QR)$: in the simple case of substituters, transducer composition is {\em exactly} the same as multiplication of substitution matrices.
In the more general case, transducer composition is always analogous to matrix multiplication,
though (unlike matrices) transducers tend in general to require more representational storage space when multiplied together
(not the case here, but see e.g. \figref{tkf91-tkf91}).}

As a simple example of transducer composition, we turn to the simple substituting transducers
in \figref{substituter} and \figref{substituter2}.  
Composing these two in series models two consecutive branches 
 $x \stackrel{Q}{\to} y \stackrel{R}{\to} z$, with 
the action of each branch modeled by a different substitution matrix, $Q$ and $R$.  

Constructing the $T \cdot U$ composite machine 
 (shown in \figref{substituter-substituter2})
simply involves constructing a machine whose Match state I/O function
 is the matrix product of the component substitution matrices, $QR$.  
If $x$ denotes the input symbol to the two-transducer ensemble (input into the Q-substituter),
$y$ denotes the output symbol from the two-transducer ensemble (output from the R-substituter),
and $z$ denotes the unobserved intermediate symbol (output from Q and input to R),
then the I/O weight function for the composite state QR in the two-transducer ensemble is
\[
(QR)_{xy} = \sum_z Q_{xz} R_{zy}
\]

\subsubsection{Multiplying two TKF91 models}
\seclabel{tkf91-tkf91}

\widepdffig{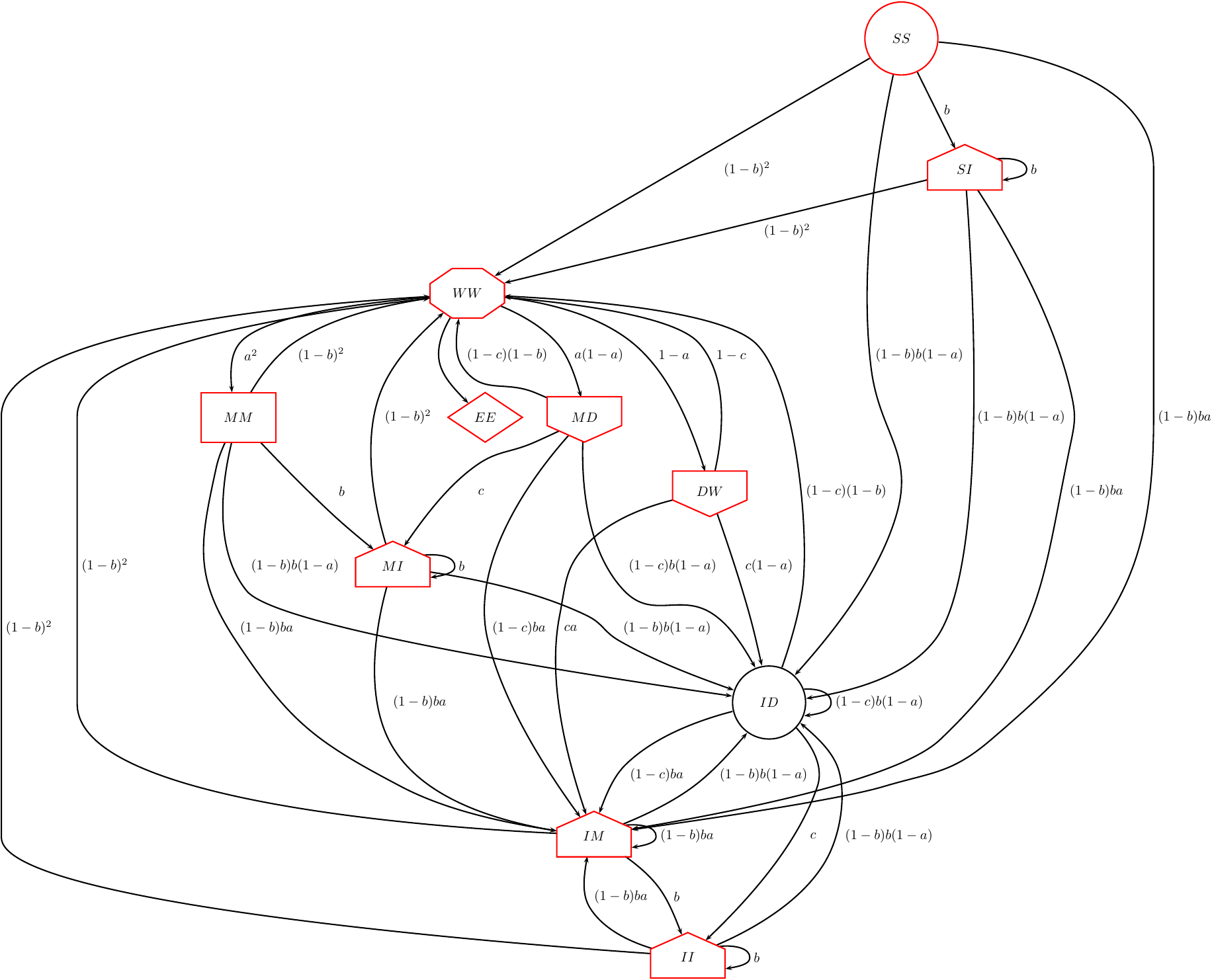}{Transducer $\tkf \cdot \tkf$, the composition of the 
TKF91 model (\figref{tkf91-labeled}) with itself, representing the evolutionary drift
 over two consecutive branches of the phylogenetic tree.
Each composite state is labeled with the states of the two component transducers.  
I/O weight functions, state labels, and meanings of transition parameters are explained
in \tabref{tkf91-tkf91} and \secref{tkf91-tkf91}.
\formaldefs
The algorithmic construction of the composite transducer is described in \secref{Composition}.
}

For a more complex example, consider the composition of a TKF91 transducer with itself.  
This is again like two consecutive branches $x \to y \to z$, 
but now TKF91 is acting along each branch, rather than a simple substitution process. 
An input sequence is fed into first TKF91; 
the output of this first TKF91 transducer is an intermediate sequence
 that is fed into the input of the second TKF91;
output of this second TKF91 is the output of the entire ensemble.

The composite machine inputs a sequence and outputs a sequence, just like the machine
in  \figref{substituter-substituter2}, but these sequences may differ by insertions and deletions
as well as substitutions. 
The intermediate sequence (emitted by the first TKF91 and absorbed by the second) is unobserved,
and whenever we sum over state paths through the composite machine, we are essentially summing over
values for this intermediate sequence. 

\figref{tkf91-tkf91} shows the state space of this transducer.  
The meaning of the various states in this model are shown in \tabref{tkf91-tkf91}.

\begin{table}
\begin{tabular}{c|p{.65\textwidth}|p{.15\textwidth}}
$b_1 b_2$ & Meaning & I/O fn. \\
\hline
$SS$ & Both transducers are in the Start state.  No characters are absorbed or emitted. &  \\
$EE$ & Both transducers are in the End state.  No characters are  absorbed or emitted. &  \\
$WW$ & Both transducers are in the Wait state.  No characters are absorbed or emitted, but both are poised to absorb sequences from their input tapes. &  \\
$SI$ & The second transducer inserts a character $y$ while the first remains in the Start state.  & $\pi_y$ \\
$IM$ & The first transducer  emits a character (via the Insert state) which the second transducer absorbs, mutates and re-emits as character $y$ (via a Match state).  & $(\pi \exp(Rt))_y$ \\
$MM$ & The first transducer absorbs character $x$ (from the input tape), mutates and emits it (via the Match state); the second transducer absorbs the output of the first, mutates it again, and emits it as $y$ (via a Match state).  & $\exp(2Rt)_{xy}$ \\
$ID$ & The first transducer  emits a character (via the Insert state) which the second transducer absorbs and deletes (via its Delete state).  &  \\
$MD$ & The first transducer  absorbs a character $x$, mutates and emits it (via the Match state) to the second transducer, which absorbs and deletes it (via its Delete state).  & $1$ \\
$DW$ & The first transducer absorbs and deletes a character $x$, whereas the second transducer idles in a Wait state (since it has recieved no input from the first tranducer).   & $1$ \\
$II$ & The second transducer inserts a character $y$ while the first remains in the Insert state from a previous insertion.  Only the second transducer is changing state (and thus emitting a character) here---the first is resting in an Insert state from a previous transition.   & $\pi_y$ \\
$MI$ & The second transducer inserts a character $y$ while the first remains in the Match state from a previous match.  Only the second transducer is changing state (and thus emitting a character) here---the first is resting in a Match state from a previous transition. & $\pi_y$ \\
\end{tabular}
\caption{ \tablabel{tkf91-tkf91} Meanings (and, where applicable, I/O functions) of all states in
transducer $\tkf \cdot \tkf$ (\figref{tkf91-tkf91}),
the composition of two TKF91 transducers
(an individual TKF91 transducer is shown in \figref{tkf91-labeled}).
Every state corresponds to a tuple $(b_1,b_2)$
where
$b_1$ is the state of the first TKF91 transducer and
$b_2$ is the state of the second TKF91 transducer.
 }
\end{table}

The last two states in \tabref{tkf91-tkf91}, $II$ and $MI$, appear to contradict the co-ordination of the machines.
Consider, specifically, the state $MI$:
the first transducer is in a state that produces output ($M$)
while the second transducer is in a state which does not receive input ($I$).
The solution to this paradox is that the first transducer has not emitted a symbol, despite being in an $M$ state,
because symbols are only emitted on transitions {\em into} $M$ states;
and inspection of the composite machine (\figref{tkf91-tkf91}) reveals that transitions into state $MI$ only occur from states $MD$ or $MM$.
Only the second transducer changes state during these transitions; the $M$ state of the first transducer is a holdover from a previous emission.
The interpretation of such transitions is well-specified by our formal construction (\secref{Composition}).

In a specific sense (summing over paths)
this composite transducer is equivalent to the single transducer in \figref{tkf91},
but with the time parameter doubled ($t \to 2t$).
We can write this statement as $\tkf(t) \cdot \tkf(t) \equiv \tkf(2t)$.
This statement is, in fact, equivalent to a form of the Chapman-Kolmogorov equation,
$B(t)B(t') = B(t+t')$,
for transition probability matrices $B(t)$ of a stationary continuous-time Markov chain.
In fact TKF91 is currently the only nontrivial transducer known to have this property
(by ``nontrivial'' we mean including all types of state, and so excluding substitution-only models such as \figref{substituter},
which are essentially special limits of TKF91 where the indel rates are zero).
An open question is whether there are any transducers for affine-gap versions of TKF91 which
have this property
(excluding TKF92 from this since it does not technically operate on strings, but rather
sequences of strings (fragments) with immovable boundaries).  
The Chapman-Kolmogorov equation for transducers
is stated in \secref{ChapmanKolmogorov}.

\subsubsection{Constraining the input to the substitution model}
\seclabel{mf-substituter}
\widepdffig{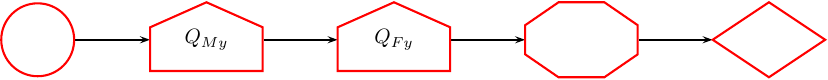}{Transducer $\generate(MF) \cdot \substitute(Q)$
corresponds to the generation of sequence MF (by the transducer in \figref{moore-mf-generator})
and its subsequent mutation via substitutions (by the transducer in \figref{substituter}).
Since no gaps are involved, and the initial sequence length is specified (by \figref{moore-mf-generator}),
there is only one state path through this transducer.
State types are indicated by the shape (as per \figref{legend}), and the I/O functions of the two Insert states are indicated.
For more information see
\secref{mf-substituter}.
}

\figref{mf-substituter} is the composition of the MF-generator (\figref{moore-mf-generator}) with the $Q$-substituter (\figref{substituter}). 
This is quite similar to a probabilistic weight matrix trained on the single sequence MF,
as each of the Insert states has its own I/O probability weights.

%

\subsubsection{Constraining the input to the TKF91 model}
\seclabel{mf-tkf91}
\widepdffig{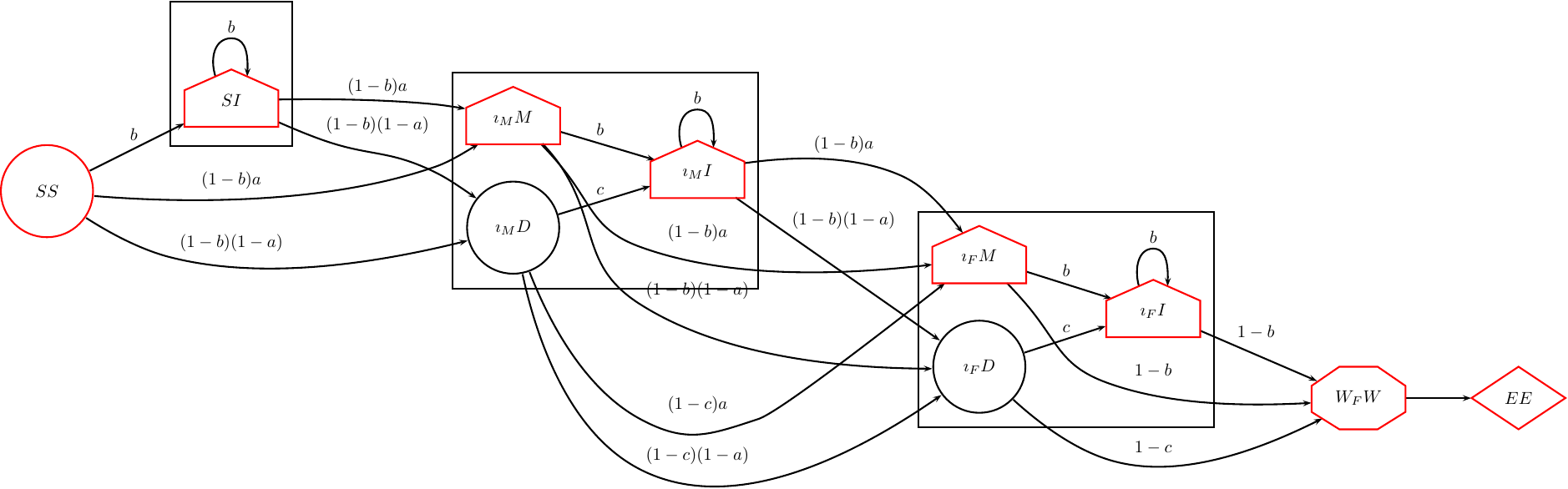}{Transducer $\generate(MF) \cdot \tkf$, the composition of the MF-generator (\figref{moore-mf-generator}) with the TKF91 transducer (\figref{tkf91-labeled}).
This can be viewed as an HMM that generates sequences sampled from the distribution of TKF91-mutated descendants of ancestor MF.
See \secref{mf-tkf91} and \tabref{mf-tkf91} for the meaning of each state. }

\figref{mf-tkf91} is the composition of the MF-generator (\figref{moore-mf-generator}) with TKF91 (\figref{tkf91-labeled}).  
In contrast to \figref{mf-substituter}, this machine generates samples
from the distribution of descendants of a known ancestor (MF) via indels and substitutions.  It is conceptually similar to a profile HMM
trained on the single sequence MF (though without the Dirichlet prior distributions that are typically used when training a profile HMM).

The meaning of the various states in this machine are shown in \tabref{mf-tkf91}.

\begin{table}
\begin{tabular}{c|p{.65\textwidth}|p{.15\textwidth}}
$i b$ & Meaning & I/O fn. \\
\hline
$SS$ & Both transducers are in the Start state.  No characters are  absorbed or emitted. &  \\
$EE$ & Both transducers are in the End state.  No characters are  absorbed or emitted. &  \\
$W_FW$ & Both transducers are in their respective Wait states.  No characters are absorbed or emitted, and both are poised to enter the End state.  &  \\
$SI$ & The TKF91 transducer inserts a character $y$ while the generator remains in the start state.  & $\pi_y$ \\
$\imath_MM$ & The  generator emits the M character, which TKF91 absorbs via its Match state and emits character $y$. & $\exp(Rt)_{My}$ \\
$\imath_MD$ & The  generator emits the M character, which TKF91 absorbs via its Delete state. &  \\
$\imath_MI$ & TKF91 inserts a character $y$ while the generator remains in the Insert state from which it emitted its last character (M).  & $\pi_y$ \\
$\imath_FM$ & The  generator emits the F character, which TKF91 absorbs via its Match state and emits character $y$. & $\exp(Rt)_{Fy}$ \\
$\imath_FD$ & The  generator emits the F character, which TKF91 absorbs via its Delete state. &  \\
$\imath_FI$ & TKF91 inserts a character $y$ while the generator remains in the Insert state from which it emitted its last character (F).  & $\pi_y$ \\
\end{tabular}
\caption{ \tablabel{mf-tkf91} Meanings (and, where applicable, I/O functions) of all states in
transducer $\generate(MF) \cdot \tkf$ (\figref{mf-tkf91}),
the composition of the MF-generator (\figref{moore-mf-generator})
with the TKF91 transducer (\figref{tkf91-labeled}).
Since this is an ensemble of two transducers, every state corresponds to a pair $(i,b)$
where
$i$ is the state of the generator transducer and
$b$ is the state of the TKF91 transducer.
 }
\end{table}

As noted in \secref{Tutorial.Composition}, a generator composed with any transducer is still a generator.
That is, the {\em composite} machine contains no states of
 type Match or Delete: it accepts null input (since its immediately-upstream
transducer, the MF-generator, accepts null input).  
Even when the TKF91 is in a state that accepts input symbols (Match or Delete),
the input symbol was inserted by the MF-generator;
the MF-generator does not itself accept any input,
so the entire ensemble accepts no input.  
Therefore the {\em entire composite machine is a generator}, since it accepts null input
(and produces non-null output).

\subsubsection{Constraining the output of the substitution model}
\figref{substituter-mf}  and \figref{substituter-cs}  show the composition of the substituter with (respectively) the MF-recognizer and the CS-recognizer.  
These machines are similar to the one in
\figref{mf-substituter},
but instead of the substituter taking its input from a generator,
the order is reversed: we are now feeding the substituter's output to a recognizer.
The composite machines accept sequence on input, and emit null output.

\widepdffig{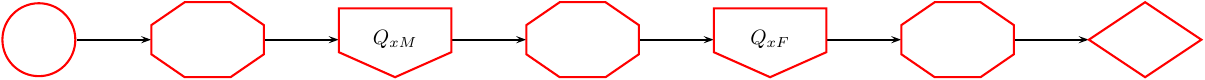}{Transducer $\substitute(Q) \cdot \recognize(MF)$ `recognizes' sequences ancestral to MF:
it computes the probability that a given input sequence will mutate into MF, assuming a point substitution model.
In contrast to the machine in \figref{mf-substituter}, this machine accepts an ancestral sequence as input and emits null output.  Note that a sequence of any length besides 2 will have recognition weight zero. }

\widepdffig{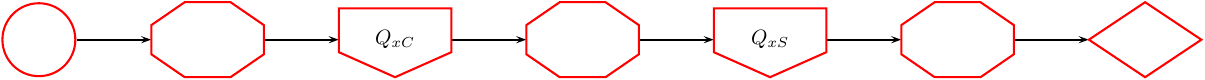}{Transducer $\substitute(Q) \cdot \recognize(CS)$ `recognizes' sequences ancestral to CS:
it computes the probability that a given input sequence will mutate into CS, assuming a point substitution model.
In contrast to the machine in \figref{mf-substituter}, this machine accepts an ancestral sequence as input and emits null output.  Note that a sequence of any length besides 2 will have recognition weight zero. }






\subsubsection{Constraining the output of the TKF91 model}
\seclabel{tkf91-mf}

\widepdffig{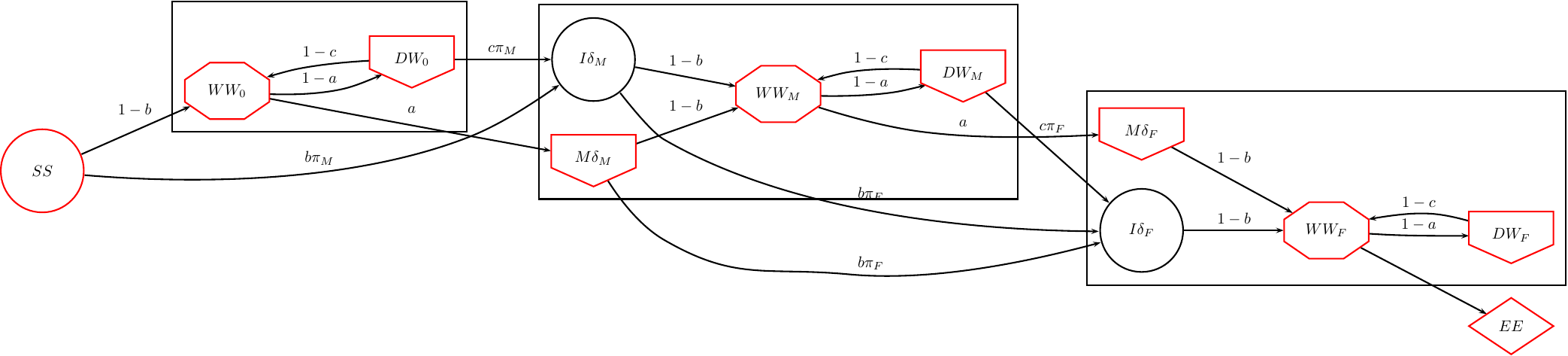}{Transducer $\tkf \cdot \recognize(MF)$
computes the probability that a given ancestral input sequence will evolve (via the TKF91 model) into the specific descendant MF.  
It can therefore be thought of as a {\em recognizer} for the ancestor of MF.
This machine absorbs sequence on its input and has null output due to 
the recognizer's null output.
See \secref{tkf91-mf} and \tabref{tkf91-mf} for more information.
}

\figref{tkf91-mf} shows the composition of a TKF91 transducer with the MF-recognizer.  
It is worth comparing the recognizer in \figref{tkf91-mf}
with the analogous generator in \figref{mf-tkf91}. 
While similar, the generator and the recognizer are not the same;
for a sequence $S$,
\figref{tkf91-mf} with $S$ as input computes $P(MF|S)$
(probability that descendant of $S$ is MF), 
whereas \figref{mf-tkf91} with $S$ as output computes $P(S|MF)$
(probability that descendant of MF is $S$).

The meaning of the various states in this model are shown in \tabref{tkf91-mf}.

\begin{table}
\begin{tabular}{c|p{.65\textwidth}|p{.15\textwidth}}
$b d$ & Meaning & I/O fn. \\
\hline
$SS$ & Both transducers are in the Start state.  No characters are  absorbed or emitted. &  \\
$EE$ & Both transducers are in the End state.  No characters are  absorbed or emitted. &  \\
$WW_0$ & Both transducers are in Wait states.  No characters are absorbed or emitted; both are poised to accept input.  &  \\
$DW_0$ & TKF91 absorbs a character $x$ and deletes it; recognizer remains in the initial Wait state ($W_0$).   & $1$ \\
$I\delta_M$ & TKF91 emits a character (M) via an Insert state and it is recognized by the $\delta_M$ state of the recognizer.   &  \\
$M\delta_M$ & TKF91 absorbs a character $x$ via a Match state, then emits an M, which is recognized by the $\delta_M$ state of the recognizer.   & $\exp(Rt)_{xM}$ \\
$WW_M$ & Both transducers are in Wait states: TKF91 in its only Wait state and the recognizer in the Wait state following the M character.  &  \\
$DW_M$ & TKF91 absorbs a character $x$ and deletes it; recognizer remains in the Wait state ($W_M$).   & $1$ \\
$I\delta_F$ & TKF91 emits a character (F) via an Insert state and it is recognized by the $\delta_F$ state of the recognizer.   &  \\
$M\delta_F$ & TKF91 absorbs a character $x$ via a Match state, then emits an F, which is recognized by the $\delta_F$ state of the recognizer.   & $\exp(Rt)_{xF}$ \\
$WW_F$ & Both transducers are in Wait states: TKF91 in its only Wait state and the recognizer in the Wait state following the F character.  &  \\
$DW_F$ & TKF91 absorbs a character $x$ and deletes it; recognizer remains in the Wait state ($W_F$).   & $1$ \\
\end{tabular}
\caption{ \tablabel{tkf91-mf} Meanings (and, where applicable, I/O functions) of all states in
transducer $\tkf \cdot \recognize(MF)$ (\figref{tkf91-mf}),
the composition of the TKF91 model (\figref{tkf91-labeled}) with the MF-recognizer (\figref{mf-labeled}).
Since this is an ensemble of two transducers, every state corresponds to a pair $(b,d)$
where
$b$ is the state of the TKF91 transducer and
$d$ is the state of the recognizer transducer.
 }
\end{table}

\subsubsection{Specifying input and output to the substitution model}

\widepdffig{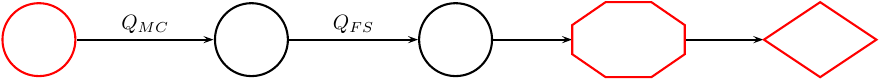}{Transducer $\generate(MF) \cdot \substitute(Q) \cdot \recognize(CS)$ is a pure Markov chain (with no I/O) whose single Start$\to$End state path describes the transformation of MF to CS via substitutions only. }

\figref{mf-substituter-cs} shows the composition of the MF-generator, substituter, and CS-recognizer.  
This machine is a pure Markov chain (with no inputs or outputs) which can be used to compute 
the probability of transforming MF to CS.
Specifically, this probability is equal to the weight of the single path through \figref{mf-substituter-cs}.

Note that composing $\generate(MF) \cdot \substitute(Q) \cdot \recognize(LIV) $
would result in a state graph with no valid paths from Start to End, since $\substitute(Q)$ cannot change sequence length.  
This is correct: the total path weight of zero, corresponding to the probability of transforming MF to LIV by point substitutions alone.

\subsubsection{Specifying input and output to the TKF91 model}
\seclabel{mf-tkf91-liv}

\widepdffig{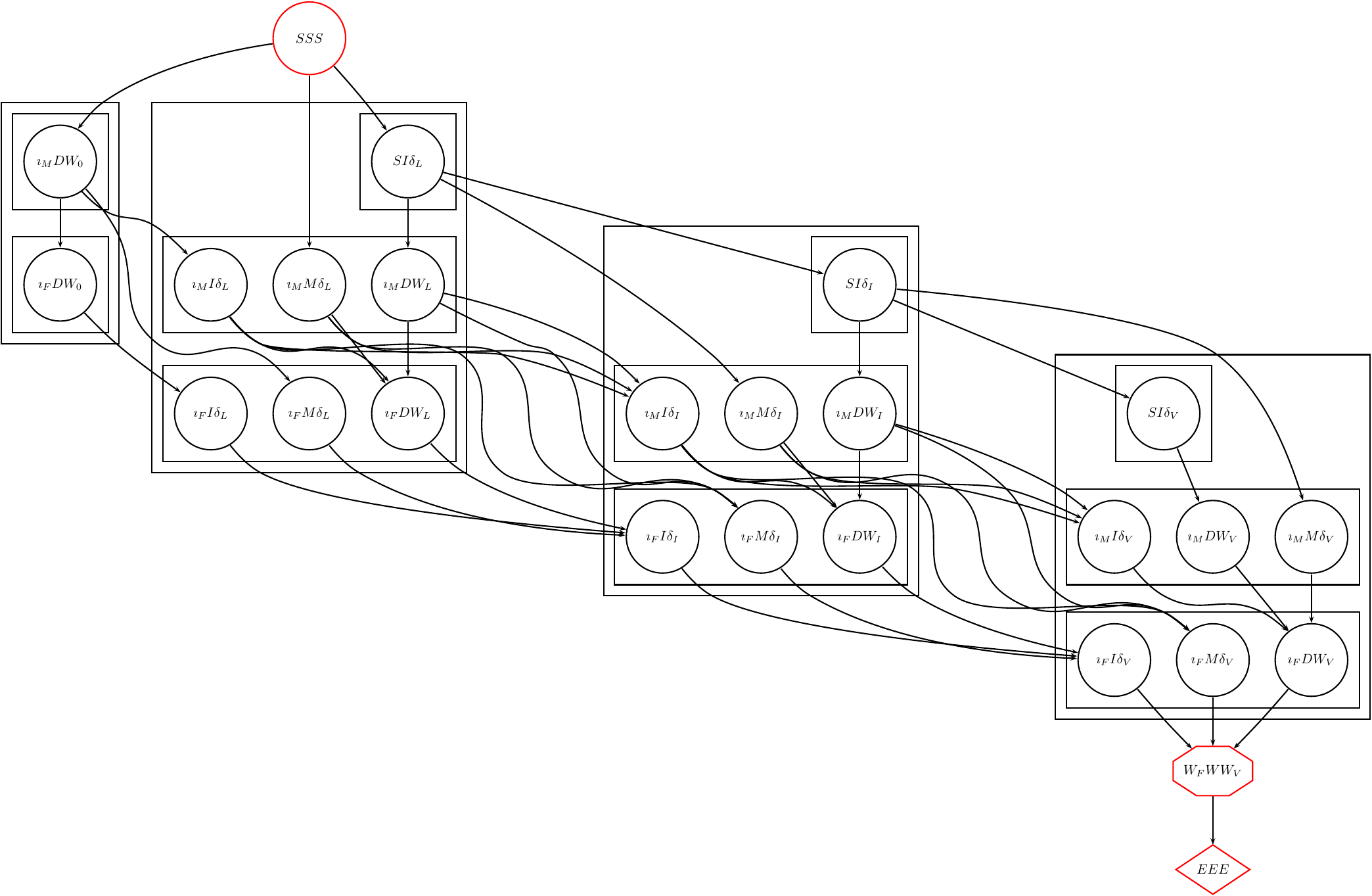}{Transducer $\generate(MF) \cdot \tkf \cdot \recognize(LIV)$ is a Markov chain
 (that is to say, a state machine, but one with no input or output characters)
 wherein every Start$\to$End state path describes an indel history
via which sequence MF mutates to sequence LIV. Note that the structure of this Markov chain is very similar to a dynamic programming matrix for pairwise sequence alignment.
The ``rows'' and ``cells'' of this matrix are shown as boxes.
Computation of the total Start$\to$End path weight, using a recursion that visits all the states in topological sort order,
is similar to the Forward algorithm of HMMs.
See \secref{mf-tkf91-liv} and \tabref{mf-tkf91-liv} for more information on the states of this model.
}

\figref{mf-tkf91-liv} shows the composition of three transducers, the MF-generator, TKF91 and LIV-recognizer
(transitions are omitted for clarity).
This is one step beyond the machine in \figref{mf-substituter-cs}, as it allows 
insertions and deletions to separate the generated (MF) and recognized (CS) sequences.  

The meaning of the various states in this model are shown in \tabref{mf-tkf91-liv}.

\begin{table}
\begin{tabular}{c|p{.8\textwidth}}
$i b d$ & Meaning \\
\hline
$SSS$ & All transducers are in the Start state.  No characters are emitted or absorbed. 
\\
$W_FWW_V$ & All transducers are in their Wait states which precede the End states for each transducer.  No characters are emitted or absorbed. 
\\
$EEE$ & All transducers are in their End states.  No characters are emitted or absorbed. 
\\
$SI\delta_y$ &
 The component TKF91 machine emits a character $y$ via the Insert state ($I$) which is read by the $\delta_y$ state of the recognizer; the generator remains in the Start state ($S$), not yet having emitted any characters. 
\\
$\imath_x I\delta_y$ &
 The component  TKF91 machine emits a character $y$ via the Insert state ($I$) which is read by the $\delta_y$ state of the recognizer; the generator remains in an Insert state ($\imath_x$) corresponding to the last character $x$ which it generated. 
\\
$\imath_x M\delta_y$ & 
The generator emits character $x$ via an Insert state ($\imath_x$); TKF91 absorbs the $x$ and emits a $y$ via a Match state ($M$);  
the recognizer absorbs the $y$ character via state $\delta_y$.  
\\
$\imath_x D W_y$ & 
The generator emits character $x$ via an Insert state ($\imath_x$), and TKF91 absorbs and deletes it via a Delete state ($D$).  
The recognizer  remains in $W_y$, the Wait  state following character $y$ (or $W_0$ if the recognizer  has not yet read any characters). 
\end{tabular}
\caption{ \tablabel{mf-tkf91-liv} Meanings of states in
transducer $\generate(MF) \cdot \tkf \cdot \recognize(LIV)$ (\figref{mf-tkf91-liv}),
the composition of the MF-generator (\figref{moore-mf-generator}),
TKF91 (\figref{tkf91-labeled}) and
the LIV-recognizer (\figref{liv-labeled}).
Since this is an ensemble of three transducers, every state corresponds to a triple $(i,b,d)$
where
$i$ is the state of the generator transducer,
$b$ is the state of the TKF91 transducer and
$d$ is the state of the recognizer transducer.
Where states are labeled with $x$ or $y$ suffices
(e.g. $SI\delta_y$),
then the definition is valid
$\forall\ x \in \{M,F\}$ and $\forall y \in \{L,I,V\}$.
 }
\end{table}

\figref{mf-tkf91-liv}, like \figref{mf-substituter-cs}, contains only null states (no match, insert, or delete states), making it a pure Markov chain with no I/O.  
Such Markov models can be viewed as a special case of an input/output machine where the input and output are both null.
(As noted previously, a {\em hidden} Markov model corresponds to  the special case of a transducer with null input, i.e. a generator.)

Probability $P(Y=\mbox{LIV}|X=\mbox{MF})$ for the TKF91 model
is equal to sum of all path weights from start to end in the Markov model of \figref{mf-tkf91-liv}. 
The set of paths corresponds to the set of valid evolutionary transformations relating
sequences MF and LIV (valid meaning those allowed under the model, namely non-overlapping single-residue indels and substitutions).  

This is directly analogous to computing a pairwise alignment (e.g. using the Needlman-Wunsch algorithm), and the structure of the Markov model shown in \figref{mf-tkf91-liv} suggests the familiar structure of a pairwise dynamic programming matrix.  

\subsection{Removal of null states}

In \figref{tkf91-tkf91}, the state $ID$ is of type Null:
it corresponds to an insertion by the first TKF91 transducer
that is then immediately deleted by the second TKF91 transducer,
so there is no net insertion or deletion.

It is often the case that Null states are irrelevant to the final analysis.
(For example, we may not care about all the potential, but unlikely,
insertions that were immediately deleted and never observed.)
It turns out that is often useful to eliminate these states;
i.e. transform the transducer into an equivalent transducer that does not contain null states.
(Here ``equivalent'' means a transducer that defines the same weight for every pair of I/O sequences;
this is made precise in \secref{Transducer}.)
Fortunately we can often find such an equivalent transducer
using a straightforward matrix inversion \cite{BradleyHolmes2009}.  
Null state removal is described in
more detail in \secref{Qn} and \secref{NullStateElim}.

\subsection{Intersection of transducers}
\seclabel{Tutorial.Intersection}

Our second operation for connecting two transducers
involves feeding the same input tape into both transducers in parallel,
and is called ``intersection''.

As with a transducer composition,
an intersection is constructed by taking the Cartesian product of two transducers' state spaces.
From two transducers $T$ and $U$,
we make a new transducer $T \fork U$
wherein every state corresponds to a pair $(t,u)$ of $T$- and $U$-states,
and whose output tape constitutes a {\em pairwise alignment} of the outputs of $T$ and $U$.

A property of intersection is that
if $T$ and $U$ are both recognizers,
then there {\em is} no output, and so $T \fork U$ is a recognizer also.

Conversely, if either $T$ or $U$ is {\em not} a recognizer, then $T \fork U$ will have an output;
and if neither $T$ or $U$ is a recognizer, then the output of $T \fork U$ will be a pairwise alignment
of $T$'s output with $U$'s output
(equivalently, we may regard $T \fork U$ as having two output tapes).
Transducer $Q_n$ in \secref{ConstrainedExpanded} is an example of such a two-output transducer.
Indeed, if we do further intersections (such as $(T \fork U) \fork V$ where $V$ is another transducer)
then the resultant transducer may have several output tapes
(as in the general multi-sequence HMMs defined in \cite{HolmesBruno2001}).
The models denoted $F_n$ in \secref{ForwardModel} are of this nature.

A formal definition of transducer intersection,
together with an algorithm for constructing the intersected transducer $T \fork U$,
is presented in \secref{Fork}. 
Like the construction of $TU$, this algorithmic construction of $T \fork U$
serves both as a proof of the existence of $T \fork U$, and an upper bound on its complexity.
It is also essential as a formal means of verifying our later results.

\subsubsection{Composition, intersection and Felsenstein's pruning algorithm}
\seclabel{Felsenstein}

If a composition is like matrix multiplication
(i.e. the operation of evolution along a contiguous branch of the phylogenetic tree),
then an intersection is like a {\em bifurcation} at a node in the phylogenetic tree,
where a branch splits into two child branches.

Intersection corresponds to the pointwise multiplication step in the Felsenstein pruning algorithm ---
i.e. the calculation
\[
P(\mbox{descendants}|\mbox{parent}) =
P(\mbox{left child and its descendants}|\mbox{parent})
P(\mbox{right child and its descendants}|\mbox{parent})
\]

Specifically, in the Felsenstein algorithm,
we define $G^{(n)}(x)$ to be the probability of all observed descendants of node $n$,
conditional on node $n$ having been in state $x$.
Let us further suppose that $M^{(n)}$ is the conditional substitution matrix
for the branch above node $n$ (coming from $n$'s parent), so
\[
M^{(n)}_{ij}=P(\mbox{node $n$ is in state $j$}|\mbox{parent of node $n$ is in state $i$})
\]
Then we can write the core recursion of Felsenstein's pruning algorithm in matrix form;
$G^{(n)}$ is a column vector, $M^{(n)}$ is a matrix, and the core recursion is
\[
G^{(n)} = \left( M^{(l)} \cdot G^{(l)} \right) \fork \left( M^{(r)} \cdot G^{(r)} \right)
\]
where $(l,r)$ are the left- and right-children of node $n$,
``$\cdot$'' denotes matrix multiplication,
and ``$\fork$'' denotes the {\em pointwise product} (also called the {\em Hadamard product}),
defined as follows for two vectors $A$ and $B$:
\[
(A \fork B)_i = A_i B_i,
\quad \quad \quad
A \fork B = \left( \begin{array}{c}
A_1 B_1 \\ A_2 B_2 \\ A_3 B_3 \\ \ldots \\ A_K B_K
\end{array} \right)
\]

Thus the two core steps of Felsenstein's algorithm (in matrix notation)
are (a) matrix multiplication and (b) the pointwise product.
Composition provides the transducer equivalent of matrix multiplication;
intersection provides the transducer equivalent of the pointwise product.

Note also that $G^{(n)}(x)$ is the probability of node $n$'s observed descendants
{\em conditional on} $x$, the state of node $n$.
Thus $G^{(n)}$ is similar to a recognition profile,
where the computed weight for a sequence $S$ represents
the probability of some event (recognition) conditional on having $S$ as input,
i.e. a probability of the form $P(\ldots|S)$
(as opposed to a probability distribution of the form $P(S|\ldots)$ where the sequence $S$ is the output,
as is computed by generative profiles).

Finally consider the initialization step of the Felsenstein algorithm.
Let $n$ be a leaf node and $y$ the observed character at that node.
The initialization step is
\[
G^{(n)}(x) = \delta(x=y)
\]
i.e. we initialize $G^{(n)}$ with a unit column vector, when $n$ is a leaf.
This unit vector is, in some sense, equivalent to our exact-match recognizer.
In fact, generators and recognizers are analogous to (respectively) row-vectors and column-vectors
in our infinite-dimensional vector space (where every dimension represents a different sequence).
The exact-match recognizer $\recognize(S)$ resembles a unit column-vector in the $S$-dimension,
while the exact-match generator $\generate(S)$ resembles a unit row-vector in the $S$-dimension.

\subsubsection{Recognizer for a common ancestor under the substitution model}
\seclabel{fork-subcs-submf}

\widepdffig{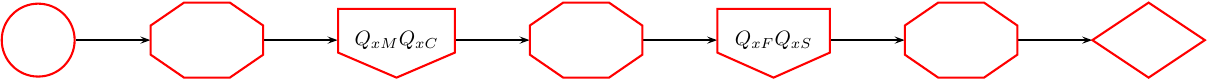}{The transducer $(\substitute(Q) \cdot \recognize(CS))\fork(\substitute(Q) \cdot \recognize(MF))$  recognizes the common ancestor of MF and CS.}

\figref{fork-subcs-submf} shows the intersection of \figref{substituter-mf} and \figref{substituter-cs}.  
This is an ensemble of four transducers.
Conceptually, what happens when a sequence is input is as follows.
First, the input sequence is duplicated;
one copy is then fed into a substituter (\figref{substituter}),
whose output is fed into the exact-matcher for CS (\figref{cs-labeled});
the other copy of the input sequence is fed into a separate substituter (\figref{substituter2}),
whose output is fed into exact-matcher for MF (\figref{mf-labeled}).

The composite transducer is a recognizer (the only I/O states are Deletes; there are no Matches or Inserts), since it absorbs input 
but the recognizers (exact-matchers) have null output.  
Note also that the I/O weight functions in the two Delete states in \figref{fork-subcs-submf}
are equivalent to the $G^{(n)}$ probability vectors in Felsenstein's algorithm (\secref{Felsenstein}).

Since this is an ensemble of four transducers, every state corresponds to a tuple $(b_1,d_1,b_2,d_2)$
where
$b_1$ is the state of the substituter transducer on the CS-branch (\figref{substituter}),
$d_1$ is the state of the exact-matcher for sequence CS (\figref{cs-labeled}),
$b_2$ is the state of the substituter transducer on the MF-branch (\figref{substituter2}),
$d_1$ is the state of the exact-matcher for sequence MF (\figref{mf-labeled}).

The I/O label for the general case where $(b_1,d_1,b_2,d_2) = (Q,\delta_A,R,\delta_B)$
denotes the vector whose elements are given by $Q_{xA} R_{xB}$.
So, for example,
the I/O label for state $(Q,\delta_C,R,\delta_M)$
is the vector whose $x$'th entry is the probability that an input symbol $x$
would mutate to C on the $Q$-branch and M on the $R$-branch.

\subsubsection{The Felsenstein likelihood for a two-branch tree using the substitution model}

\widepdffig{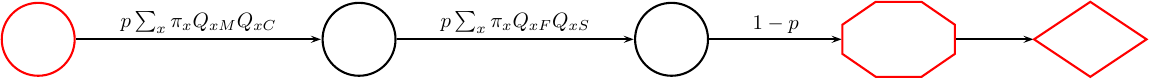}{Transducer $\nullmodel \cdot ((\substitute(Q) \cdot \recognize(CS))\fork(\substitute(Q) \cdot \recognize(MF)))$ 
allows computing  the final Felsenstein probability for the two sequence MF and CS descended from an unobserved ancestor
by character substitutions.  Since characters emitted by the $\nullmodel$ transducer are
all absorbed by the two recognizers, the composite machine has null input and output.
The probabilities (involving the geometric parameter $p$, the prior $\pi$ and the substitution matrix $Q$)
are all encoded on the transitions.
The use of summation operators in the weights of these transitions reflects the fact that multiple transitions have been collapsed into when
by our composition algorithm, which marginalizes I/O characters that are not directly observed.
Note that this machine is equivalent to the composition of a null model (\figref{null-model}) with the machine in \figref{fork-subcs-submf}.
}

As noted, the previous machine (\figref{fork-subcs-submf}) can be considered to compute the Felsenstein probability vector $G^{(n)}$ at an internal
node $n$ of a tree.  
When $n$ is the root node of the tree, this must be multiplied by a prior over sequences if we are to compute the final Felsenstein
probability,
\[
P(\mbox{sequences}|\mbox{tree}) = \pi G^{(1)} = \sum_x \pi_x G_x^{(1)}
\]

In transducer terms, this ``multiplication by a prior for the root'' is a composition: we connect a root generator to the input of the machine.
Composing \figref{null-model} (a simple prior over root sequences: geometric length distribution and IID with frequencies $\pi$)
with the transducer of \figref{fork-subcs-submf} (which represents $G^{(1)}$)
yields a pure Markov chain
whose total path weight from Start to End is the final Felsenstein probability (\figref{root-fork-subcs-submf}).  
Furthermore, sampling traceback paths through this machine provides an easy
way to  sample from the  posterior distribution over ancestral sequences relating MF and CS.  

\subsubsection{Recognizer for a common ancestor under the TKF91 model}

\widepdffig{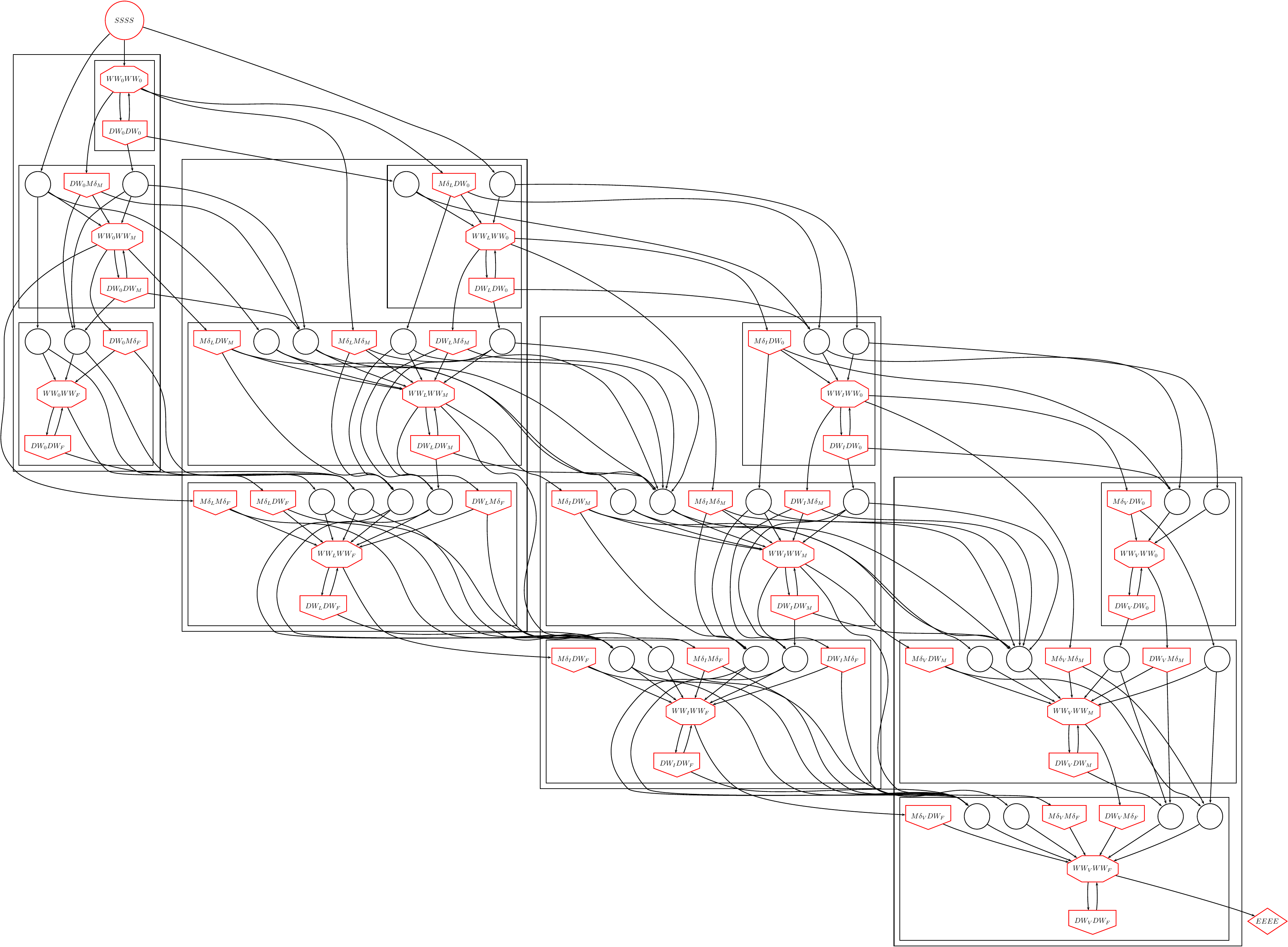}{Transducer $(\tkf \cdot \recognize(LIV))\fork(\tkf \cdot \recognize(MF))$ recognizes the ancestor of LIV and MF, assuming common descent under the TKF91 model.  
As with \figref{mf-tkf91-liv}, the structure of this machine is very similar to a dynamic programming matrix for pairwise sequence alignment; again, the ``rows'' and ``cells'' of this matrix have been shown as boxes.
In contrast to \figref{mf-tkf91-liv}, this machine is not a Markov model (ancestral sequence already encoded in the state space), but a recognizer (ancestral sequence read from input tape).
This Figure, along with several others in this manuscript, was autogenerated using phylocomposer \cite{BradleyHolmes2007} and graphviz \cite{GraphViz}.
\formaldefs
This type of machine is denoted $G_n$ in \secref{EvidenceExpandedModel}.
The algorithms for constructing composite and intersected transducer state graphs, are reviewed in \secref{Composition} and \secref{Fork}.
} 

The transducer shown in \figref{fork-tkf91liv-tkf91mf}, is the recognition profile for the TKF91-derived common ancestor of LIV and MF.
LIV and MF may each differ from the ancestor by insertions, deletions, and substitutions;
 a particular path through this machine represents one such explanation of differences
 (or, equivalently, an alignment of LIV, MF, and their ancestor).  
This type of machine is denoted $G_n$ in \secref{EvidenceExpandedModel}.

\figref{fork-tkf91liv-tkf91mf} is an ensemble of four transducers.
Conceptually, what happens to an input sequence is as follows.
First, the input sequence is duplicated;
one copy of the input is fed into TKF91 (\figref{tkf91-labeled}),
whose output is fed into the exact-matcher for LIV (\figref{liv-labeled});
the other copy of the input is fed into a separate TKF91 machine (\figref{tkf91-labeled}),
whose output is fed into the exact-matcher for MF (\figref{mf-labeled}).

Since this is an ensemble of four transducers, every state corresponds to a tuple $(b_1,d_1,b_2,d_2)$
where
$b_1$ is the state of the TKF91 transducer on the LIV-branch (\figref{tkf91-labeled}),
$d_1$ is the state of the exact-matcher for sequence LIV (\figref{liv-labeled}),
$b_2$ is the state of the TKF91 transducer on the MF-branch (\figref{tkf91-labeled}),
and $d_1$ is the state of the exact-matcher for sequence MF (\figref{mf-labeled}).

Note that, as with \figref{mf-tkf91-liv},
the underlying structure of this state graph is somewhat like a DP matrix,
with rows, columns and cells.
In fact, 
(modulo some quirks of the automatic graph layout performed by graphviz's `dot' program)
\figref{fork-tkf91liv-tkf91mf} and \figref{mf-tkf91-liv} are structurally quite similar.
However, compared to \figref{mf-tkf91-liv},
\figref{fork-tkf91liv-tkf91mf} has more states in each ``cell'',
because this transducer tracks events on two separate branches
(whereas \figref{mf-tkf91-liv} only tracks one branch).

Since this machine is a recognizer, it has Delete states but no Match or Insert states.
The delete states present do not necessarily correspond to  deletions by the TKF91 transducers:
all characters are ultimately deleted by the exact-match recognizers for LIV and MF,
so even if the two TKF91 transducers allow symbols to pass through undeleted,
they will still be deleted by the exact-matchers.

In fact, this transducer's states distinguish between
deletion events on one branch {\em vs} insertion events on the other:
this is significant because a ``deleted'' residue is homologous to a residue in the ancestral sequence,
while an ``inserted'' residue is not.
There are, in fact, four delete states in each ``cell'' of the matrix,
corresponding to four fates of the input symbol after it is duplicated:
\begin{enumerate}
\item Both copies of the input symbol pass successfully through respective TKF91 transducers
and are then deleted by respective downstream exact-matchers for sequences LIV and MF
 (e.g. $M D_L M D_F$);
\item One copy of the input symbol is deleted by the TKF91 transducer on the LIV-branch,
leaving the downstream LIV-matcher idling in a Wait state;
the other copy of the input symbol passes through the TKF91 transducer on the MF-branch and is then deleted by the downstream MF-matcher;
 (e.g. $D W_0 M D_F$);
\item One copy of the input symbol passes through the TKF91 transducer on the LIV-branch
and is then deleted by the downstream LIV-matcher;
the other copy is deleted by TKF91 transducer on MF-branch,
leaving the downstream MF-matcher idling in a Wait state;
 (e.g. $M D_L D W_0$);
\item Both copies of the input symbol are deleted by the respective TKF91 transducers,
while the downstream exact-matchers idle in Wait states without seeing any input
 (e.g. $D W_0 D W_0$).
\end{enumerate}
The other states in each cell of the matrix are Null states
(where the symbols recognized by the LIV- and MF-matchers originate from insertions by the TKF91 transducers,
rather than as input symbols)
and Wait states (where the ensemble waits for the next input symbol).

\subsubsection{The Felsenstein likelihood for a two-branch tree using the TKF91 model}

\widepngfig{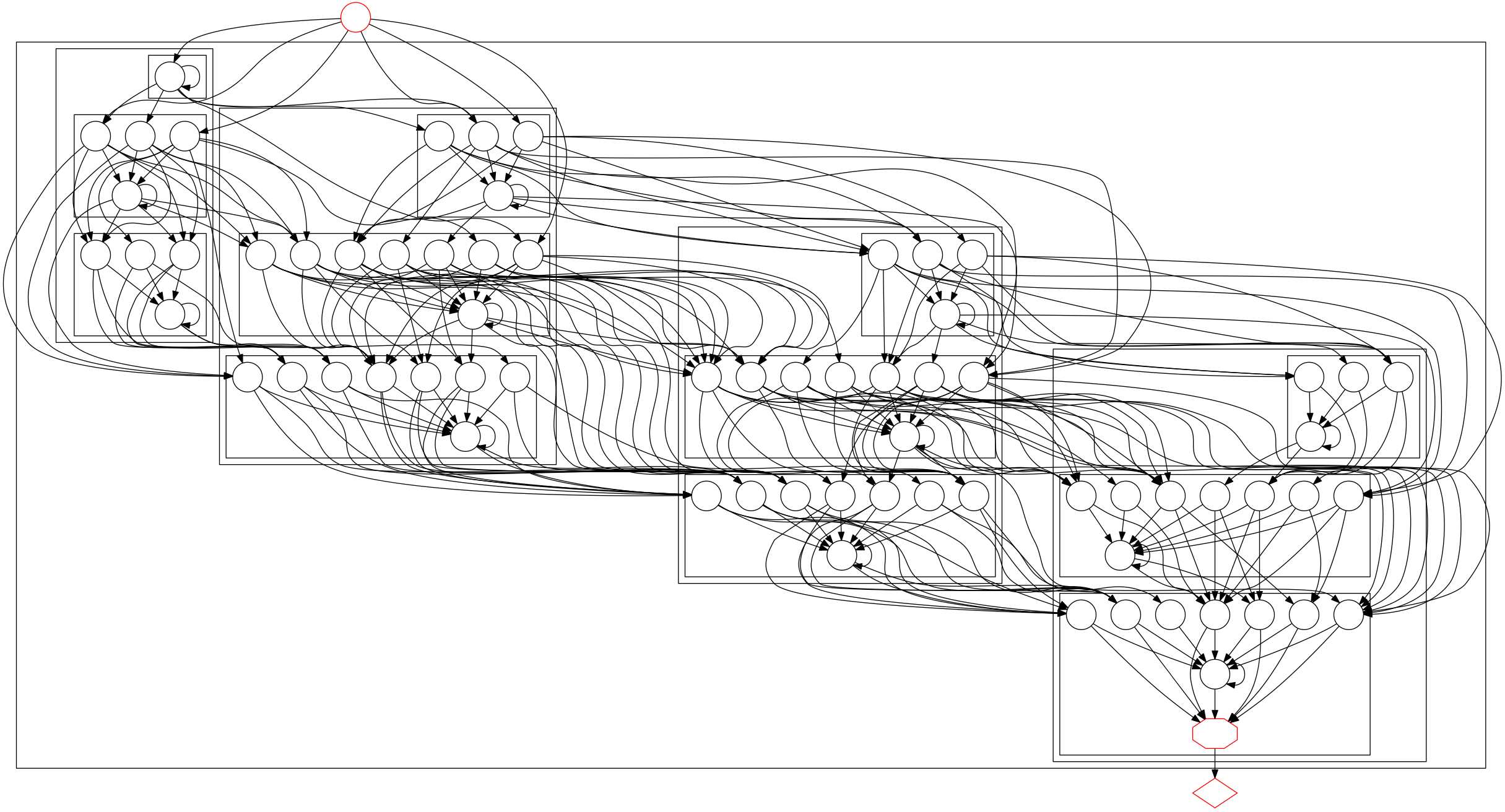}{Transducer $\tkfroot \cdot ((\tkf \cdot \recognize(LIV))\fork(\tkf \cdot \recognize(MF)))$ models the generation of an ancestral sequence
which is then duplicated; the two copies are mutated (in parallel) by two TKF91 transducers into (respectively) LIV and MF.  
This machine is equivalent to the generator in \figref{tkf91-root} coupled to the input of the recognizer in \figref{fork-tkf91liv-tkf91mf}.
Because it is a generator coupled to a recognizer,
there is no net input or output, and so we can think of this as a straightforward Markov model,
albeit with some probability ``missing'': the total sum-over-paths weight from Start$\to$End is less than one.
Indeed, the total sum-over-paths weight through this Markov model
corresponds to the joint likelihood of the two sibling sequences, LIV and MF.
\formaldefs
This type of machine is denoted $M_n$ in \secref{ConstrainedExpanded}.  
}

If we want to compute the joint marginal probability of LIV and MF as siblings under the TKF91 model, marginalizing the unobserved common ancestor,
we have to use the same trick that we used to convert the recognizer in \figref{fork-subcs-submf} into the Markov model of \figref{root-fork-subcs-submf}.
That is, we must connect a generator to the input of \figref{fork-tkf91liv-tkf91mf},
where the generator emits sequences from the root prior (equilibrium) distribution of the TKF91 model.
Using the basic generator shown in \figref{tkf91-root}, we construct the machine shown
in \figref{root-fork-tkf91liv-tkf91mf}.  
(This type of machine is denoted $M_n$ in \secref{ConstrainedExpanded}.  )

Summing over all Start$\to$End paths in \figref{root-fork-tkf91liv-tkf91mf},
the total weight is the final Felsenstein probability, i.e. the joint likelihood $P(LIV,MF|\mbox{tree},\mbox{TKF91})$.
Sampling traceback paths through \figref{root-fork-tkf91liv-tkf91mf} yields samples from the  posterior distribution 
over common ancestors of LIV and MF.  
The sum-over-paths is computed via a form of the standard Forward dynamic programming
algorithm, described in \secref{DynamicProgramming}.  

This ability to sample paths will allow us to constrain
the size of the state space when we move from  pairs of 
sequences to entire phylogenetic trees.  
Tracing paths back through the state graph according to their posterior probability
is straightforward once the Forward matrix is filled; 
the algorithmic internals are detailed in \secref{Traceback}.

\subsubsection{Maximum likelihood ancestral reconstruction under the TKF91 model}

\widepngfig{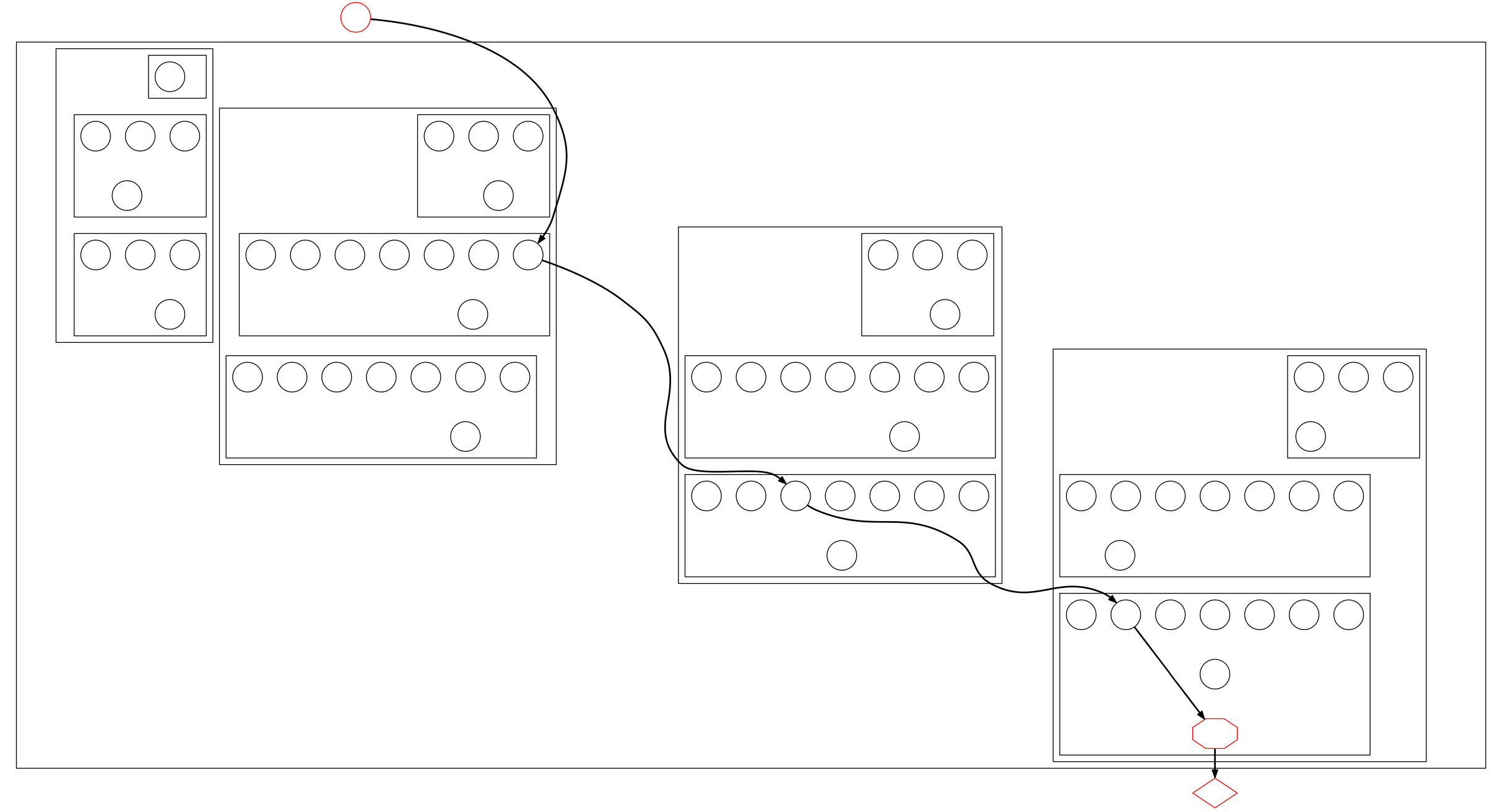}{The highest-weight path through
the machine of \figref{root-fork-tkf91liv-tkf91mf}
 corresponds to the most likely evolutionary history relating the two sequences.  
Equivalently, this path corresponds to an alignment of the two sequences and their 
ancestors.
\formaldefs
The subset of the state graph corresponding to this path is denoted $M'_n$ in \secref{ConstrainedExpanded}.
}

In \figref{viterbi-root-fork-tkf91liv-tkf91mf}, the highest-weight (Viterbi) traceback path
is highlighted.  
This path, via the significances of each of the visited states, corresponds to 
an ancestral alignment relating LIV and MF: an alignment of the sequences and their ancestor.

e.g.
\begin{tabular}{lccc}
Ancestral sequence & * & * & * \\
Sequence 1         & L & V & I \\
Sequence 2         & M & F & -
\end{tabular}

\widepdffig{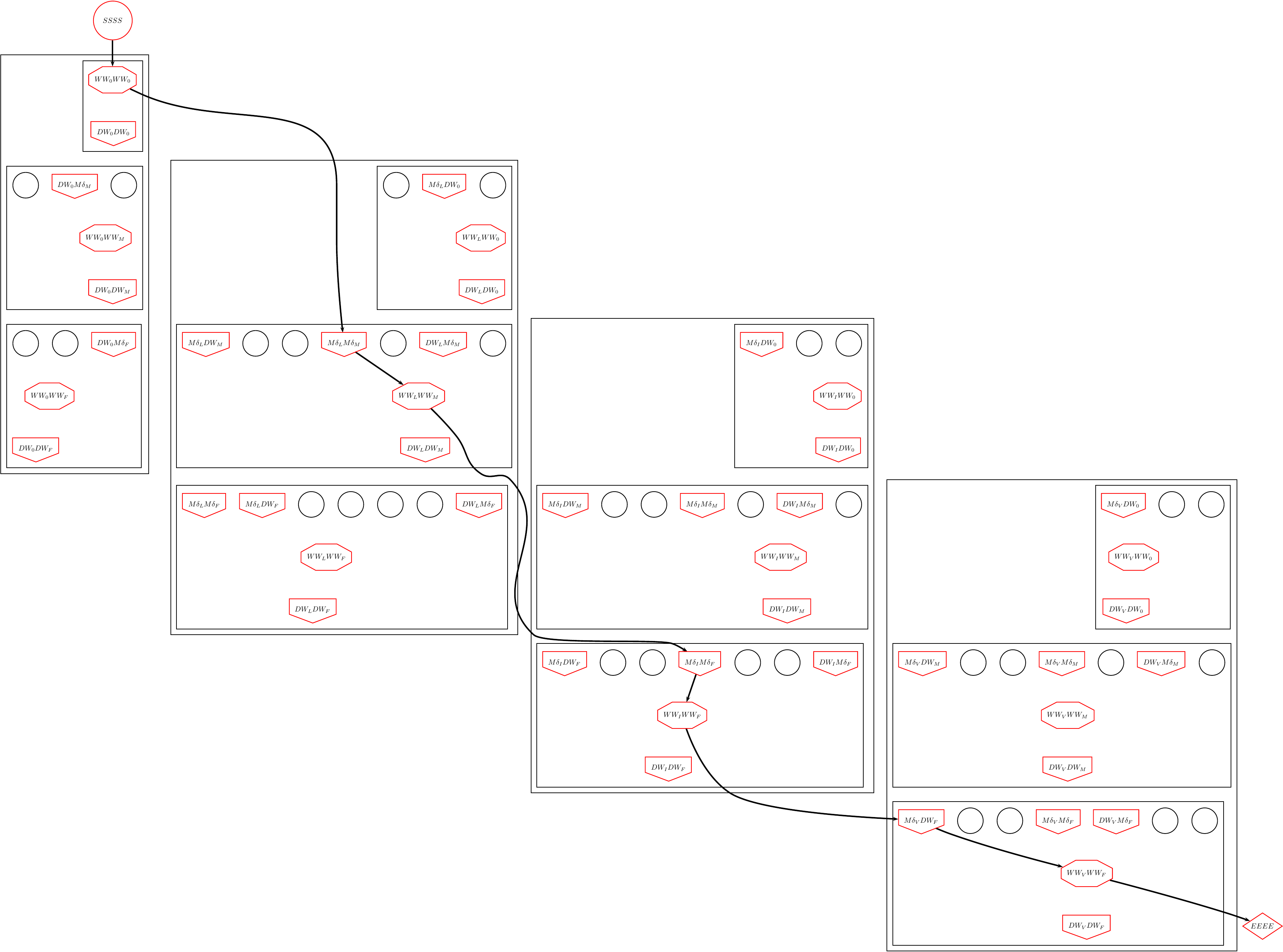}{Removing the generator from \figref{viterbi-root-fork-tkf91liv-tkf91mf} leaves a 
recognition profile for the ancestral sequence relating MF and LIV. 
\formaldefs
This and related transformations to sampled state paths are described in \secref{Mn2En}.}

Computing this alignment/path is straightforward, and essentially amounts to choosing
the highest-probability possibility (as opposed to sampling) 
in each step of the traceback detailed in
section \secref{Traceback}.  

If we remove  the root generator, we  obtain a linear recognizer profile
for the sequence at the ancestral node (\figref{viterbi-profile}),
as might be computed by progressive alignment.  
This can be thought of as a profile HMM trained on an alignment 
of the two sequences MF and LIV.  
It is a machine of the same form as $E_n$ in \secref{ConstrainedExpanded}.  

\widepdffig{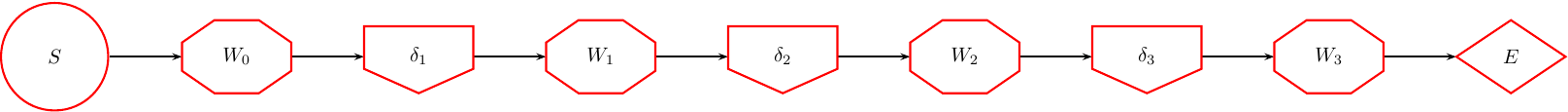}{Transducer $\profile_1$ is a linear recognition profile for the 
ancestor relating MF and LIV, created by taking the states visited by the Viterbi path shown in \figref{viterbi-fork-tkf91liv-tkf91mf}.
\formaldefs
This machine, and the one in \figref{forward2-profile}, are examples of the recognition profiles $E_n$ described in \secref{ConstrainedExpanded}.  }

\subsubsection{Sampling ancestral reconstructions under the TKF91 model}

\widepngfig{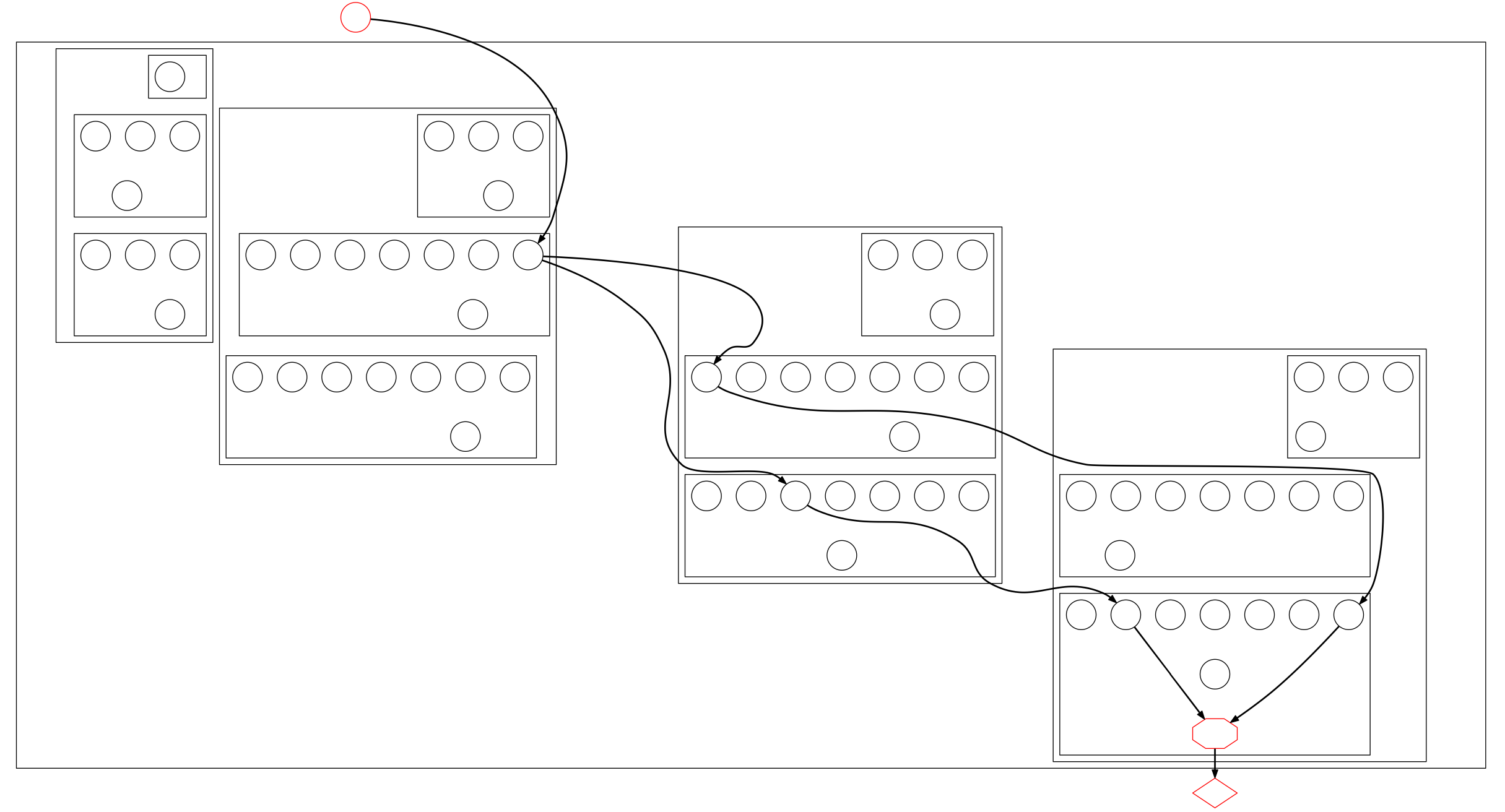}{
As well as finding just the highest-weight path through
the machine of \figref{root-fork-tkf91liv-tkf91mf}
(as in \figref{viterbi-root-fork-tkf91liv-tkf91mf}),
it is possible to sample suboptimal paths proportional to their posterior probability.  
Here, two sampled paths through the state graph are shown. 
\formaldefs
The subset of the state graph covered by the set of sampled paths is denoted $M'_n$ in \secref{ConstrainedExpanded}.
The mathematical detail of sampling paths is described in \secref{Traceback}.}

Instead of only saving the single highest-weight path through
the machine of \figref{root-fork-tkf91liv-tkf91mf}
(as in \figref{viterbi-root-fork-tkf91liv-tkf91mf}),
we can sample several paths from the posterior probability distribution over paths.  
In \figref{forward2-root-fork-tkf91liv-tkf91mf}, two sampled paths through the 
state graph are shown.  
Tracing paths back through the state graph according to their posterior probability
is straightforward once the Forward matrix is filled; 
the algorithmic internals are detailed in \secref{Traceback}.

It is possible that many paths through the state graph have weights only slightly less than the 
Viterbi path.
By sampling suboptimal paths according to their weights, it is possible to retain and 
propogate  this uncertainty when applying this method to progressive alignment. 
Intuitively, this corresponds to storing multiple evolutionary histories of a subtree
as  progressive alignment climbs the phylogenetic tree.  

\noindent For instance, in addition to sampling this path

\begin{tabular}{lccc}
Ancestral sequence & * & * & * \\
Sequence 1         & L & V & I \\
Sequence 2         & M & F & -
\end{tabular}
\\ \\ 
we also have this path

\begin{tabular}{lccc}
Ancestral sequence & * & * & * \\
Sequence 1         & L & V & I \\
Sequence 2         & M & - & F
\end{tabular}

\widepdffig{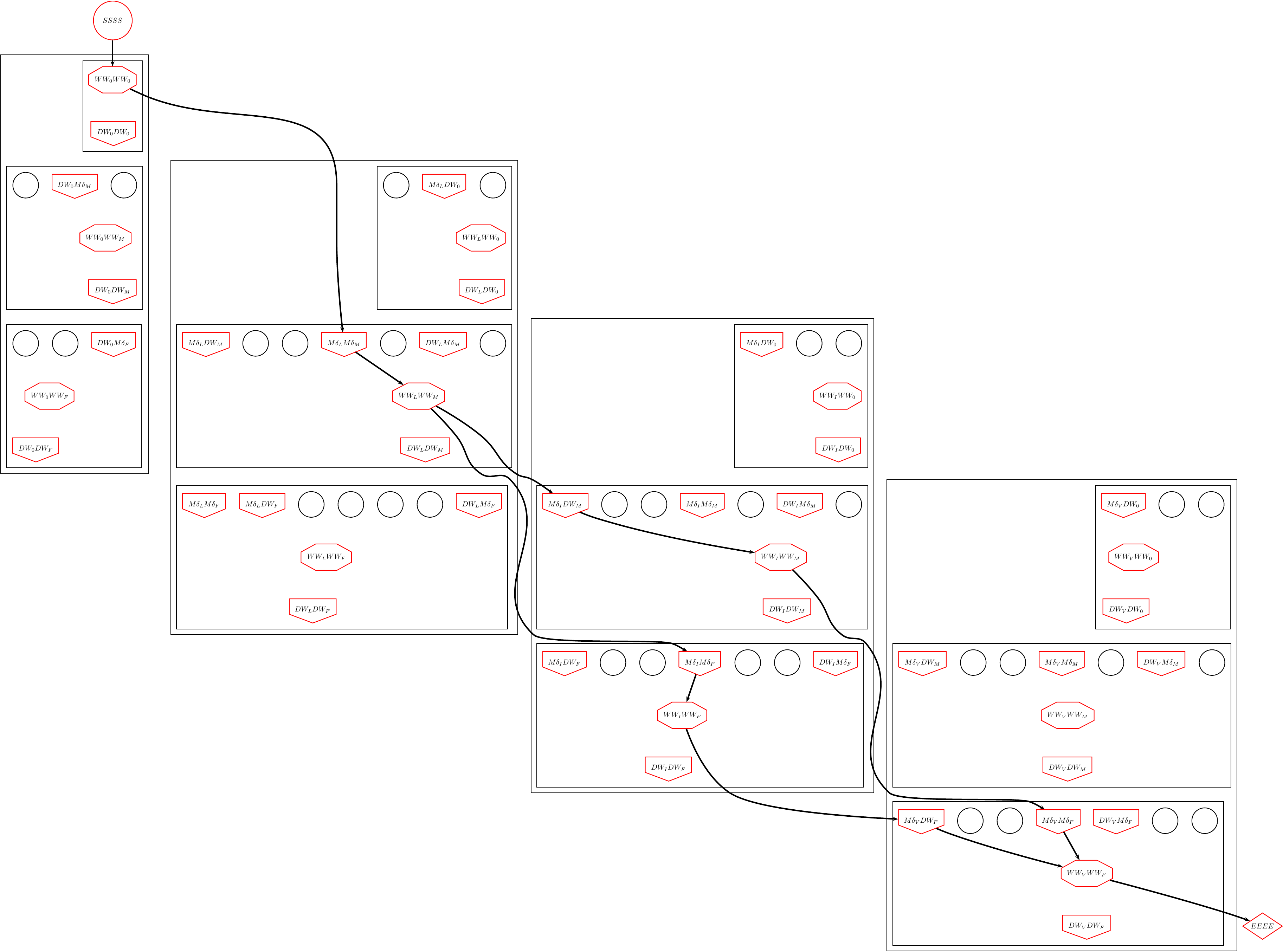}{Two sample paths through the 
machine in \figref{root-fork-tkf91liv-tkf91mf}, representing
possible evolutionary histories relating MF and LIV.
These are the same two paths as in \figref{forward2-root-fork-tkf91liv-tkf91mf},
but we have removed the root generator (as we did to transform \figref{viterbi-root-fork-tkf91liv-tkf91mf} into \figref{viterbi-fork-tkf91liv-tkf91mf}).
Paths are sampled according to their posterior probability, 
allowing us to select a high-probability subset of the state graph.  
The mathematical details of sampling paths is described in \secref{Traceback}.}

Combining these paths into a single graph and relabeling again, 
we still have a recognition profile, but it is now branched,
reflecting the possible uncertainty in our alignment/ancestral prediction, shown in \figref{forward2-profile}
(which is a transducer of the form $E_n$ in \secref{ConstrainedExpanded}).
The exact series of transformations required to convert \figref{forward2-root-fork-tkf91liv-tkf91mf} into \figref{forward2-profile}
(removing the root generator, eliminating null states, and adding wait states to restore the machine to Moore normal form)
is detailed in \secref{Mn2En}.  

Note that these are all still approximations to the full
recognition profile for the ancestor, which is \figref{fork-tkf91liv-tkf91mf}.  
While we could retain this entire profile, progressively climbing up a tree
would add so many states to the graph that inference would quickly become an intractable
problem.  
Storing a subset allows a flexible way in which to retain many high-probability solutions
while still allowing for a strict bound on the size of the state space. 

\widepdffig{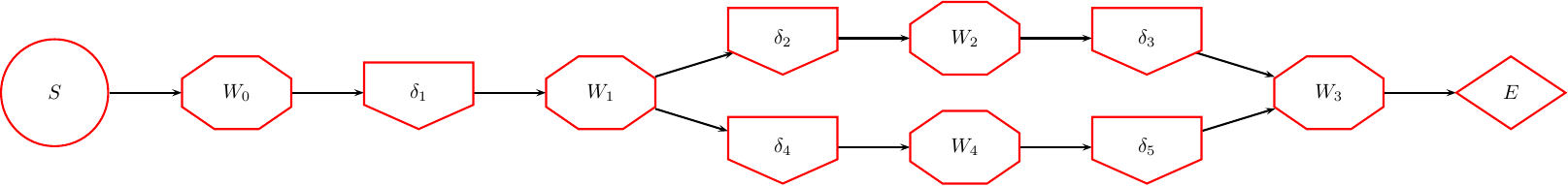}{Transducer $\profile_2$ is a branched recognizer for the 
ancestor common to MF and LIV.  The branched structure results from sampling
multiple paths through the state graph in \figref{root-fork-tkf91liv-tkf91mf}, 
mapping the paths back to \figref{fork-tkf91liv-tkf91mf},
and retaining only the subset of the graph visited by a sampled path.
\formaldefs
This machine, and the one in \figref{viterbi-profile}, are examples of the recognition profiles $E_n$ described in \secref{ConstrainedExpanded}.  }

\subsubsection{Ancestral sequence recognizer on a larger TKF91 tree}

\figref{fork3-tkf91liv-tkf91mf-tkf91cs} 
shows  the full recognition profile for the 
root-ancestral sequence in \figref{cs-mf-liv-tree}, 
representing all the possible evolutionary histories relating the three sequences.  
For clarity, many transitions in this diagram have been removed or collapsed.
(The recognition transducer for the root profile is denoted $G_1$ in \secref{EvidenceExpandedModel}.)

\widepngfig{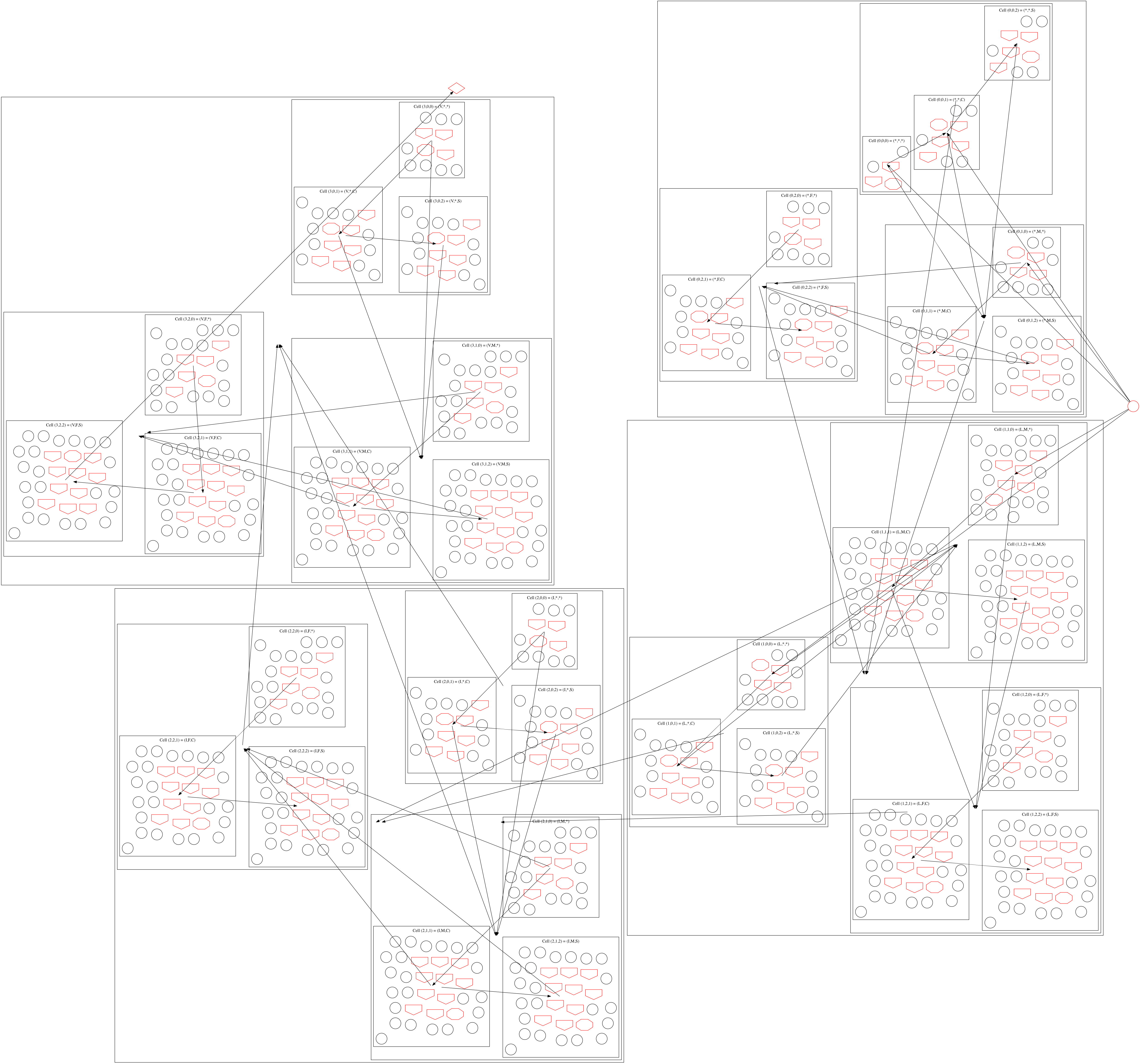}
{Transducer $(\tkf \cdot (\tkf \cdot \recognize(LIV))\fork(\tkf \cdot \recognize(MF)))\fork(\tkf \cdot \recognize(CS))$ 
is the full recognition profile for the root-ancestral sequence in \figref{cs-mf-liv-tree}, 
representing all the possible evolutionary histories relating the three sequences.  
For clarity, many transitions in this diagram have been removed or collapsed.
\formaldefs
This transducer, which recognizes the root sequence in \figref{cs-mf-liv-tree}, is denoted $G_1$ in \secref{EvidenceExpandedModel}.
}

\subsubsection{Felsenstein probability on a larger TKF91 tree}

We have now seen the individual steps of the 
transducer version of the Felsenstein recursion.
Essentially, composition replaces matrix multiplication, intersection replaces the 
pointwise product, and the initiation/termination steps involve (respectively) recognizers and generators.

The full recursion (with the same ${\cal O}(L^N)$ complexity as the Sankoff algorithm \cite{SankoffCedergren83}
for simultaneously aligning $N$ sequences of length $L$, ignoring secondary structure)
involves starting with exact-match recognition profiles at the leaves (\figref{mf-labeled}, \figref{liv-labeled}),
using those to construct recognition profiles for the parents (\figref{fork-tkf91liv-tkf91mf}),
and progressively climbing the tree toward the root,
constructing ancestral recognition profiles for each internal node. 
At the root, compose the root generator with the root recognition profile,
and the Forward probability can be computed. 

\subsubsection{Progressive alignment version of Felsenstein recursion}

The ``progressive alignment'' version,
equivalent to doing Felsenstein's pruning recursion on a single alignment
found using the progressive alignment algorithm,
involves sampling the single best linear recognition profile of the parent
 at each internal node, 
as in \figref{viterbi-fork-tkf91liv-tkf91mf}.  

\widepdffig{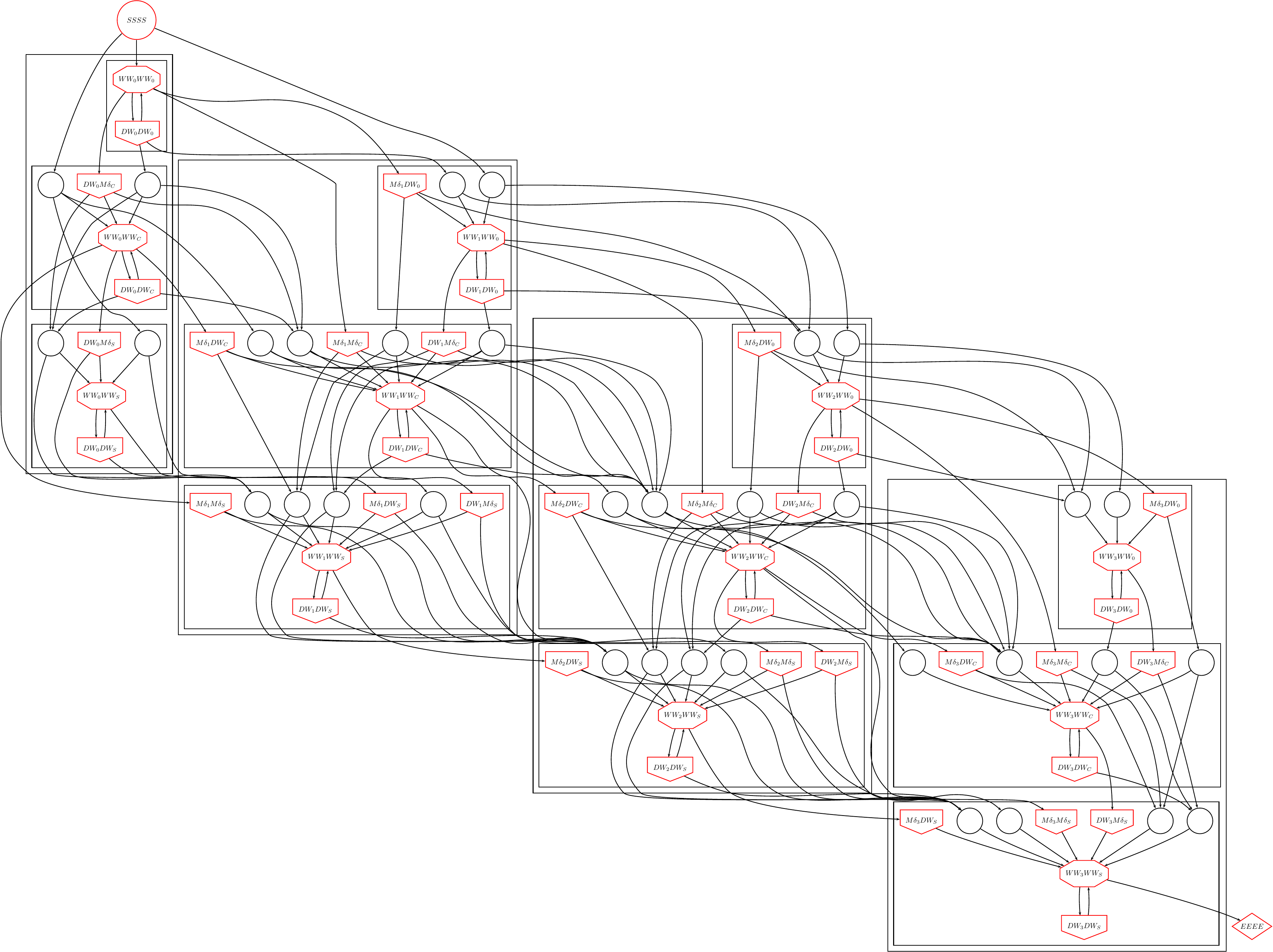}{Transducer $(\tkf \cdot \profile_1)\fork(\tkf \cdot \recognize(CS))$ recognizes the common ancestor of CS and $\profile_1$.
Transducer $\profile_1$, shown in \figref{viterbi-profile}, itself models the common ancestor of MF and LIV.
Using profile $\profile_1$, which is essentially a best-guess reconstructed ancestral profile, represents the most resource-conservative form of progressive alignment: only the 
maximum-likelihood indel reconstruction is kept during each step of the Felsenstein pruning recursion.
\formaldefs
This type of transducer is denoted $H_n$ in \secref{ConstrainedExpanded}.
}

We then repeat the process at the next level up in the tree, aligning the parent profile to its sibling,
as shown in \figref{fork-tkf91viterbi-tkf91cs}.
The machine in \figref{fork-tkf91viterbi-tkf91cs} is an example of the transducer $H_n$
in \secref{ConstrainedExpanded}.
(Technically, \figref{fork-tkf91liv-tkf91mf} is also an example of $H_n$, as well as being an example of $G_n$.
The difference is that $G_n$ describes all possible histories below node $n$,
since it is made by combining the two transducers $G_l$ and $G_r$ for the two children $(l,r)$ of node $n$.
By contrast, $H_n$ only describes a {\em subset} of such histories,
since it is made by combining the two transducers $E_l$ and $E_r$,
which are subsets of the corresponding $H_l$ and $H_r$;
just as \figref{viterbi-profile} and \figref{forward2-profile} are subsets of
\figref{fork-tkf91liv-tkf91mf}.)

This method can be recognized as a form of ``sequence-profile''
 alignment as familiar from progressive alignment,
except that we don't really make a distinction between a sequence and a profile
(in that observed sequences are converted into exact-match recognition
 profiles in the very first steps of the procedure).

The ``progressive'' algorithm proceeds in the exact same way as the ``full'' version, 
except that at each internal node a linear profile is created from the Viterbi
path through the state graph.  This ``best guess'' of the alignment of each subtree
is likely to work well in cases where the alignment is unambiguous, but under
certain evolutionary parameters the alignment of a subtree may not be clear
until more sequences are observed. 

\subsection{Stochastic lower bound version of Felsenstein recursion}

Our stochastic lower bound version is intermediate to the progressive alignment (Viterbi-like) algorithm and the full Sankoff algorithm.
Rather than just sampling the best linear profile for each parent, as in progressive alignment
(\figref{viterbi-fork-tkf91liv-tkf91mf}),
we sample some fixed number of such paths
(\figref{forward2-fork-tkf91liv-tkf91mf}).
This allows us to account for some amount of alignment uncertainty,
while avoiding the full complexity of the complete ancestral profile
(\figref{fork3-tkf91liv-tkf91mf-tkf91cs}).

By sampling a fixed number of traceback paths,
we can construct a recognition profile for the ancestral sequence
that is linearly bounded in size and offers a stochastic ``lower bound''
on the probability computed by the full Felsenstein transducer in \figref{root-fork-tkf91liv-tkf91mf}
\hl{that (if we include the Viterbi path along with the sampled paths) is guaranteed to improve on the Viterbi lower-bound for the full Felsenstein probability.}

\widepngfig{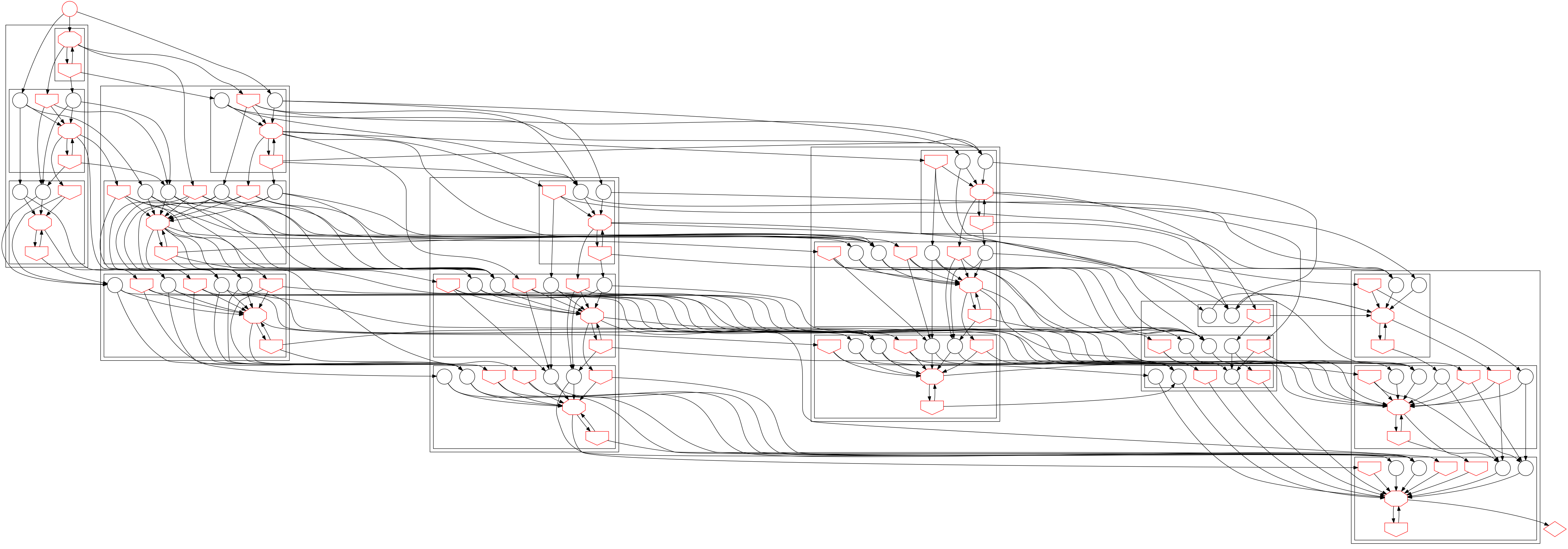}{Transducer $(\tkf \cdot \profile_2)\fork(\tkf \cdot \recognize(CS))$ shows the alignment of the sequence CS with the sampled profile $\profile_2$.
Transducer $\profile_2$, shown in \figref{forward2-profile}, is a branched profile whose different paths represent alternate ancestries of sibling sequences MF and LIV.
\formaldefs
The type of transducer shown in this Figure is denoted $H_n$ in \secref{ConstrainedExpanded}.
}

\figref{fork-tkf91forward2-tkf91cs} shows the intersection of $\profile_2$ (the ancestor of MF and LIV) with sequence CS, with TKF91 models accounting for the differences.
 Again this is a ``sequence-profile'' alignment, though it can also be called ``profile-profile'' alignment,
since CS can be considered to be a (trivial) linear profile.  
Unlike in traditional progressive alignment
 (but quite like e.g. partial order alignment \cite{LeeGrassoSharlow2002}),
one of the profiles is now branched
 (because we sampled more than one path to construct it in \figref{forward2-fork-tkf91liv-tkf91mf}), allowing a tunable (by modulating how many paths are sampled in the traceback step)
way to account for alignment uncertainty.  

Just like \figref{fork-tkf91viterbi-tkf91cs},
the machine in \figref{fork-tkf91forward2-tkf91cs}
is an example of the transducer $H_n$
in \secref{ConstrainedExpanded}.


Finally we compose the root generator (\figref{tkf91-root})
with \figref{fork-tkf91forward2-tkf91cs}.
(If there were more than two internal nodes in this tree,
we would simply continue the process of aligning siblings and sampling a recognition profile for the parent,
iterating until the root node was reached.)
The sum of all path weights through the state graph of the final machine
 represents the stochastic lower-bound
on the final Felsenstein probability for this tree.
Since we have omitted some states at each progressive step,
the computed probability does not sum over all possible histories
relating the sequences, hence it is a lower bound on the true Felsenstein probability.
(Each time we create a branched profile from a subset of the complete state graph,
we discard some low-probability states and therefore leak a little bit of probability, but hopefully not much.)
The hope is that, in practice, by sampling sufficiently many paths we are able to 
recover the maximum likelihood reconstruction at the root level, and so the
lower bound is a good one.
\hl{Furthermore, as long as we include the Viterbi path in with the retained sampled paths at every progressive step,
then the final state machine will be a superset of the machine that Viterbi progressive alignment will have constructed,
so we are guaranteed that the final likelihood will be greater than the Viterbi likelihood, while still representing a lower bound.}

\hl{
In terms of accuracy at estimating indel rates, we find that our method performs significantly better than Viterbi progressive alignment and 
approaches the accuracy of the true alignment. 
We simulated indel histories under 5 indel rates
 (0.005, 0.01, 0.02, 0.04, and 0.08 indels per unit time), and analyzed the unaligned leaf
sequences using PRANK} \cite{LoytynojaGoldman2008} 
\hl{ and an implementation of our algorithm, ProtPal.  
Insertion and deletion rates were computed using basic event counts normalized by branch lengths 
(more details are provided in Methods). 
PRANK provides an excellent comparison since it is phylogenetically-oriented, but
 uses progressive Viterbi reconstruction rather than  ensembles/profiles at internal nodes. 
Comparing indel rates computed using  the true alignment gives
 an idea how well PRANK and ProtPal compare to a  ``perfect''
method.  Note that this does not necessarily correspond to sampling
 infinitely many paths in our method, 
since the maximum likelihood reconstruction may not always be the true alignment.  }

\figref{estimates_biases_condensed} \hl{shows the error distributions 
of estimated insertion and deletion
rates using the PRANK, ProtPal, and true reconstructions.  The root mean squared error (RMSE),  
a measure of the overall distance from the `true' value ($\frac{inferred}{true} = 1$, indicated
with a vertical dashed line), is shown 
to the right of each distribution.  
ProtPal's RMSE values lie between  PRANK's and the true alignment, indicating using alignment
profiles at internal nodes allows for inferring more accurate alignments, or at least alignments
which allow for more accurately computing summary statistics such as the indel rate.  
ProtPal's distributions are also more symmetric than PRANK's between insertions and deletions, 
indicating that it is more able to avoid the bias towards deletions
 described in} \cite{LoytynojaGoldman2008}. 

\begin{center}
\begin{figure}[h!]
\includegraphics[width=1\textwidth]{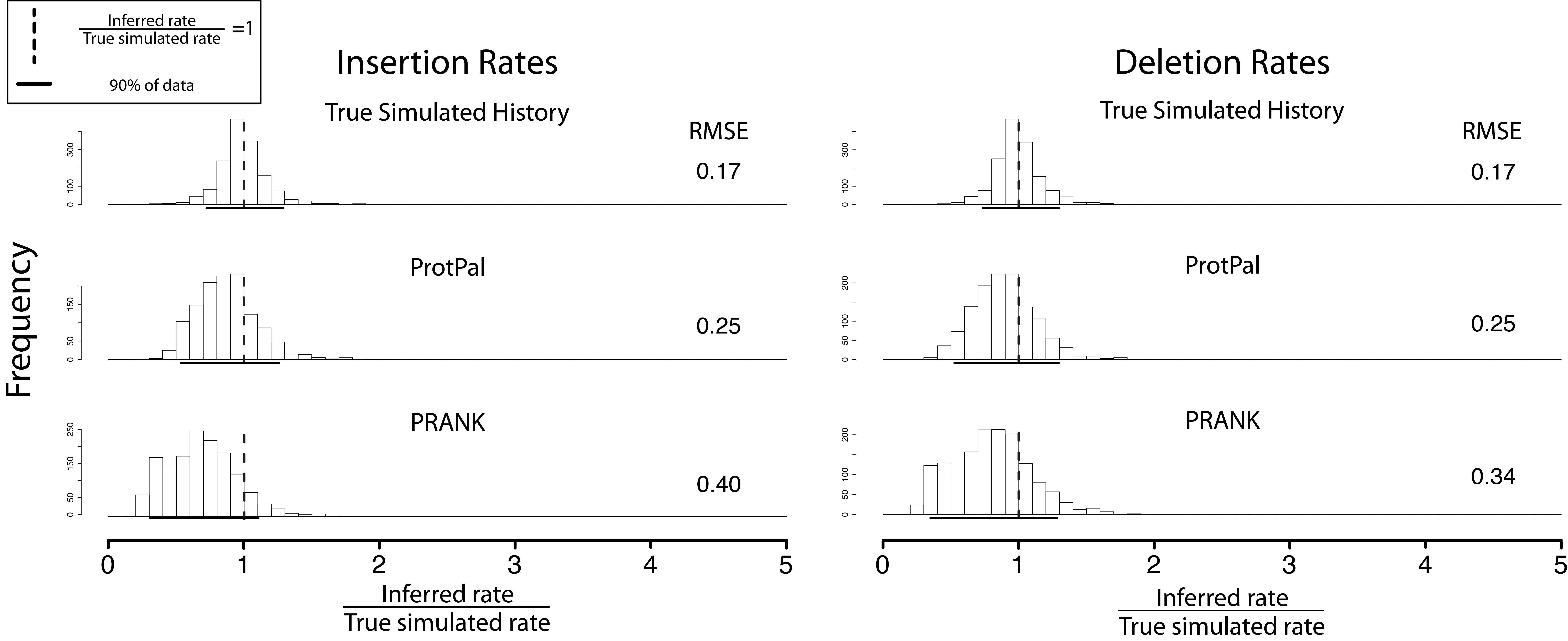}
\caption{In a  simulation experiment, an implementation of our algorithm
outperforms PRANK \cite{LoytynojaGoldman2008} and approaches the accuracy of the true indel history.
Alignments were simulated over range of evolutionary parameters, and unaligned leaf sequences
along with a phylogeny were fed to PRANK and ProtPal, which each produced a predicted indel 
reconstruction. From each reconstruction, insertion and deletion rates were calculated by event
counts normalized by branch lengths.  The plot shows the ratio of $\frac{inferred}{true}$ rates
pooled over all evolutionary parameters, with root mean squared error (RMSE) and middle 90\% 
quantiles appearing to the right and below each histogram. 
 }
\figlabel{estimates_biases_condensed}
\end{figure}
\end{center}

\subsection{A Pair HMM for aligning siblings}

While we have constructed our hierarchy of phylogenetic transducers by using recognizers,
it is also sometimes useful (and perhaps more conventional in bioinformatics) to think in terms of generators.
For example, we can describe the multi-sequence HMM that simultaneously emits an alignment of all of the sequences;
this is a generator, and in fact is the model $F_n$ described in \secref{ForwardModel}.
(An animation of such a multi-sequence generator emitting a small alignment can be viewed at
\url{http://youtu.be/EcLj5MSDPyM}.)

If we take the intersection of two TKF91 transducers, we obtain a transducer that has one input sequence and two output sequences
(or, equivalently, one output tape that encodes a pairwise alignment of two sequences).
This transducer is shown in \figref{fork-tkf91-tkf91}.
If we then connect a TKF91 equilibrium generator (\figref{tkf91-root}) to the input of \figref{fork-tkf91-tkf91},
we get \figref{root-fork-tkf91-tkf91}: a generator with two output tapes, i.e. a Pair HMM.
(To avoid introducing yet more tables and cross-references, we have confined the descriptions
 of the states in \figref{root-fork-tkf91-tkf91} and \figref{fork-tkf91-tkf91}
 to the respective Figure captions.)

The Pair HMM in \figref{root-fork-tkf91-tkf91} is particularly useful, as it crops up in our algorithm
whenever we have to align two recognizers. Specifically, 
\figref{root-fork-tkf91liv-tkf91mf}---which looks a bit like a pairwise dynamic programming matrix
for aligning sequence LIV to sequence MF---is essentially
\figref{root-fork-tkf91-tkf91} with one output tape connected to the LIV-recognizer (\figref{liv-labeled})
and the other output tape connected to the MF-recognizer (\figref{mf-labeled}).
In computing the state space of machines like \figref{root-fork-tkf91liv-tkf91mf}
(which are called $M_n$ in \secref{ConstrainedExpanded}),
it is useful to precompute the state space of the component machine in \figref{root-fork-tkf91-tkf91}
(called $Q_n$ in \secref{ConstrainedExpanded}).
This amounts to a run-time optimization, though it also helps us verify correctness of the model.

\widepdffig{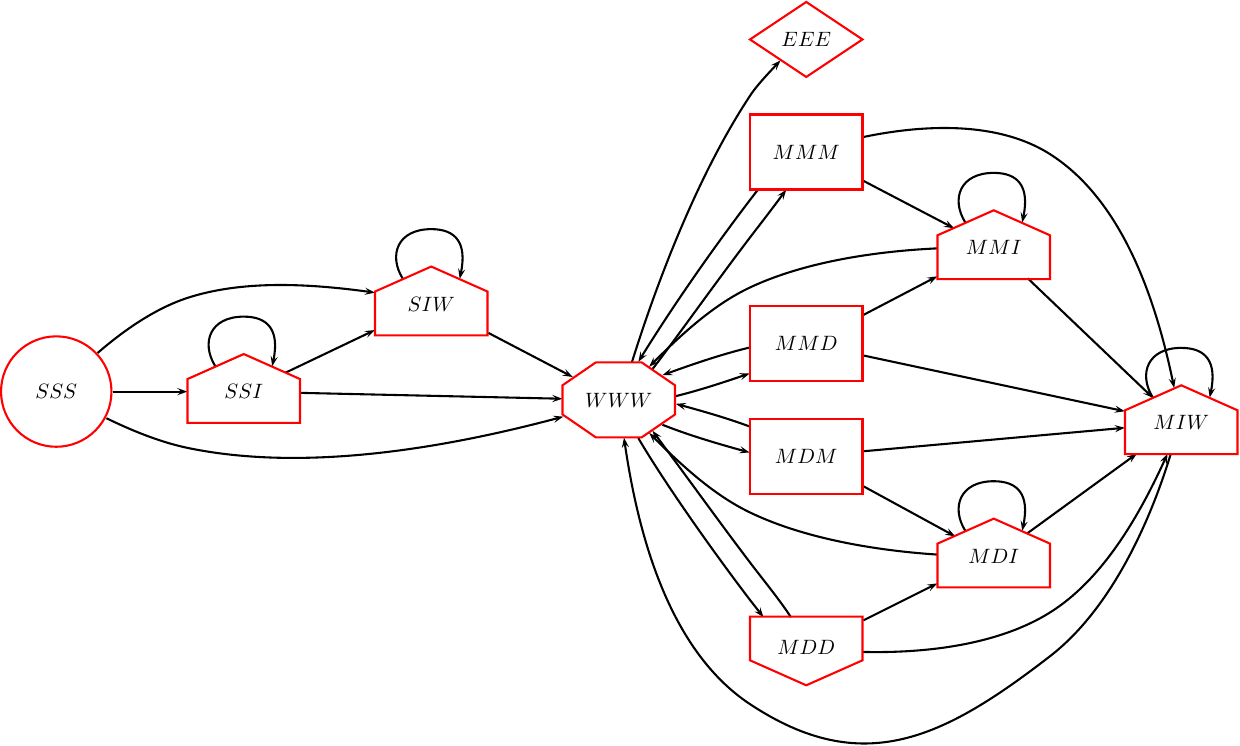}{
Transducer $\idfork \circ (\tkf \fork \tkf)$ is a bifurcation of two TKF91 transducers.
It can be viewed as a transducer with one input tape and two output tapes.
Each state has the form $\upsilon b_1 b_2$ where
 $\upsilon$ is the state of the bifurcation transducer (\secref{Identity}),
 $b_1$ is the state of the first TKF91 machine (\figref{tkf91-labeled}) and $b_2$ is the state of the second TKF91 machine.
The meaning of I/O states (Match, Insert, Delete) is subtle in this model, because there are two output tapes.
Dealing first with the Inserts:
in states $SIW$ and $MIW$, the first TKF91 transducer is inserting symbols to the first output tape,
while in states $SSI$, $MMI$ and $MDI$, the second TKF91 transducer is emitting symbols to the second output tape.
Dealing now with the Matches and Deletes:
the four states that can receive an input symbol
are $MMM$, $MMD$, $MDM$ and $MDD$.
Of these,
$MMM$ emits a symbol to both output tapes (and so is a Match);
$MMD$ only emits a symbol to the first output tape (and so qualifies as a Match because it has input and output);
$MDM$ only emits a symbol to the second output tape (and so qualifies as a Match);
and $MDD$ produces no output at all (and is therefore the only true Delete state).
}

\widepdffig{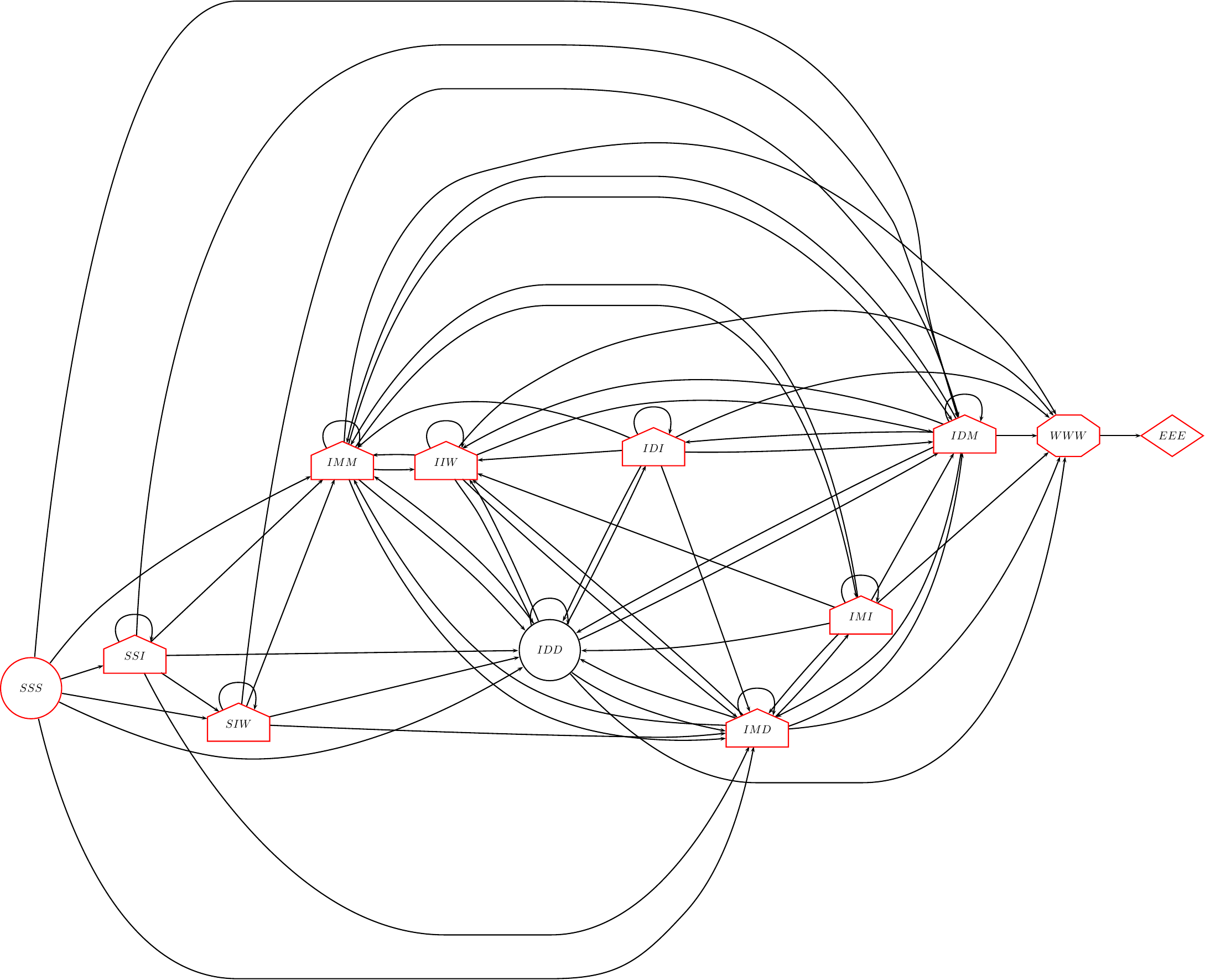}{
Transducer $\tkfroot \circ (\tkf \fork \tkf)$ represents the operation
of sampling a sequence from the TKF91 equilibrium distribution
and then feeding that sequence independently into two TKF91 transducers.
Equivalently, it is the composition of the TKF91 equilibrium generator (\figref{tkf91-root})
with a bifurcation of two TKF91 transducers (\figref{fork-tkf91-tkf91}).
It can be viewed as a generator with two output tapes; i.e. a Pair HMM.
Each state has the form $\rho b_1 b_2$ where
 $\rho$ is the state of the generator,
 $b_1$ is the state of the first TKF91 machine and $b_2$ is the state of the second.
As in \figref{fork-tkf91-tkf91},
the meaning of I/O states is subtle in this model, because there are two output tapes.
We first deal with Insert states where one of the TKF91 transducers is responsible for the insertion.
In states $SIW$ and $IIW$, the first TKF91 transducer is emitting symbols to the first output tape;
in states $SSI$, $IDI$ and $IMI$, the second TKF91 transducer is emitting symbols to the second output tape.
The remaining states (excluding $SSS$ and $EEE$) all involve symbol emissions by the generator,
that are then processed by the two TKF91 models in various ways.
These four states that involve emissions by the generator
are $IMM$, $IMD$, $IDM$ and $IDD$.
Of these,
$IMM$ emits a symbol to both output tapes (and so qualifies as an Insert state);
$IMD$ only emits a symbol to the first output tape (and is an Insert state);
$IDM$ only emits a symbol to the second output tape (and is an Insert state);
and $IDD$ produces no output at all (and is therefore a Null state).
Note that \figref{root-fork-tkf91liv-tkf91mf},
which looks a bit like a pairwise dynamic programming matrix,
is essentially
this Pair HMM with one output tape connected to the LIV-recognizer (\figref{liv-labeled})
and the other output tape connected to the MF-recognizer (\figref{mf-labeled}),
\formaldefs
This type of transducer is called $Q_n$ in \secref{ConstrainedExpanded}.
When both its outputs are connected to recognizers ($H_l$ and $H_r$),
then one obtains a transducer of the form $M_n$.
}

\subsection{A note on the relationship between this tutorial and the formal definitions section}
\seclabel{FormalTutorialRelationship}

As noted throughout the tutorial, the example phylogeny and transducers can be directly related to the ``Hierarchy of phylogenetic transducers''
described in \secref{ModelStructure} onwards.

Consider the tree of \figref{cs-mf-liv-tree}.
Let the nodes of this tree be numbered as follows:
(1) Ancestral sequence,
(2) Intermediate sequence,
(3) Sequence LIV,
(4) Sequence MF,
(5) Sequence CS.

Some of the transducers defined for this tree in \secref{ModelStructure} include
\[
\begin{array}{rcll}
R & = & \tkfroot & \mbox{\figref{tkf91-root}} \\
\forall n \in \{ 2, 3, 4, 5 \} : \quad B_n & = & \tkf & \mbox{\figref{tkf91-labeled}} \\
E_5 = H_5 = G_5 & = & \recognize(CS) & \mbox{\figref{cs-labeled}} \\
E_4 = H_4 = G_4 & = & \recognize(MF) & \mbox{\figref{mf-labeled}} \\
E_3 = H_3 = G_3 & = & \recognize(LIV) & \mbox{\figref{liv-labeled}} \\
G_2 & = & (B_3 \cdot G_3) \fork (B_4 \cdot G_4) & \\
& = & ( \tkf \cdot \recognize(LIV) ) \fork ( \tkf \cdot \recognize(MF) ) & \mbox{\figref{fork-tkf91liv-tkf91mf}} \\
H_2 & = & (B_3 \cdot E_3) \fork (B_4 \cdot E_4) & \\
& = & ( \tkf \cdot \recognize(LIV) ) \fork ( \tkf \cdot \recognize(MF) ) & \mbox{\figref{fork-tkf91liv-tkf91mf} again} \\
M_2 & = & R \cdot H_2 & \\
    & = & \tkfroot \cdot ( \tkf \cdot \recognize(LIV) ) \fork ( \tkf \cdot \recognize(MF) ) & \mbox{\figref{root-fork-tkf91liv-tkf91mf}} \\
E_2 & \subseteq & H_2 & \mbox{\figref{viterbi-profile}, \figref{forward2-profile}} \\
G_1 & = & (B_2 \cdot G_2) \fork (B_5 \cdot G_5) & \\
& = & (\tkf \cdot (\tkf \cdot \recognize(LIV))\fork(\tkf \cdot \recognize(MF)))\fork(\tkf \cdot \recognize(CS)) & \mbox{\figref{fork3-tkf91liv-tkf91mf-tkf91cs}} \\
H_1 & = & (B_2 \cdot E_2) \fork (B_5 \cdot E_5) & \mbox{\figref{fork-tkf91viterbi-tkf91cs}, \figref{fork-tkf91forward2-tkf91cs}} \\
G_0 & = & R \cdot G_1 & \\
& = & \tkfroot \cdot (\tkf \cdot (\tkf \cdot \recognize(LIV))\fork(\tkf \cdot \recognize(MF)))\fork(\tkf \cdot \recognize(CS)) & \mbox{\figref{cs-mf-liv-machines}} \\
\forall n \in \{1,2\}: \quad Q_n & = & \tkfroot \fork (\tkf \cdot \tkf) & \mbox{\figref{root-fork-tkf91-tkf91}} \\
\end{array}
\]

\pagebreak
\section{Formal definitions}
\seclabel{Formal}
This report makes our transducer-related definitions precise,
 including notation for state types, weights (i.e. probabilities),
 transducer composition, etc.

Notation relating to mundane manipulations of sequences (sequence length, sequence concatenation, etc.) is deferred to the end of the document,
 so as not to interrupt the flow.

We first review the letter transducer $T$,
 transduction weight $\wtrans{\weight}{x}{T}{y}$ and
 equivalence $T \transequiv T'$.

We then define two operations for combining transducers:
 composition ($T \compose U$) which unifies $T$'s output with $U$'s input,
and
 intersection ($T \fork U$) which unifies $T$'s and $U$'s input.

We define our ``normal'' form for letter transducers,
 partitioning the states and transitions into types $\{S,M,D,I,N,W,E\}$ based on their input/output labeling.
(These types stand for Start, Match, Delete, Insert, Null, Wait, End.)
This normal form is common in the bioinformatics literature \cite{Durbin98}
 and forms the basis for our previous constructions of phylogenetic transducers \cite{Holmes2003,BradleyHolmes2009}.

We define exact-match and identity transducers, and give constructions of these.

We define our hierarchy of phylogenetic transducers, and give constructions and inference algorithms,
including the concept of ``alignment envelopes'' for size-limiting of transducers.

\subsection{Input-output automata}
\seclabel{Transducer}
{\em The letter transducer} is a tuple $T = (\Omega_I, \Omega_O, \States, \startstate, \laststate, \Transitions, \weight)$
where
 $\Omega_I$ is an input alphabet,
 $\Omega_O$ is an output alphabet,
$\States$ is a set of states,
$\startstate \in \States$ is the start state,
$\laststate \in \States$ is the end state,
$\Transitions \subseteq \States \times \gappedalphabet{I} \times \gappedalphabet{O} \times \States$ is the transition relation, and
$\weight:\Transitions \to [0,\infty)$ is the transition weight function.

{\em Transition paths:}
The transitions in $\Transitions$ correspond to the edges of a labeled multidigraph over states in $\States$.
Let $\Pi \subset \Transitions^\ast$ be the set of all labeled transition paths from $\startstate$ to $\laststate$.

{\em I/O sequences:}
Let
$S_I:\Pi \to \Omega_I^\ast$ and
$S_O:\Pi \to \Omega_O^\ast$
be functions returning the input and output sequences of a transition path,
obtained by concatenating the respective transition labels.

{\em Transduction weight:}
For a transition path $\pi \in \Pi$,
define the path weight $\weight(\pi)$ and
(for sequences $x \in \Omega_I^\ast, y \in \Omega_O^\ast$)
the transduction weight $\wtrans{\weight}{x}{T}{y}$

\begin{eqnarray*}
\weight(\pi) & = & \prod_{\Transitions \in \pi} \weight(\Transitions) \\
\wtrans{\weight}{x}{T}{y} & = & \sum_{\pi \in \Pi, S_I(\pi)=x, S_O(\pi)=y} \weight(\pi)
\end{eqnarray*}

{\em Equivalence:}
If for transducers $T,T'$ it is true that $\wtrans{\weight}{x}{T}{y}=\wtrans{\weight'}{x}{T'}{y}\ \forall x,y$ then the transducers are equivalent in weight, $T \transequiv T'$.

\subsection{State types and normal forms}
\seclabel{StateTypes}
{\em Types of state and transition:}
If there exists a state type function, $\statetype:\States \to {\cal T}$, mapping states to types in ${\cal T} = \{S,M,D,I,N,W,E\}$,
and functions $\transweightfun{}: \States^2 \to [0,\infty)$ and $\emitweightfun{}: \gappedpair{I}{O} \times \States \to [0,\infty)$,
such that
\begin{eqnarray*}
\States_U & = & \{ \phi: \phi\in\States, \statetype(\phi) \in U \subseteq {\cal T} \} \\
\statesoftype{S} & = & \{ \startstate \} \\
\statesoftype{E} & = & \{ \laststate \} \\
\States & \equiv & \statesoftype{SMDINWE} \\   
\Transitions_M & \subseteq & \statesoftype{W} \times \Omega_I \times \Omega_O \times \statesoftype{M} \\
\Transitions_D & \subseteq & \statesoftype{W} \times \Omega_I \times \{\epsilon\} \times \statesoftype{D} \\
\Transitions_I & \subseteq & \statesoftype{SMDIN} \times \{\epsilon\} \times \Omega_O \times \statesoftype{I} \\
\Transitions_N & \subseteq & \statesoftype{SMDIN} \times \{\epsilon\} \times \{\epsilon\} \times \statesoftype{N} \\
\Transitions_W & \subseteq & \statesoftype{SMDIN} \times \{\epsilon\} \times \{\epsilon\} \times \statesoftype{W} \\
\Transitions_E & \subseteq & \statesoftype{W} \times \{\epsilon\} \times \{\epsilon\} \times \statesoftype{E} \\
\Transitions & = & \Transitions_M \cup \Transitions_D \cup \Transitions_I \cup \Transitions_N \cup \Transitions_W \cup \Transitions_E \\
\weight(\phi_{\mbox{src}},\omega_{\mbox{in}},\omega_{\mbox{out}},\phi_{\mbox{dest}}) & \equiv & \transweightfun{}(\phi_{\mbox{src}},\phi_{\mbox{dest}}) \emitweightfun{}(\omega_{\mbox{in}},\omega_{\mbox{out}},\phi_{\mbox{dest}})
\end{eqnarray*}
then the transducer is in {\em (weak) normal form}.
If, additionally, $\statesoftype{N} = \emptyset$, then the transducer is in {\em strict normal form}.
The above transition and I/O constraints are summarized graphically in \figref{transitions}.

{\em Interpretation:}
A normal-form transducer can be thought of as associating inputs and outputs with states, rather than transitions.
(Thus, it is like a Moore machine.)
The state types are
 start ($S$) and end ($E$);
 wait ($W$), in which the transducer waits for input;
 match ($M$) and delete ($D$), which process input symbols;
 insert ($I$), which writes additional output symbols; and
 null ($N$), which has no associated input or output.
All transitions also fall into one of these types, via the destination states;
thus, $\Transitions_M$ is the set of transitions ending in a match state, etc.
The transition weight ($\weight$) factors into a term that is independent of the input/output label ($\transweightfun{}$)
and a term that is independent of the source state ($\emitweightfun{}$).

{\em Universality:}
For any weak-normal form transducer $T$ there exists an equivalent in strict-normal form which can be found by applying the state-marginalization algorithm to eliminate null states.
For any transducer, there is an equivalent letter transducer in weak normal form, and therefore, in strict normal form.

\subsection{Moore and Mealy machines}

The following terms in common usage relate approximately to our definitions:

{\em Mealy machines} are transducers with I/O occurring on transitions, as with our general definition of the letter transducer.

{\em Moore machines} are transducers whose I/O is associated with states, as with our normal form.
The difference between these two views is illustrated via a small example in
\figref{condensed-emission} and \figref{fanned-emission}.  

\subsection{Composition ($T\compose U$) unifies output of $T$ with input of $U$}
\seclabel{Composition}

{\em Transducer composition:}
Given letter transducers
 $T = (\Omega_X, \Omega_Y, \States, \startstate, \laststate, \Transitions, \weight)$ and
 $U = (\Omega_Y, \Omega_Z, \States', \startstate', \laststate', \Transitions', \weight')$,
there exists a letter transducer $T\compose U = (\Omega_X, \Omega_Z, \States'' \ldots \weight'')$ 
such that $\forall x \in \Omega_X^\ast, z \in \Omega_Z^\ast$:
\[
\wtrans{\weight''}{x}{T\compose U}{z} = \sum_{y\in\Omega_Y^\ast} \wtrans{\weight}{x}{T}{y} \wtrans{\weight'}{y}{U}{z}
\]

{\em Example construction:}
Assume without loss of generality that $T$ and $U$ are in strict normal form.
Then $\States'' \subset \States \times \States'$,
$\startstate''=(\startstate,\startstate')$, $\laststate''=(\laststate,\laststate')$
and
\begin{eqnarray*}
\lefteqn{\weight''((t,u),\omega_x,\omega_z,(t',u')) =} & & \\
& & \left\{ \begin{array}{ll}
\delta(t=t') \delta(\omega_x=\epsilon) \weight'(u,\epsilon,\omega_z,u') & \mbox{if $\statetype(u) \neq W$} \\
\weight(t,\omega_x,\epsilon,t') \delta(\omega_z=\epsilon) \delta(u=u') & \mbox{if $\statetype(u) = W,\statetype(t') \notin \{M,I\}$} \\
\displaystyle
\sum_{\omega_y \in \Omega_Y} \weight(t,\omega_x,\omega_y,t') \weight'(u,\omega_y,\omega_z,u') & \mbox{if $\statetype(u) = W,\statetype(t') \in \{M,I\}$} \\
0 & \mbox{otherwise}
\end{array} \right.
\end{eqnarray*}
The resulting transducer is in weak-normal form (it can be converted to a strict-normal form transducer by eliminating null states).

In the tutorial section, many examples of simple and complex compositions are shown in \secref{Tutorial.Composition}, for instance \figref{tkf91-tkf91}, \figref{mf-tkf91} and \figref{mf-tkf91-liv}.  
See also \appref{MealyComposition} for the Mealy machine formulation.

\subsection{Intersection ($T\fork U$) unifies input of $T$ with input of $U$}
\seclabel{Fork}

{\em Transducer intersection:}
Given letter transducers
 $T = (\Omega_X, \Omega_T, \States, \startstate, \laststate, \Transitions, \weight)$ and
 $U = (\Omega_X, \Omega_U, \States', \startstate', \laststate', \Transitions', \weight')$,
there exists a letter transducer $T\fork U = (\Omega_X, \Omega_V, \States'' \ldots \weight'')$
where $\Omega_V \subseteq \gappedpair{T}{U}$
such that $\forall x \in \Omega_X^\ast, t \in \Omega_T^\ast, u \in \Omega_U^\ast$:
\[
\wtrans{\weight}{x}{T}{t} \wtrans{\weight'}{x}{U}{u} = \wtrans{\weight''}{x}{T\fork U}{(t,u)}
\]
where the term on the right is defined as follows
\[
\wtrans{\weight''}{x}{T\fork U}{(t,u)} = \sum_{v \in \Omega_V^\ast, S_1(v)=t, S_2(v)=u } \wtrans{\weight''}{x}{T\fork U}{v}
\]
Here $\Omega_V$ is the set of all possible pairwise alignment columns,
$v \in \Omega_V^\ast$ is a pairwise alignment and
$S_1(v)$ and $S_2(v)$ are the sequences in (respectively) the first and second rows of $v$.

{\em Example construction:}
Assume without loss of generality that $T$ and $U$ are in strict normal form.
Then $\States'' \subset \States \times \States'$,
$\startstate''=(\startstate,\startstate')$, $\laststate''=(\laststate,\laststate')$
and
\begin{eqnarray*}
\lefteqn{\weight''((t,u),\omega_x,(\omega_y,\omega_z),(t',u')) =} & & \\
 & & \left\{ \begin{array}{ll}
\delta(t=t') \delta(\omega_x=\omega_y=\epsilon) \weight'(u,\epsilon,\omega_z,u') & \mbox{if $\statetype(u) \neq W$} \\
\weight(t,\epsilon,\omega_x,t') \delta(\omega_x=\omega_z=\epsilon) \delta(u=u') & \mbox{if $\statetype(u) = W,\statetype(t) \neq W$} \\
\weight(t,\omega_x,\omega_y,t') \weight'(u,\omega_x,\omega_z,u') & \mbox{if $\statetype(t) = \statetype(u) = W$} \\
0 & \mbox{otherwise}
\end{array} \right.
\end{eqnarray*}
The resulting transducer is in weak-normal form (it can be converted to a strict-normal form transducer by eliminating null states).
In the tutorial section, many examples of simple and complex intersections are shown in \secref{Tutorial.Intersection}, for instance \figref{fork-subcs-submf} and  \figref{root-fork-tkf91liv-tkf91mf}. 
See also \appref{MealyFork} for the Mealy machine formulation.

\subsection{Identity and bifurcation transducers ($\identity$, $\idfork$)}
\seclabel{Identity}

{\em Identity:}
There exists a transducer $\identity=(\Omega,\Omega\ldots)$ that copies its input identically to its output.
An example construction (not in normal form) is
\begin{eqnarray*}
\identity & = & (\Omega,\Omega,\{\phi\},\phi,\phi,\transitionsof{\identity},1) \\
\transitionsof{\identity} & = & \left\{(\phi,\omega,\omega,\phi):\omega\in\Omega\right\}
\end{eqnarray*}

{\em Bifurcation:}
There exists a transducer $\idfork=(\Omega,\Omega^2\ldots)$ that duplicates its input in parallel.
That is, for input $x_1 x_2 x_3 \ldots$ it gives output $\dup{x_1}\dup{x_2}\dup{x_3}\ldots$.
An example construction (not in normal form) is
\begin{eqnarray*}
\idfork & = & (\Omega,\Omega^2,\{\phi\},\phi,\phi,\transitionsof{\idfork},1) \\
\transitionsof{\idfork} & = & \left\{\left(\phi,\omega,\dup{\omega},\phi\right):\omega\in\Omega\right\}
\end{eqnarray*}
It can be seen that $\idfork \transequiv \identity \fork \identity$.

An intersection $T \fork U$ may be considered a parallel composition of $\idfork$ with $T$ and $U$.
We write this as  $\forkfun{T}{U}$ or, diagrammatically,
\begin{parsetree}
 ( .. ( .$\idfork$. ( .$T$. ~ ) ( .$U$. ~ ) ) )
\end{parsetree}

We use the notation $\forkfun{T}{U}$ in several places, when it is convenient to have a placeholder transducer $\idfork$ at a bifurcating node in a tree.

\subsection{Exact-match recognizers ($\recognize(S)$)}
\seclabel{ExactMatch}

{\em Recognition profiles:}
A transducer $T$ is a {\em recognizer} if it has a null output alphabet, and so generates no output except the empty string.

{\em Exact-match recognizer:}
For $S \in \Omega^\ast$, there exists a transducer $\recognize(S) = (\Omega,\emptyset\ldots \weight)$
that accepts the specific sequence $S$ with weight one, but rejects all other input sequences
\[
\wtrans{\weight}{x}{\recognize(S)}{\epsilon} = \delta(x=S)
\]
Note that $\recognize(S)$ has a null output alphabet, so its only possible output is the empty string, and it is a recognizer.

In general, if $T=(\Omega_X,\Omega_Y \ldots \weight')$ is any transducer then $\forall x \in \Omega_X^\ast, y \in \Omega_Y^\ast$
\[
\wtrans{\weight'}{x}{T}{y} \equiv \wtrans{\weight}{x}{T \compose \recognize(y)}{\epsilon}
\]

An example construction (not in normal form) is
\begin{eqnarray*}
\recognize(S) & = & (\Omega,\emptyset,\mathbb{Z}_{\seqlen{S}+1},0,\seqlen{S},\transitionsof{\recognize},1) \\
\transitionsof{\recognize} & = & \left\{\left(n,\charat{S}{n+1},\epsilon,n+1\right):n \in \mathbb{Z}_{\seqlen{S}}\right)\}
\end{eqnarray*}
where $\mathbb{Z}_N$ is the set of integers modulo $N$,
 and $\charat{S}{k}$ is the $k$'th position of $S$ (for $1 \leq k \leq \seqlen{S}$).
Note that this construction has $\seqlen{S}+1$ states.

For later convenience it is useful to define the function
\begin{eqnarray*}
\profTrans{\recognize(S)}(i,j) & = & \transweightfun{\recognize(S)}(i,j) \\
& = & \delta(i+1 = j)
\end{eqnarray*}

\figref{liv-small} shows a small example of an exact-match transducer for sequence LIV,
while \figref{liv-labeled} shows an equivalent exact-match transducer in normal form.

\subsection{Generators}
\seclabel{GenSinglet}
{\em Generative transducers:}
A transducer $T$ is generative (or ``a {\em generator}'') if it has a null input alphabet, and so rejects any input except the empty string.
Then $T$ may be regarded as a state machine that generates an output, equivalent to a Hidden Markov Model.
Define the probability (weight) distribution over the output sequence
\[
P(x|T) \equiv \wtrans{\weight}{\epsilon}{T}{x}
\]

\figref{moore-mf-generator} and \figref{tkf91-labeled} are both examples of generative transducers.
\figref{moore-mf-generator} is a specific generator that only emits one sequence (with probability 1),
while \figref{tkf91-labeled} can potentially emit (and defines a probability for) any output sequence.


\subsection{Algorithmic complexities}
\begin{eqnarray*}
\numberofstates{T \compose U} & = & \order{\numberofstates{T} \numberofstates{U}} \\
\numberofstates{T \fork U} & = & \order{\numberofstates{T} \numberofstates{U}} \\
\numberofstates{\recognize(S)} & = & \order{\seqlen{S}}
\end{eqnarray*}

The complexity of computing $\wtrans{\weight}{x}{T}{y}$ is similar to the Forward algorithm:
the time complexity is $\order{\numberoftransitions{T} \seqlen{x} \seqlen{y}}$ and
the memory complexity is $\order{\numberofstates{T} \min\left(\seqlen{x},\seqlen{y}\right)}$.
Memory complexity rises to $\order{\numberofstates{T} \seqlen{x} \seqlen{y}}$ if a traceback is required.
Analogously to the Forward algorithm, there are checkpointing versions which trade memory complexity for time complexity.

\subsection{Chapman-Kolmogorov equation}
\seclabel{ChapmanKolmogorov}

If $T_t$ is a transducer parameterized by a continuous time parameter $t$,
modeling the evolution of a sequence for time $t$ under a continuous-time Markov process,
then the Chapman-Kolmogorov equation \cite{KarlinTaylor75} can be expressed as a transducer equivalence
\[
T_t \compose T_u \transequiv T_{t+u}
\]

The TKF91 transducers, for example, have this property.
Furthermore, for TKF91, $T_{t+u}$ has the same number of states and transitions as $T_t$,
so this is a kind of self-similarity.
TKF91 composed with itself is shown in \figref{tkf91-tkf91}. 

In this paper, we have deferred the difficult problem of finding time-parameterized transducers that solve this equation
(and so may be appropriate for Felsenstein recursions).
For studies of this problem the reader is referred to previous work \cite{ThorneEtal91,ThorneEtal92,KnudsenMiyamoto2003,MiklosLunterHolmes2004,Rivas05}.

\subsection{Hierarchy of phylogenetic transducers}
\seclabel{ModelStructure}
\subsubsection{Phylogenetic tree ($n$, $\leaves$)}
Suppose we have an evolutionary model defined on a rooted binary phylogenetic tree,
and a set of $\numberofleaves$ observed sequences associated with the leaf nodes of the tree.

The nodes are numbered in preorder, with internal nodes $(1 \ldots \numberofinternalnodes)$ preceding leaf nodes $\leaves = \{ \leafnode{1} \ldots \numberofnodes \}$. Node $1$ is the root.

\subsubsection{Hidden and observed sequences ($\outputn{n}$)}

Let $\outputn{n} \in \Omega^\ast$ denote the sequence at node $n$ and
let $\outputs = \{ \outputn{n}, n \in \leaves \}$ denote the observed leaf-node sequences.

\subsubsection{Model components ($B_n$, $R$)}

Let $B_n=(\Omega,\Omega,\ldots)$ be a transducer modeling the evolution on the branch to node $n>1$, from $n$'s parent.
Let $R=(\emptyset,\Omega,\ldots)$ be a generator modeling the distribution of ancestral sequences at the root node.

\figref{tkf91-labeled} and \figref{protpal} are examples of $B_n$ transducers.
\figref{tkf91-root} is an example of an $R$ transducer.

\subsubsection{The forward model ($F_n$)}
\seclabel{ForwardModel}

If $n \geq 1$ is a leaf node, define $F_n = \identity$.
Otherwise, let $(l,r)$ denote the left and right child nodes, and define
\[
F_n = (B_l \compose F_l) \fork (B_r \compose F_r)
\]

which we can represent as
\begin{parsetree}
 ( .. ( .$\idfork$. ( .$B_l$. ( .$F_l$. ~ ) ) ( .$B_r$. ( .$F_r$. ~ ) )  ) )
\end{parsetree}
(recall that $\idfork$ is the bifurcation transducer).
\\

The complete, generative transducer is
$F_0 = R \compose F_1$

The output alphabet of $F_0$ is $\gappedalphabet{}^\numberofleaves$ where $\numberofleaves$ is the number of leaf sequences.
Letting $S_n:\Transitions^\ast \to \Omega^\ast$ denote the map from a transition path $\pi$ to the $n$'th output leaf sequence (with gaps removed),
we define the output distribution
\[
P(\outputs|F_0) = \wtrans{\weight}{\epsilon}{F_0}{\outputs} = \sum_{\pi: S_n(\pi)=\outputn{n} \forall n \in \leaves_n} \weight(\pi)
\]
where $\leaves_n$ denotes the set of leaf nodes that have $n$ as a common ancestor.

Note that $\numberofstates{F_0} \simeq \prod_n^{\numberofnodes} \numberofstates{B_n}$ where $\numberofnodes$ is the number of nodes in the tree.
So the state space grows exponentially with the size of the tree---and this is before we have even introduced any sequences.
We seek to avoid this with our hierarchy of approximate models, which will have state spaces that are bounded in size.

First, however, we expand the state space even more, by introducing the observed sequences explicitly into the model.

\subsubsection{The evidence-expanded model ($G_n$)}
\seclabel{EvidenceExpandedModel}

Inference with stochastic grammars often uses a dynamic programming matrix (e.g. the Inside matrix)
to track the ways that a given evidential sequence can be produced by a given grammar.

For our purposes it is useful to introduce the evidence in a different way,
by transforming the model to incorporate the evidence directly.
We augment the state space so that the model is no longer capable of generating any sequences {\em except} the observed $\{\outputn{n}\}$,
by composing $F_0$ with exact-match transducers that will only accept the observed sequences.
This yields a model whose state space is very large and, in fact, is directly analogous to the Inside dynamic programming matrix.

If $n \geq 1$ is a leaf node, define $G_n = \recognize(\outputn{n})$.
The number of states is $\numberofstates{G_n} = \order{\outseqlen{n}}$.

Otherwise, let $(l,r)$ denote the left and right child nodes, and define
\[
G_n = (B_l \compose G_l) \fork (B_r \compose G_r)
\]

which we can represent as
\begin{parsetree}
 ( .. ( .$\idfork$. ( .$B_l$. .$G_l$. ) ( .$B_r$. .$G_r$. )  ) )
\end{parsetree}
\\

\figref{fork-tkf91liv-tkf91mf} and \figref{fork3-tkf91liv-tkf91mf-tkf91cs} are examples of $G_n$-transducers
for the tree of \figref{cs-mf-liv-tree}.

The complete evidence-expanded model is
$G_0 = R \compose G_1$.
(In our tutorial example, the state graph of this transducer has too many transitions to show, but it is
the configuration shown in \figref{cs-mf-liv-machines}.)

The probability that the forward model $F_0$ generates the evidential sequences $\outputs$
is identical to the probability that the evidence-expanded model $G_0$ generates the empty string
\[
P(\outputs | F_0) = \wtrans{\weight}{\epsilon}{F_0}{\outputs} = \wtrans{\weight}{\epsilon}{G_0}{\epsilon}
\]

Note the astronomical number of states in $G_0$
\[
\numberofstates{G_0} \simeq \left( \prod_{n=1}^{\numberofleaves} \outseqlen{n} \right) \left( \prod_{n=1}^{\numberofnodes} \numberofstates{B_n} \right)
\]
This is even worse than $F_0$; in fact, it is the same as the number of cells in the Inside matrix for computing $P(\outputs|F_0)$.
The good news is we are about to start constraining it.

\subsubsection{The constrained-expanded model ($H_n$, $E_n$, $M_n$, $Q_n$)}
\seclabel{ConstrainedExpanded}
We now introduce a progressive series of approximating constraints to make inference under the model more tractable.

If $n \geq 1$ is a leaf node, define $H_n = \recognize(\outputn{n}) \equiv G_n$.
The number of states is $\numberofstates{H_n} \simeq \outseqlen{n}$, just as with $G_n$.

Otherwise, let $(l,r)$ denote the left and right child nodes, and define
\[
H_n = (B_l \compose E_l) \fork (B_r \compose E_r)
\]

where $\statesof{E_n} \subseteq \statesof{H_n}$.

We can represent $H_n$ diagramatically as
\begin{parsetree}
 ( .. ( .$\idfork$. ( .$B_l$. .$E_l$. ) ( .$B_r$. .$E_r$. )  ) )
\end{parsetree}
\\
\figref{fork-tkf91liv-tkf91mf}, \figref{fork-tkf91viterbi-tkf91cs} and \figref{fork-tkf91forward2-tkf91cs} are examples of $H_n$ transducers.

Transducer $E_n$, which is what we mean by the ``constrained-expanded model'', is effectively a profile of sequences that might plausibly appear at node $n$, given the observed descendants of that node.
\figref{viterbi-profile}  and \figref{forward2-profile} are examples of such  transducers for
the ``intermediate sequence'' in the tree of \figref{cs-mf-liv-tree}.
(\figref{viterbi-fork-tkf91liv-tkf91mf} and \figref{forward2-fork-tkf91liv-tkf91mf} show the relationship to the corresponding $H_n$ transducers).

The profile is constructed as follows.

The general idea is to generate a set of candidate sequences at node $n$, by sampling from the posterior distribution of such sequences {\bf given only the descendants of node $n$,}
ignoring (for the moment) the nodes outside the $n$-rooted subtree.
To do this, we need to introduce a prior distribution over the sequence at node $n$.
This prior is an approximation to replace the true (but as yet unknown) posterior distribution due to nodes outside the $n$-rooted subtree
(including $n$'s parent, and ancestors all the way back to the root, as well as siblings, cousins etc.)

A plausible choice for this prior, equivalent to assuming {\em stationarity} of the underlying evolutionary process, is the same prior that we use for the root node; that is, the generator model $R$.
We therefore define
\begin{eqnarray*}
M_n & = & R \compose H_n \\
& = & R \compose ((B_l \compose E_l) \fork (B_r \compose E_r))
\end{eqnarray*}

We can represent $M_n$ diagramatically as
\begin{parsetree}
 ( .$R$. ( .$\idfork$. ( .$B_l$. .$E_l$. ) ( .$B_r$. .$E_r$. )  ) )
\end{parsetree}
\\

\figref{root-fork-tkf91liv-tkf91mf} is an example of the $M_n$ type of model.

The transducer $Q_n = R \compose (B_l \fork B_r)$, which forms the comparison kernel of $M_n$, is also useful.
It can be represented as
\begin{parsetree}
 ( .$R$. ( .$\idfork$. ( .$B_l$. .. ) ( .$B_r$. .. ) ) )
\end{parsetree}
\\

Conceptually, $Q_n$ is a generator with two output tapes (i.e. a Pair HMM).
These tapes are generated by sampling a sequence from the root generator $R$,
making two copies of it, and feeding the two copies into $B_l$ and $B_r$ respectively.
The outputs of $B_l$ and $B_r$ are the two outputs of the Pair HMM.
The different states of $Q_n$ encode information about how each output symbol originated
(e.g. by root insertions that were then matched on the branches, {\em vs} insertions on the branches).
\figref{root-fork-tkf91-tkf91} shows an example of a $Q_n$-like transducer.

Transducer $M_n$ can be thought of as the dynamic programming matrix that we get
if we use the Pair HMM $Q_n$ to align the recognition profiles $E_l$ and $E_r$.

\subsubsection{Component state tuples $\mstate$}
Suppose that $a \in \statesof{A}, b \in \statesof{B}, \upsilon \in \statesof{\idfork}$.
Our construction of composite transducers allows us to represent any state in $A \fork B = \forkfun{A}{B}$ as a tuple $(\upsilon, a,b)$.
Similarly, any state in $A \compose B$ can be represented as $(a,b)$.
Each state in $M_n$ can thus be written as a tuple $\mstate$ of component states, where
\begin{itemize}
\item $\rho$ is the state of the generator transducer $R$
\item $\upsilon$ is the state of the bifurcation transducer $\idfork$
\item $b_l$ is the state of the left-branch transducer $B_l$
\item $e_l$ is the state of the left child profile transducer $E_l$
\item $b_r$ is the state of the right-branch transducer $B_l$
\item $e_r$ is the state of the right child profile transducer $E_r$
\end{itemize}
Similarly, each state in $H_n$ (and $E_n$) can be written as a tuple $\hstate$.

\subsubsection{Constructing $E_n$ from $H_n$}
The construction of $E_n$ as a sub-model of $H_n$ proceeds as follows:
\begin{enumerate}
\item sample a set of $K$ paths from $P(\pi | M_n) = \weight(\pi) / \wtrans{\weight}{\epsilon}{M_n}{\epsilon}$;
\item identify the set of $M_n$-states $\{\mstate\}$ used by the sampled paths;
\item strip off the leading $\rho$'s from these $M_n$-states to find the associated set of $H_n$-states $\{\hstate\}$;
\item the set of $H_n$-states so constructed is the subset of $E_n$'s states that have type $D$ (wait states must be added to place it in strict normal form).
\end{enumerate}

Here $K$ plays the role of a bounding parameter.
For the constrained-expanded transducer, $\numberofstates{E_n} \simeq KL$, where $L = \max_n \outseqlen{n}$.
Models $H_n$ and $M_n$, however, contain $\order{b^2 K^2 L^2}$ states,
where $b = \max_n \numberofstates{B_n}$,
as they are constructed by intersection of two $\order{bKL}$-state transducers ($B_l \compose E_l$ and $B_r \compose E_r$).

\subsection{Explicit construction of $Q_n$}
\seclabel{Qn}
\begin{eqnarray*}
Q_n & = & R \compose (B_l \fork B_r) \\
& = & (\emptyset,\gapsquared,\statesof{Q_n},\startstateof{Q_n},\laststateof{Q_n},\transitionsof{Q_n},\weightfunof{Q_n}) \\
\startstateof{Q_n} & = & (\startstateof{R},\startstateof{\idfork},\startstateof{B_l},\startstateof{B_r}) \\
\laststateof{Q_n} & = & (\laststateof{R},\laststateof{\idfork},\laststateof{B_l},\laststateof{B_r})
\end{eqnarray*}

\subsubsection{States of $Q_n$}

Define $\statetype(\phi_1,\phi_2,\phi_3 \ldots) = (\statetype(\phi_1), \statetype(\phi_2), \statetype(\phi_3) \ldots)$.

Let $q=\qstate\in\statesof{Q_n}$.
We construct $\statesof{Q_n}$ from classes, adopting the convention that each class of states is defined by its associated types:
\[
\stateset{class} = \{ q: \statetype\qstate \in \typeset{class} \}
\]

The state typings are
\begin{eqnarray*}
\matchtypes & = & \{ (I,M,M,M) \}  \\
\rightdeletetypes & = & \{ (I,M,M,D) \}  \\
\leftdeletetypes & = & \{ (I,M,D,M) \}  \\
\nulltypes & = & \{ (I,M,D,D) \}  \\
\rightinserttypes & = & \{ (S,S,S,I),\ (I,M,M,I),\ (I,M,D,I) \}  \\
\leftinserttypes & = & \{ (S,S,I,W),\ (I,M,I,W) \}  \\
\qwaittypes & = & \{ (W,W,W,W) \} \\
\rightemittypes & = & \leftdeletetypes\ \cup\ \rightinserttypes \\
\leftemittypes & = & \leftinserttypes\ \cup\ \rightdeletetypes
\end{eqnarray*}

The state space of $Q_n$ is
\[
\statesof{Q_n} = \{ \startstateof{Q_n},\ \laststateof{Q_n} \}\ \cup\ \matchstates\ \cup\ \leftemitstates\ \cup\ \rightemitstates\ \cup\ \nullstates\ \cup\ \qwaitstates
\]

It is possible to calculate transition and I/O weights of $Q_n$
by starting with the example constructions given for $T \compose U$ and $T \fork U$,
then eliminating states that are not in the above set.
This gives the results described in the following sections.

\subsubsection{I/O weights of $Q_n$}

Let $(\omega_l,\omega_r)\ \in \gapsquared$.

The I/O weight function for $Q_n$ is
\[
\emitweightfun{Q_n}(\epsilon,(\omega_l,\omega_r),q) = \left\{
\begin{array}{ll}
\displaystyle
\sum_{\omega \in \Omega} \emitweightfun{R}(\epsilon,\omega,R)
 \emitweightfun{B_l}(\omega,\omega_l,b_l)
 \emitweightfun{B_r}(\omega,\omega_r,b_r)
 & \mbox{if $q \in \matchstates$} \\
\displaystyle
\sum_{\omega \in \Omega} \emitweightfun{R}(\epsilon,\omega,R)
 \emitweightfun{B_r}(\omega,\omega_r,b_r)
 & \mbox{if $q \in \leftdeletestates$} \\
\displaystyle
\sum_{\omega \in \Omega} \emitweightfun{R}(\epsilon,\omega,R)
 \emitweightfun{B_l}(\omega,\omega_l,b_l)
 & \mbox{if $q \in \rightdeletestates$} \\
 \emitweightfun{B_l}(\epsilon,\omega_l,b_l)
 & \mbox{if $q \in \leftinsertstates$} \\
 \emitweightfun{B_r}(\epsilon,\omega_r,b_r)
 & \mbox{if $q \in \rightinsertstates$} \\
1
 & \mbox{otherwise}
\end{array}
\right.
\]

\subsubsection{Transition weights of $Q_n$}
The transition weight between two states
$q=\qstate$ and
$q'=\qstatedest$
always takes the form
\[
\transweightfun{Q_n}(q,q') \equiv \sumoverpaths{R} . \sumoverpaths{\idfork} . \sumoverpaths{B_l} . \sumoverpaths{B_r}
\]
where $\sumoverpaths{T}$ represents a sum over a set of paths through component transducer $T$.
The allowed paths $\{\pi_T\}$ are constrained by the types of $q,q'$ as shown in Table~\ref{QTransitionTypes}.
\begin{table}
\[
\begin{array}{ll|cccc}
\statetype\qstate & \statetype\qstatedest & |\pi_R| & |\pi_{\idfork}| & |\pi_{B_l}| & |\pi_{B_r}| \\
\hline
(S,S,S,*) & (S,S,S,I) & 0 & 0 & 0 & 1 \\
          & (S,S,I,W) & 0 & 0 & 1 & 1 \\
          & (I,M,M,M) & 1 & 2 & 2 & 2 \\
          & (I,M,M,D) & 1 & 2 & 2 & 2 \\
          & (I,M,D,M) & 1 & 2 & 2 & 2 \\
          & (I,M,D,D) & 1 & 2 & 2 & 2 \\
          & (W,W,W,W) & 1 & 1 & 1 & 1 \\
\hline
(S,S,I,W) & (S,S,I,W) & 0 & 0 & 1 & 0 \\
          & (I,M,M,M) & 1 & 2 & 2 & 1 \\
          & (I,M,M,D) & 1 & 2 & 2 & 1 \\
          & (I,M,D,M) & 1 & 2 & 2 & 1 \\
          & (I,M,D,D) & 1 & 2 & 2 & 1 \\
          & (W,W,W,W) & 1 & 1 & 1 & 0 \\
\hline
(I,M,M,*) & (I,M,M,I) & 0 & 0 & 0 & 1 \\
          & (I,M,I,W) & 0 & 0 & 1 & 1 \\
          & (I,M,M,M) & 1 & 2 & 2 & 2 \\
          & (I,M,M,D) & 1 & 2 & 2 & 2 \\
          & (I,M,D,M) & 1 & 2 & 2 & 2 \\
          & (I,M,D,D) & 1 & 2 & 2 & 2 \\
          & (W,W,W,W) & 1 & 1 & 1 & 1 \\
\hline
(I,M,D,*) & (I,M,D,I) & 0 & 0 & 0 & 1 \\
          & (I,M,I,W) & 0 & 0 & 1 & 1 \\
          & (I,M,M,M) & 1 & 2 & 2 & 2 \\
          & (I,M,M,D) & 1 & 2 & 2 & 2 \\
          & (I,M,D,M) & 1 & 2 & 2 & 2 \\
          & (I,M,D,D) & 1 & 2 & 2 & 2 \\
          & (W,W,W,W) & 1 & 1 & 1 & 1 \\
\hline
(I,M,I,W) & (I,M,I,W) & 0 & 0 & 1 & 0 \\
          & (I,M,M,M) & 1 & 2 & 2 & 1 \\
          & (I,M,M,D) & 1 & 2 & 2 & 1 \\
          & (I,M,D,M) & 1 & 2 & 2 & 1 \\
          & (I,M,D,D) & 1 & 2 & 2 & 1 \\
          & (W,W,W,W) & 1 & 1 & 1 & 0 \\
\hline
(W,W,W,W) & (E,E,E,E) & 1 & 1 & 1 & 1 \\
\end{array}
\]
\caption[]{ \label{QTransitionTypes} Transition types of $Q_n$, the transducer described in  \secref{Qn}
 This transducer requires its input to be empty: it is
 `generative'. It jointly models a parent sequence (hidden) and a pair
 of sibling sequences (outputs), and is somewhat analogous to a Pair 
 HMM. It is used during progressive reconstruction.}
\end{table}
Table~\ref{QTransitionTypes} uses the following conventions:
\begin{itemize}
\item A $0$ in any column means that the corresponding state must remain unchanged.
For example, if $|\pi_{B_l}|=0$ then
\[
\sumoverpaths{B_l} \equiv \delta(b_l=b'_l)
\]
\item A $1$ in any column means that the corresponding transducer makes a single transition.
For example, if $|\pi_{B_l}|=1$ then
\[
\sumoverpaths{B_l} \equiv \transweightfun{B_l}(b_l,b'_l)
\]
\item A $2$ in any column means that the corresponding transducer makes two transitions, via an intermediate state.
For example, if $|\pi_{B_l}|=2$ then
\[
\sumoverpaths{B_l} \equiv \sum_{b''_l \in \statesof{B_l}} \transweightfun{B_l}(b_l,b''_l) \transweightfun{B_l}(b''_l,b'_l)
\]
(Since the transducers are in strict normal form, and given the context in which the $2$'s appear,
it will always be the case that the intermediate state $b''_l$ has type $W$.)
\item An asterisk ($*$) in a type-tuple is interpreted as a wildcard; for example, $(S,S,S,*)$ corresponds to $\{ (S,S,S,S),\ (S,S,S,I) \}$.
\item If a transition does not appear in the above table, or if any of the $\transweightfun{}$'s are zero,
then $\nexists (h,\omega,\omega',h') \in \transitionsof{H_n}$.
\end{itemize}

\subsubsection{State types of $Q_n$}

\[
\statetype(q) = \left\{ \begin{array}{ll}
S & \mbox{if $q = \startstateof{Q_n}$} \\
E & \mbox{if $q = \laststateof{Q_n}$} \\
W & \mbox{if $q \in \qwaitstates$} \\
I & \mbox{if $q \in \matchstates\ \cup\ \leftemitstates\ \cup\ \rightemitstates$} \\
N & \mbox{if $q \in \nullstates$}
\end{array} \right.
\]

Note that $Q_n$ contains null states ($\nullstates$) corresponding to coincident deletions on branches $n \to l$ and $n \to r$.
These states have $\statetype\qstate=(I,M,D,D)$.
There are transition paths that go through these states, including paths that cycle indefinitely among these states.

We need to eliminate these states before constructing $M_n$.
Let $Q'_n \transequiv Q_n$ denote the transducer obtained from $Q_n$ by marginalizing $\nullstates$
\[
\statesof{Q'_n} = \{ \startstateof{Q_n},\ \laststateof{Q_n} \}\ \cup\ \matchstates\ \cup\ \leftemitstates\ \cup\ \rightemitstates\ \cup\ \qwaitstates
\]

The question arises, how to restore these states when constructing $E_n$?
Ortheus samples them randomly,
but (empirically) a lot of samples are needed before there is any chance of guessing the right number,
and in practice it makes little difference to the accuracy of the reconstruction.
In principle it might be possible to leave them in as self-looping delete states in $E_n$, but this would make $E_n$ cyclic.

\subsection{Explicit construction of $M_n$ using $Q'_n$, $E_l$ and $E_r$}
\seclabel{Mn}
Refer to the previous section for definitions pertaining to $Q'_n$.

\begin{eqnarray*}
M_n & = & R \compose ((B_l \compose E_l) \fork (B_r \compose E_r)) \\
Q'_n & \transequiv & R \compose (B_l \fork B_r)
\end{eqnarray*}

\subsubsection{States of $M_n$}
The complete set of $M_n$-states is
\begin{eqnarray*}
\statesof{M_n} & = & \{ \startstateof{M_n},\ \laststateof{M_n} \} \\
& & \quad \cup\ \{ \mstate: \qstate \in \matchstates,\ \statetype(e_l)=\statetype(e_r)=D \} \\
& & \quad \cup\ \{ \mstate: \qstate \in \leftemitstates,\ \statetype(e_l)=D, \statetype(e_r)=W \} \\
& & \quad \cup\ \{ \mstate: \qstate \in \rightemitstates,\ \statetype(e_l)=W, \statetype(e_r)=D \} \\
& & \quad \cup\ \{ \mstate: \qstate \in \waitstates,\ \statetype(e_l)=\statetype(e_r)=W \}
\end{eqnarray*}

\subsubsection{I/O weights of $M_n$}
Let $m = \mstate$ be an $M_n$-state and $q = \qstate$ the subsumed $Q'_n$-state.

Similarly, let $m' = \mstatedest$ and $q' = \qstatedest$.

The I/O weight function for $M_n$ is
\[
\emitweightfun{M_n}(\epsilon,\epsilon,m) = \left\{
\begin{array}{ll}
\displaystyle
\sum_{\omega_l \in \Omega}
\sum_{\omega_r \in \Omega}
 \emitweightfun{Q'_n}(\epsilon,(\omega_l,\omega_r),q)
. \emitweightfun{E_l}(\omega_l,\epsilon,e_l)
. \emitweightfun{E_r}(\omega_r,\epsilon,e_r)
 & \mbox{if $q \in \matchstates$} \\
\displaystyle
\sum_{\omega_l \in \Omega}
 \emitweightfun{Q'_n}(\epsilon,(\omega_l,\epsilon),q)
. \emitweightfun{E_l}(\omega_l,\epsilon,e_l)
 & \mbox{if $q \in \leftemitstates$} \\
\displaystyle
\sum_{\omega_r \in \Omega}
 \emitweightfun{Q'_n}(\epsilon,(\epsilon,\omega_r),q)
. \emitweightfun{E_r}(\omega_r,\epsilon,e_r)
 & \mbox{if $q \in \rightemitstates$} \\
1
 & \mbox{otherwise}
\end{array}
\right.
\]

\subsubsection{Transitions of $M_n$}

As before,
\begin{eqnarray*}
q & = & \qstate \\
m & = & \mstate \\
q' & = & \qstatedest \\
m' & = & \mstatedest
\end{eqnarray*}
An ``upper bound'' (i.e. superset) of the transition set of $M_n$ is as follows
\begin{eqnarray*}
\transitionsof{M_n} & \subseteq & \{ (m,\epsilon,\epsilon,m'): q'\in\matchstates,\statetype(q)\in\{S,I\},\statetype(e'_l,e'_r)=(D,D) \} \\
 & & \ \cup\ \{ (m,\epsilon,\epsilon,m'): q'\in\leftemitstates,\statetype(q)\in\{S,I\},\statetype(e'_l,e'_r)=(D,W) \} \\
 & & \ \cup\ \{ (m,\epsilon,\epsilon,m'): q'\in\rightemitstates,\statetype(q)\in\{S,I\},\statetype(e'_l,e'_r)=(W,D) \} \\
 & & \ \cup\ \{ (m,\epsilon,\epsilon,m'): q'\in\waitstates,\statetype(q)\in\{S,I\},\statetype(e'_l,e'_r)=(W,W) \} \\
 & & \ \cup\ \{ (m,\epsilon,\epsilon,m'): \statetype(q,e_l,e_r)=(W,W,W),\statetype(q',e'_l,e'_r)=(E,E,E) \}
\end{eqnarray*}
More precisely, $\transitionsof{M_n}$ contains the transitions in the above set for which the transition weight (defined in the next section) is nonzero.
(This ensures that the individual transition paths $q \to q'$, $e_l \to e'_l$ and $e_r \to e'_r$ exist with nonzero weight.)

\subsubsection{Transition weights of $M_n$}

Let $\transviawait{E_n}(e,e')$ be the weight of {\bf either} the direct transition $e \to e'$,
{\bf or} a double transition $e \to e'' \to e'$ summed over all intermediate states $e''$
\[
\transviawait{E_n}(e,e') = \left\{
\begin{array}{ll}
\displaystyle
\sum_{e'' \in \statesof{E_n}}
\transweightfun{E_n}(e,e'')
\transweightfun{E_n}(e'',e')
& \mbox{if $\statetype(e) \in \{S,D\}$} \\
\transweightfun{E_n}(e,e')
& \mbox{if $\statetype(e) = W$} \\
\end{array}
\right.
\]

Let $\transtowait{E_n}(e,e')$ be the weight of a transition (or non-transition) that leaves $E_n$ in a wait state
\[
\transtowait{E_n}(e,e') = \left\{
\begin{array}{ll}
\transweightfun{E_n}(e,e')
& \mbox{if $\statetype(e) \in \{S,D\},\ \statetype(e') = W$} \\
1
& \mbox{if $e=e',\ \statetype(e') = W$} \\
0
& \mbox{otherwise} \\
\end{array}
\right.
\]

The transition weight function for $M_n$ is
\[
\transweightfun{M_n}(m,m') = \transweightfun{Q'_n}(q,q') \times \left\{
\begin{array}{ll}
\transviawait{E_l}(e_l,e'_l) \transviawait{E_r}(e_r,e'_r)
 & \mbox{if $q' \in \matchstates $} \\
\transviawait{E_l}(e_l,e'_l) \transtowait{E_r}(e_r,e'_r)
 & \mbox{if $q' \in \leftemitstates$} \\
\transtowait{E_l}(e_l,e'_l) \transviawait{E_r}(e_r,e'_r)
 & \mbox{if $q' \in \rightemitstates$} \\
\transtowait{E_l}(e_l,e'_l) \transtowait{E_r}(e_r,e'_r)
 & \mbox{if $q' \in \waitstates$} \\
\transweightfun{E_l}(e_l,e'_l) \transweightfun{E_r}(e_r,e'_r)
 & \mbox{if $q' = \laststateof{Q'_n}$}
\end{array}
\right.
\]

\subsection{Explicit construction of $H_n$}
\seclabel{Hn}
This construction is somewhat redundant, since we construct $M_n$ from $Q_n$, $E_l$ and $E_r$, rather than from $R \compose H_n$.
It is retained for comparison.

\begin{eqnarray*}
H_n & = & (B_l \compose E_l) \fork (B_r \compose E_r) \\
& = & (\Omega,\emptyset,\statesof{H_n},\startstateof{H_n},\laststateof{H_n},\transitionsof{H_n},\weightfunof{H_n}) \\
\end{eqnarray*}

Assume $B_l,B_r,E_l,E_r$ in strict-normal form.

\subsubsection{States of $H_n$}

Define $\statetype(\phi_1,\phi_2,\phi_3 \ldots) = (\statetype(\phi_1), \statetype(\phi_2), \statetype(\phi_3) \ldots)$.

Let $h=\hstate\in\statesof{H_n}$.
We construct $\statesof{H_n}$ from classes, adopting the convention that each class of states is defined by its associated types:
\[
\stateset{class} = \{ h: \statetype\hstate \in \typeset{class} \}
\]

Define $\externalcascades \subset \statesof{H_n}$ to be the subset of $H_n$-states that follow {\em externally-driven cascades}
\begin{eqnarray*}
\externaltypes & = & \{ (M,M,D,M,D),\ (M,M,D,D,W), \\
& & \quad (M,D,W,M,D),\ (M,D,W,D,W) \}
\end{eqnarray*}

Define $\internalcascades \subset \statesof{H_n}$ to be the subset of $H_n$-states that follow {\em internal cascades}
\begin{eqnarray*}
\internaltypes & = & \lefttypes \cup \righttypes \\
\lefttypes & = & \{ (S,I,D,W,W),\ (M,I,D,W,W) \} \\
\righttypes & =  & \{ (S,S,S,I,D),\ (M,M,D,I,D),\ (M,D,W,I,D) \}
\end{eqnarray*}

Remaining states are the start, end, and wait states:
\begin{eqnarray*}
\startstateof{H_n} & = & (\startstateof{\idfork},\startstateof{B_l},\startstateof{E_l},\startstateof{B_r},\startstateof{E_r}) \\
\laststateof{H_n} & = & (\laststateof{\idfork},\laststateof{B_l},\laststateof{E_l},\laststateof{B_r},\laststateof{E_r}) \\
\waittypes & = & \{ (W,W,W,W,W) \}
\end{eqnarray*}
The complete set of $H_n$-states is
\[
\statesof{H_n} = \{ \startstateof{H_n},\ \laststateof{H_n} \}\ \cup\ \externalcascades\ \cup\ \internalcascades\ \cup\ \waitstates
\]

It is possible to calculate transition and I/O weights of $H_n$
by starting with the example constructions given for $T \compose U$ and $T \fork U$,
then eliminating states that are not in the above set.
This gives the results described in the following sections.

\subsubsection{I/O weights of $H_n$}

Let $\omega,\omega' \in \gappedalphabet{}$.

Let $C_n(b_n,e_n)$ be the I/O weight function for $B_n \compose E_n$ on a transition into composite state $(b_n,e_n)$ where $\statetype(b_n,e_n)=(I,D)$
\[
C_n(b_n,e_n) = \sum_{\omega \in \Omega} \emitweightfun{B_n}(\epsilon,\omega,b_n) \emitweightfun{E_n}(\omega,\epsilon,e_n)
\]

Let $D_n(\omega,b_n,e_n)$ be the I/O weight function for $B_n \compose E_n$ on a transition into composite state $(b_n,e_n)$
 where $\statetype(b_n,e_n) \in \{(M,D),(D,W)\}$ with input symbol $\omega$ 
\[
D_n(\omega,b_n,e_n) = \left\{
\begin{array}{ll}
\displaystyle
\sum_{\omega' \in \Omega} \emitweightfun{B_n}(\omega,\omega',b_n) \emitweightfun{E_n}(\omega',\epsilon,e_n)
 & \mbox{if $\statetype(b_n,e_n)=(M,D)$} \\
\emitweightfun{B_n}(\omega,\epsilon,b_n)
 & \mbox{if $\statetype(b_n,e_n)=(D,W)$}
\end{array}
\right.
\]

The I/O weight function for $H_n$ is
\[
\emitweightfun{H_n}(\omega,\epsilon,h) = \left\{
\begin{array}{ll}
D_l(\omega,b_l,e_l) D_r(\omega,b_r,e_r)
 & \mbox{if $h \in \externalcascades$} \\
C_l(b_l,e_l)
 & \mbox{if $h \in \leftcascades$} \\
C_r(b_r,e_r)
 & \mbox{if $h \in \rightcascades$} \\
1
 & \mbox{otherwise}
\end{array}
\right.
\]

\subsubsection{Transition weights of $H_n$}

The transition weight between two states
$h=\hstate$ and
$h'=\hstatedest$
always takes the form
\[
\transweightfun{H_n}(h,h') \equiv \sumoverpaths{\idfork} . \sumoverpaths{B_l} . \sumoverpaths{E_l} . \sumoverpaths{B_r} . \sumoverpaths{E_r}
\]
where the RHS terms again represent sums over paths, with the allowed paths depending on the types of $h,h'$ as shown in Table~\ref{HTransitionTypes}.
\begin{table}
\[
\begin{array}{ll|ccccc}
\statetype\hstate & \statetype\hstatedest & |\pi_{\idfork}| & |\pi_{B_l}| & |\pi_{E_l}| & |\pi_{B_r}| & |\pi_{E_r}| \\
\hline
(S,S,S,S,S) & (S,S,S,I,D) & 0 & 0 & 0 & 1 & 2 \\
            & (S,I,D,W,W) & 0 & 1 & 2 & 1 & 1 \\
            & (W,W,W,W,W) & 1 & 1 & 1 & 1 & 1 \\
\hline
(S,S,S,I,D) & (S,S,S,I,D) & 0 & 0 & 0 & 1 & 2 \\
            & (S,I,D,W,W) & 0 & 1 & 2 & 1 & 1 \\
            & (W,W,W,W,W) & 1 & 1 & 1 & 1 & 1 \\
\hline
(S,I,D,W,W) & (S,I,D,W,W) & 0 & 1 & 2 & 0 & 0 \\
            & (W,W,W,W,W) & 1 & 1 & 1 & 0 & 0 \\
\hline
(W,W,W,W,W) & (M,M,D,M,D) & 1 & 1 & 1 & 1 & 1 \\
            & (M,M,D,D,W) & 1 & 1 & 1 & 1 & 0 \\
            & (M,D,W,M,D) & 1 & 1 & 0 & 1 & 1 \\
            & (M,D,W,D,W) & 1 & 1 & 0 & 1 & 0 \\
            & (E,E,E,E,E) & 1 & 1 & 1 & 1 & 1 \\
\hline
(M,M,D,M,D) & (M,M,D,I,D) & 0 & 0 & 0 & 1 & 2 \\
            & (M,I,D,W,W) & 0 & 1 & 2 & 1 & 1 \\
            & (W,W,W,W,W) & 1 & 1 & 1 & 1 & 1 \\
\hline
(M,M,D,D,W) & (M,M,D,I,D) & 0 & 0 & 0 & 1 & 1 \\
            & (M,I,D,W,W) & 0 & 1 & 2 & 1 & 0 \\
            & (W,W,W,W,W) & 1 & 1 & 1 & 1 & 0 \\
\hline
(M,D,W,M,D) & (M,D,W,I,D) & 0 & 0 & 0 & 1 & 2 \\
            & (M,I,D,W,W) & 0 & 1 & 1 & 1 & 1 \\
            & (W,W,W,W,W) & 1 & 1 & 0 & 1 & 1 \\
\hline
(M,D,W,D,W) & (M,D,W,I,D) & 0 & 0 & 0 & 1 & 1 \\
            & (M,I,D,W,W) & 0 & 1 & 1 & 1 & 0 \\
            & (W,W,W,W,W) & 1 & 1 & 0 & 1 & 0 \\
\hline
(M,M,D,I,D) & (M,M,D,I,D) & 0 & 0 & 0 & 1 & 2 \\
            & (M,I,D,W,W) & 0 & 1 & 2 & 1 & 1 \\
            & (W,W,W,W,W) & 1 & 1 & 1 & 1 & 1 \\
\hline
(M,D,W,I,D) & (M,D,W,I,D) & 0 & 0 & 0 & 1 & 2 \\
            & (M,I,D,W,W) & 0 & 1 & 1 & 1 & 1 \\
            & (W,W,W,W,W) & 1 & 1 & 0 & 1 & 1 \\
\hline
(M,I,D,W,W) & (M,I,D,W,W) & 0 & 1 & 2 & 0 & 0 \\
            & (W,W,W,W,W) & 1 & 1 & 1 & 0 & 0 \\
\end{array}
\]
\caption{
\label{HTransitionTypes}
Transition types of $H_n$, the transducer described in  \secref{Hn}
 This transducer requires non-empty input: it is
 a `recognizing profile' or `recognizer'. It  models a subtree of sequences
 conditional on an absorbed parental sequence. 
It is used during progressive reconstruction.}
\end{table}
Table~\ref{HTransitionTypes} uses the same conventions as Table~\ref{QTransitionTypes}.

\subsubsection{State types of $H_n$}

\[
\statetype(h) = \left\{ \begin{array}{ll}
S & \mbox{if $h = \startstateof{H_n}$} \\
E & \mbox{if $h = \laststateof{H_n}$} \\
W & \mbox{if $h \in \waitstates$} \\
D & \mbox{if $h \in \externalcascades$} \\
N & \mbox{if $h \in \internalcascades$}
\end{array} \right.
\]

Since $H_n$ contains states of type $N$ (the internal cascades),
it is necessary to eliminate these states from $E_n$ (after sampling paths through $M_n$),
so as to guarantee that $E_n$ will be in strict normal form.

\subsection{Explicit construction of $M_n$ using $R$ and $H_n$}

This construction is somewhat redundant, since we construct $M_n$ from $Q_n$, $E_l$ and $E_r$, rather than from $R \compose H_n$.
It is retained for comparison.

The following construction uses the fact that $M_n = R \compose H_n$ so that we can compactly define $M_n$ by referring back to the previous construction of $H_n$.
In practice, it will be more efficient to precompute $Q_n = R \compose (B_l \fork B_r)$.

Refer to the previous section (``Explicit construction of $H_n$'') for definitions of
$\externalcascades,\internalcascades,\waitstates,\transweightfun{H_n}(h,h'),\emitweightfun{H_n}(\omega,\omega',h)$.

Assume that $R$ is in strict normal form.

\begin{eqnarray*}
M_n & = & R \compose H_n \\
& = & R \compose ((B_l \compose E_l) \fork (B_r \compose E_r)) \\
& = & (\emptyset,\emptyset,\statesof{M_n},\startstateof{M_n},\laststateof{M_n},\transitionsof{M_n},\weightfunof{M_n}) \\
\startstateof{M_n} & = & (\startstateof{R},\startstateof{\idfork},\startstateof{B_l},\startstateof{E_l},\startstateof{B_r},\startstateof{E_r}) \\
\laststateof{M_n} & = & (\laststateof{R},\laststateof{\idfork},\laststateof{B_l},\laststateof{E_l},\laststateof{B_r},\laststateof{E_r})
\end{eqnarray*}

\subsubsection{States of $M_n$}
The complete set of $M_n$-states is
\begin{eqnarray*}
\statesof{M_n} & = & \{ \startstateof{M_n},\ \laststateof{M_n} \} \\
& & \quad \cup\ \{ \mstate: \hstate \in \externalcascades,\ \statetype(\rho)=I \} \\
& & \quad \cup\ \{ \mstate: \hstate \in \waitstates,\ \statetype(\rho)=W \} \\
& & \quad \cup\ \{ \mstate: \hstate \in \internalcascades,\ \statetype(\rho)=\statetype(\upsilon)=S \} \\
& & \quad \cup\ \{ \mstate: \hstate \in \internalcascades,\ \statetype(\rho)=I,\ \statetype(\upsilon)=M \}
\end{eqnarray*}

\subsubsection{I/O weights of $M_n$}
Let $m = \mstate$ be an $M_n$-state and $h = \hstate$ the subsumed $H_n$-state.

Similarly, let $m' = \mstatedest$ and $h' = \hstatedest$.

The I/O weight function for $M_n$ is
\[
\emitweightfun{M_n}(\epsilon,\epsilon,m) = \left\{
\begin{array}{ll}
\displaystyle
\sum_{\omega \in \Omega} \emitweightfun{R}(\epsilon,\omega,\rho) \emitweightfun{H_n}(\omega,\epsilon,h)
 & \mbox{if $h \in \externalcascades$} \\
\emitweightfun{H_n}(\epsilon,\epsilon,h)
 & \mbox{otherwise}
\end{array}
\right.
\]

\subsubsection{Transition weights of $M_n$}
The transition weight function for $M_n$ is
\[
\transweightfun{M_n}(m,m') = \left\{
\begin{array}{ll}
\displaystyle
\transweightfun{H_n}(h,h')
 & \mbox{if $h' \in \internalcascades$} \\
\displaystyle
\transweightfun{R}(\rho,\rho') \sum_{h'' \in \waitstates} \transweightfun{H_n}(h,h'') \transweightfun{H_n}(h'',h')
 & \mbox{if $h' \in \externalcascades$} \\
\transweightfun{R}(\rho,\rho') \transweightfun{H_n}(h,h')
 & \mbox{otherwise}
\end{array}
\right.
\]

If $E_l$ and $E_r$ are acyclic, then $H_n$ and $E_n$ will be acyclic too.
However, $M_n$ does contain cycles among states of type $(I,M,D,W,D,W)$.
These correspond to characters output by $R$ that are then deleted by both $B_l$ and $B_r$.
It is necessary to eliminate these states from $M_n$ by marginalization, and to then restore them probabilistically when sampling paths through $M_n$.

\subsection{Dynamic programming algorithms}
\seclabel{DynamicProgramming}
The recursion for $\wtrans{\weight}{\epsilon}{M_n}{\epsilon}$ is
\begin{eqnarray*}
\wtrans{\weight}{\epsilon}{M_n}{\epsilon} & = & Z(\laststate) \\
Z(m') & = & \sum_{m : (m,\epsilon,\epsilon,m') \in \Transitions} Z(m) \weightfunof{}(m,\epsilon,\epsilon,m')  \quad\quad \forall m' \neq \startstate \\
Z(\startstate) & = & 1
\end{eqnarray*}

The algorithm to fill $Z(m)$ has the general structure shown in Algorithm~\ref{ForwardTransducer}.
(Some optimization of this algorithm is desirable, since not all tuples $\mstate$ are states of $M_n$.
If $E_n$ is in strict-normal form its $W$- and $D$-states will occur in pairs
(c.f. the strict-normal version of the exact-match transducer $\recognize(S)$).
These $(D,W)$ pairs are largely redundant: the choice between $D$ and $W$ is dictated by the parent $B_n$,
as can be seen from Table~\ref{HTransitionTypes} and the construction of $\statesof{H_n}$.)

\hl{\paragraph{Time complexity}
The skeleton structure of } Algorithm~\ref{ForwardTransducer} \hl{is three nested loops,
over $\statesof{E_l}, \statesof{E_r}$, and $\statesof{Q_n}$.  
The state spaces $\statesof{E_l}, \statesof{E_r}$, and $\statesof{Q_n}$ 
are independent of each other, and so }Algorithm~\ref{ForwardTransducer} \hl{ has time complexity
 ${\cal O}(\numberofstates{E_l} \numberofstates{E_l} \numberofstates{Q_n} t_{Z(m)})$, 
where $t_{Z(m)}$ is the time required to compute $Z(m)$ for a given $m \in M_n$. 

The quantities $\numberofstates{E_*}$ can be bounded by a user-specified constant $\profileSizeLimit$  
by terminating  stochastic sampling such that $\numberofstates{E_*} \leq \profileSizeLimit$ 
as described in} \secref{Mn2En}.
\hl{$\statesof{Q_n}$ is comprised of pairs of states from transducers $B_l$ and $B_r$, (detailed in}
 \secref{Qn}
\hl{), and so it has size
${\cal O}( \numberofstates{B_l} \numberofstates{B_r})$.  
Computing $Z(m)$ (outlined in } Function~\ref{fillZ})
\hl{requires summing over all incoming states, so $t_{Z(m)}$ has time complexity
 ${\cal O}(|m:(m,\epsilon,\epsilon,m') \in \Transitions|)$.  
In typical cases, this set will be small (e.g. a linear profile will have exactly one 
incoming transition per state), though the worst-case size is ${\cal O}(p)$.  
If we assume the same branch transducer $B$ is used throughout, 
the full forward recursion has worst-case time complexity ${\cal O}(\numberofstates{B}^2 \profileSizeLimit^4)$.}

\begin{algorithm}
  Initialize $Z(\startstate) \leftarrow 1$;
  \BlankLine
  \ForEach(\tcc*[f]{topologically-sorted}){$e_l \in \statesof{E_l}$} {
    \ForEach(\tcc*[f]{topologically-sorted}){$e_r \in \statesof{E_r}$} {
      \ForEach(\tcc*[f]{topologically-sorted}){$\qstate \in \statesof{Q_n}$} {
        Let $m = \mstate$;
        \BlankLine
        \If{$m \in \statesof{M_n}$}{
          Compute $Z(m)$;
        }
      }
    }
  }
  Return $Z(\laststate)$.
\caption{\label{ForwardTransducer}
The analog of the Forward algorithm for transducer $M_n$, described in \secref{Mn}. This is used during progressive reconstruction to store the sum-over-paths likelihood up to each state in $\statesof{M_n}$.  The value of $Z(\laststateof)$ is the likelihood of sequences descended from node $n$. 
}
\end{algorithm}

For comparison, the Forward algorithm for computing the probability of two sequences $(S_l,S_r)$
being generated by a Pair Hidden Markov Model $(M)$ has the general structure shown in Algorithm~\ref{ForwardPairHMM}.

\begin{algorithm}
  Initialize cell $(0,0,\mbox{START})$;
  \BlankLine
  \ForEach(\tcc*[f]{ascending order}){$0 \leq i_l \leq \seqlen{S_l}$} {
    \ForEach(\tcc*[f]{ascending order}){$0 \leq i_r \leq \seqlen{S_r}$} {
      \ForEach(\tcc*[f]{topologically-sorted}){$\sigma \in M$} {
        Compute the sum-over-paths up to cell $(i_l,i_r,\sigma)$;
      }
    }
  }
  Return cell $(\seqlen{S_l},\seqlen{S_r},\mbox{END})$.
\caption{\label{ForwardPairHMM}
The general form of the Forward algorithm for computing the joint probability of two sequences generated by the model $M$, a  Pair HMM.  
}
\end{algorithm}

The generative transducer $Q_n \equiv R \compose (B_l \fork B_r)$
in Algorithm~\ref{ForwardTransducer} is effectively identical to the Pair HMM in Algorithm~\ref{ForwardPairHMM}.

\subsection{Pseudocode for DP recursion}

We outline a more precise version of the Forward-like DP recursion in Algorithm~\ref{ForwardTransducerFull} and the associated Function $\addToDPFunction$. 
Let $\getprofiletype(q,side)$ return the state type for the profile on $side$ which is consistent with $q$.  

\subsubsection{Transition sets}

Since  all transitions in the state spaces $Q'_n, E_l$, and $E_r$ are known, we can define the following sets :
\begin{eqnarray*}
\incomingLeftProfile{j} &=& \{i: \newTransName{l}(i,j) \neq 0 \} \\
\incomingRightProfile{j} &=& \{i: \newTransName{r}(i,j) \neq 0 \} \\
\incomingM{q'} &=& \{ q:q \in \matchstates, \transweightfun{Q'_n}(q,q') \neq 0 \}\\
\incomingL{q'} &=& \{ q:q \in \leftemitstates, \transweightfun{Q'_n}(q,q') \neq 0 \}\\
\incomingR{q'} &=& \{ q:q \in \rightemitstates, \transweightfun{Q'_n}(q,q') \neq 0 \}
\end{eqnarray*}

\begin{algorithm}
  Initialize $Z(\startstate) \leftarrow 1$;
  \BlankLine
  \ForEach{$1 \leq \rightToI \leq N_r$}{ 
    \ForEach{$\qTo \in \{ q:type(q) = (S,S,S,I)\}$}{ 
      Let $(\leftProfTo,\rightProfTo)  = (\startstate, \profiledelete{\rightToI})$; \\
      $\addToDP{\qTo}{\leftProfTo}{\rightProfTo}$;
    }
  }

  \ForEach{$1 \leq \leftToI \leq N_l$}{ 
    \ForEach{ $1 \leq \rightToI \leq N_r$}{ 
      \If{$\envelope{\leftToI}{\rightToI}$  }{
      \ForEach{$\qTo \in \matchstates $} { 
        Let $(\leftProfTo,\rightProfTo)  = (\profiledelete{\leftToI}, \profiledelete{\rightToI})$; \\
        $\addToDP{\qTo}{\leftProfTo}{\rightProfTo}$;
        }

      \ForEach{$\qTo \in \leftemitstates $} { 
        Let $(\leftProfTo, \rightProfTo) = (\profiledelete{\leftToI}, \profilewait{\rightToI})$; \\
        $\addToDP{\qTo}{\leftProfTo}{\rightProfTo}$;
    }
      \ForEach{$\qTo \in \rightemitstates$} { 
        \If{$type(\qTo) == (S,S,S,I)$}{continue}
        Let $\tau = \getprofiletype(\qTo, left)$; \\
        $(\leftProfTo, \rightProfTo) = (\profileunknown{\leftToI}, \profiledelete{\rightToI})$; \\
        $\addToDP{\qTo}{\leftProfTo}{\rightProfTo}$;
      }
      \ForEach{$\qTo \in \waitstates $} { 
        Let $(\leftProfTo, \rightProfTo) = (\profilewait{\leftToI}, \profilewait{\rightToI})$; \\
        $\addToDP{\qTo}{\leftProfTo}{\rightProfTo}$;
      }
      }
    }
  }
  \ForEach{$1 \leq \leftToI \leq N_l$}{ 
    \ForEach{$\qTo \in \leftemitstates $} { 
      Let $(\leftProfTo, \rightProfTo) = (\profiledelete{\leftToI},  \profileterminate )$; \\
      $\addToDP{\qTo}{\leftProfTo}{\rightProfTo}$;
    }
  }
    \ForEach{ $1 \leq \rightToI \leq N_r$}{ 
      \ForEach{$\qTo \in \rightemitstates $} { 
      Let $(\leftProfTo, \rightProfTo) = (\profileterminate, \profiledelete{\rightToI})$; \\
      $\addToDP{\qTo}{\leftProfTo}{\rightProfTo}$;
    }
  }

  \ForEach{$\qTo \in \waitstates $} { 
    Let $(\leftProfTo, \rightProfTo) = (\profileterminate, \profileterminate)$; \\
    $\addToDP{\qTo}{\leftProfTo}{\rightProfTo}$;
  }

  Initialize $Z(\laststate) \leftarrow 0$; \\
  \ForEach{$\qFrom \in \waitstates $ } {
    Let $\mFrom = (\qFrom,\profileterminate, \profileterminate)$; \\
    $Z(\laststate) \leftarrow Z(\laststate) + Z(\mFrom) \transweightfun{Q'_n}(\qFrom,\laststateof{Q'_n})$;
  }

\caption{\label{ForwardTransducerFull}
The full version of the analog of the Forward algorithm for transducer $M_n$, 
described in \secref{Mn}. 
This to visit each state in $\statesof{M_n}$ 
in the proper order, storing the sum-over-paths likelihood
 up to that state using \addToDPFunction(\ldots) (defined separately).
The value of $Z(\laststateof)$ is the likelihood of sequences descended from node $n$. 
}
\end{algorithm}

\begin{function}
\KwIn{$(\qTo, \leftProfTo, \rightProfTo)$.}
\KwResult{The cell in $Z$ for $m=(\qTo, \leftProfTo, \rightProfTo)$ is filled.}
\BlankLine
Let $\mTo = (\qTo,\leftProfTo, \rightProfTo)$; \\
Let $\emitProb = \emitweightfun{M_n}(\epsilon,\epsilon,\mTo)$; \\

Initialize $Z(\mTo) \leftarrow \transweightfun{Q'_n}(\startstate,\qTo) \profTrans{l}(0,\leftToI) \profTrans{r}(0,\rightToI) \emitProb$; \\
\ForEach{ $\leftFromI \in \incomingLeftProfile{\leftToI} $}{
  \ForEach{ $\rightFromI \in \incomingRightProfile{\rightToI} $} {
      Let $(\leftProfFrom, \rightProfFrom) = (\profiledelete{\leftFromI}, \profiledelete{\rightFromI})$; \\
    \ForEach{ $\qFrom \in \incomingM{\qTo} $ } {
      
      Let $\mFrom = (\qFrom,\leftProfFrom, \rightProfFrom)$; \\
      $Z(\mTo) \leftarrow Z(\mTo) + Z(\mFrom) \transweightfun{Q'_n}(\qFrom,\qTo) \profTrans{l}(\leftFromI,\leftToI)
      \profTrans{r}(\rightFromI,\rightToI) \emitProb$
     }
    }
}

\ForEach{ $\leftFromI \in \incomingLeftProfile{\leftToI} $}{
  \ForEach{ $\qFrom \in \incomingL{\qTo} $ } {
    Let $(\leftProfFrom, \rightProfFrom) = (\profiledelete{\leftFromI}, \profilewait{\rightToI})$; \\
    Let $\mFrom =( \qFrom,\leftProfFrom, \rightProfFrom)$; \\
    $Z(\mTo) \leftarrow Z(\mTo) + Z(\mFrom) \transweightfun{Q'_n}(\qFrom,\qTo) \profTrans{l}(\leftFromI,\leftToI)\emitProb$;
  }    
}

\ForEach{ $\rightFromI \in \incomingRightProfile{\rightToI} $} {
  \ForEach{ $\qFrom \in \incomingR{\qTo} $ } {
    Let $\tau = \getprofiletype(\qFrom, left)$; \\
    Let $(\leftProfFrom, \rightProfFrom) = (\profileunknown{\leftToI}, \profiledelete{\rightFromI})$; \\
    Let $\mFrom = (\qFrom,\leftProfFrom, \rightProfFrom)$; \\
    $Z(\mTo) \leftarrow Z(\mTo) + Z(\mFrom) \transweightfun{Q'_n}(\qFrom,\qTo) \profTrans{r}(\leftFromI,\leftToI)\emitProb$;
  }
}

\label{fillZ}
\caption{\addToDPFunction()   used by Algorithm \ref{ForwardTransducerFull}.  
This is used during the Forward algorithm to compute the sum-over-paths likelihood ending at a given
state.  This quantity is later used to guide stochastic sampling (Algorithms \ref{traceback.short} and \ref{traceback.full}).  
}

\end{function}

\subsubsection{Traceback}
\seclabel{Traceback}
Sampling a path from $P(\pi|M_n)$ is analogous to stochastic traceback through the Forward matrix.  The basic traceback algorithm is presented in Algorithm~\ref{traceback.short}, and a more precise version is presented in Algorithm~\ref{traceback.full}.  

Let the function $\sample(set, weights)$ input two equal-length vectors and return a randomly-chosen element of $set$, such that  $set_i$ is sampled
with probability $\frac{weights_i}{sum(weights)}$.
A state path with two sampled paths is shown in \figref{forward2-fork-tkf91liv-tkf91mf}. 

\hl{
\paragraph{Alternative sampling schemes}  The above stochastic 
sampling strategy was chosen for its
ease of presentation and implementation, but our approach is sufficiently general to 
allow any algorithm which selects  a subset of complete paths through $M_n$.  
This selection may be by random (as ours is) or deterministic means. 
Randomized algorithms are widespread in computer science} \cite{Motwani2010}, 
\hl{though deterministic algorithms may be easier to analyze mathematically. 

For instance, if an analog to the backward algorithm for HMMs was developed
for the state space of $M_n$ (e.g. Algorithm } \ref{ForwardTransducerFull} \hl{reversed), 
we could select
a set of states according to their posterior probability (e.g. the $n$ states with highest
posterior probability), and determine the most likely paths (via Viterbi paths)
 from start to end
which include these states.  Alternatively, a decision theory-based ``optimal accuracy'' 
approach could be used to 
optimize the total posterior probability of the selected states. 
These approaches require  an additional dynamic programming
recursion (the backward algorithm) at each step, and we suspect 
the improvement in accuracy may be
minimal in the limit of sampling  many paths.  
Empirically, we have observed that sampling paths is
very fast compared to filling a DP matrix, and so we have focused our attention on the 
outlined stochastic approach.
}

\begin{algorithm}
\KwIn{Z}
\KwOut{A path  $\pi$ through $M_n$ sampled proportional to $\transweightfun{M_n}(\pi)$.}
\BlankLine
Initialize $\pi \leftarrow  (\laststateof{M_n})$ \\
Initialize $\currentstate \leftarrow \laststateof{M_n}$ \\
\BlankLine
\While { $\currentstate \neq \startstateof{M_n}$ } {
Let $K = \{ k \in M_n: \transweightfun{M_n}(k,\currentstate) \neq 0 \} $ \\
With probability $\frac{Z(\newstate)\transweightfun{M_n}(\newstate, \currentstate)}
{\sum_{k \in K} Z(k)\transweightfun{M_n}(k, \currentstate)}$, prepend $\newstate$ to $\pi$.\\
Set $\currentstate \leftarrow \newstate$
}
\KwRet{$\pi$}

\label{traceback.short}
\caption{
Pseudocode for Stochastic traceback for sampling paths through the transducer $M_n$, described
in \secref{Mn}. 
Stochastic sampling is done such that a path $\pi$ through $M_n$ is visited proportional to its
likelihood weight.  
By tracing a series of paths through $M_n$ and storing the union of these paths as a sequence
profile, we are able to limit the number of solutions considered during progressive
 reconstruction, reducing time and memory complexity. 
} 
\end{algorithm}

\begin{algorithm}
\KwIn{Z}
\KwOut{A path  $\pi$ through $M_n$ sampled proportional to $\transweightfun{M_n}(\pi)$.}
\BlankLine
Initialize $\pi \leftarrow  (\laststateof{M_n})$ \\
Initialize $\currentstate \leftarrow \laststateof{M_n}$ \\
\BlankLine
\While { $\currentstate \neq \startstateof{M_n}$ } {
Initialize $\instates \leftarrow \emptyarray $ \\
Initialize $\inweights \leftarrow \emptyarray $ \\
Set $ (\qTo, \leftProfTo, \rightProfTo) \leftarrow \mTo $\\

\If {$\currentstate = \laststateof{M_n}$}{
\ForEach { $\qFrom \in \waitstates$ } {
  Add $(\qTo, \profileterminate, \profileterminate)$ to $\instates$ \\
  Add $Z((\qFrom, \profileterminate, \profileterminate))\transweightfun{Q'_n}(\qFrom, \laststateof{Q'_n})$ to $\inweights$\\
}
}
\Else{
  \If {$\transweightfun{Q'_n}(\startstateof{Q'_n}, qTo) \profTrans{l}(0, \leftToI)\profTrans{r}(0, \rightToI) \neq0$}{
    Add $(\startstateof{Q'_n}, \startstate, \startstate)$ to $\instates$ \\
    Add $\transweightfun{Q'_n}(\startstateof{Q'_n}, qTo) \profTrans{l}(0, \leftToI) \profTrans{r}(0, \rightToI)$ to $\inweights$\\
  }
  \ForEach{ $\leftFromI \in \incomingLeftProfile{\leftToI} $}{
    \ForEach{ $\rightFromI \in \incomingRightProfile{\rightToI} $} {
      Let $(\leftProfFrom, \rightProfFrom) = (\profiledelete{\leftFromI}, \profiledelete{\rightFromI})$; \\
      \ForEach{ $\qFrom \in \incomingM{\qTo} $ } {
        
        Add $(\qFrom,\leftProfFrom, \rightProfFrom)$ to $\instates$; \\
        Add $Z((\qFrom, \leftProfFrom, \rightProfFrom))\transweightfun{Q'_n}(\qFrom,\qTo) \profTrans{l}(\leftFromI,\leftToI)\profTrans{r}(\rightFromI,\rightToI)$ to $\inweights$; \\  
      }
    }
  }

\ForEach{ $\leftFromI \in \incomingLeftProfile{\leftToI} $}{
    Let $(\leftProfFrom, \rightProfFrom) = (\profiledelete{\leftFromI}, \profilewait{\rightToI})$; \\
    \ForEach{ $\qFrom \in \incomingL{\qTo} $ } {
   Add $(\qFrom,\leftProfFrom, \rightProfFrom)$ to $\instates$; \\
   Add $Z((\qFrom, \leftProfFrom, \rightProfFrom))\transweightfun{Q'_n}(\qFrom,\qTo) \profTrans{l}(\leftFromI,\leftToI)$ to $\inweights$; \\  
   }
}

\ForEach{ $\rightFromI \in \incomingRightProfile{\rightToI} $} {
  Let $\tau = \getprofiletype(\qTo, left)$; \\
  Let $(\leftProfFrom, \rightProfFrom) = (\profileunknown{\leftToI}, \profiledelete{\rightFromI})$; \\
    \ForEach{ $\qFrom \in \incomingR{\qTo} $ } {
    Add $(\qFrom,\leftProfFrom, \rightProfFrom)$ to $\instates$; \\
    Add $Z((\qFrom, \leftProfFrom, \rightProfFrom))\transweightfun{Q'_n}(\qFrom,\qTo) \profTrans{r}(\rightFromI,\rightToI)$ to $\inweights$; \\  
    }
  }
}
\BlankLine
Set $ \newstate \leftarrow \sample(\instates, \inweights)$\\
Prepend $\newstate$ to $\pi$\\
Set $\currentstate \leftarrow \newstate$\\
}

\KwRet{$\pi$}

\label{traceback.full}
\caption{
Pseudocode for stochastic traceback for sampling paths through the transducer $M_n$, 
described in \secref{Mn}. 
Stochastic sampling is done such that a path $\pi$ through $M_n$ is visited proportional to its
likelihood weight.  
By tracing a series of paths through $M_n$ and storing the union of these paths as a sequence
profile, we are able to limit the number of solutions considered during progressive
 reconstruction, reducing time and memory complexity. }
\end{algorithm}

\subsection{Alignment envelopes}
\seclabel{alignmentEnvelopes}
Note that states $e \in \statesof{E_n}$ of the constrained-expanded model, as with states $g \in \statesof{G_n}$ of the expanded model,
can be associated with a vector of subsequence co-ordinates
(one subsequence co-ordinate for each leaf-sequence in the clade descended from node $n$).
For example, in our non-normal construction of $\recognize(S)$, the state $\phi \in \mathbb{Z}_{|S|+1}$ is itself the co-ordinate.
The co-ordinate information associated with $e_l$ and $e_r$ can, therefore, be used to define some sort of {\em alignment envelope}, as in \cite{Holmes2005}.
For example, we could exclude $(e_l,e_r)$ pairs if they result in alignment cutpoints that are too far from the main diagonal of the sequence co-ordinate hypercube.  

Let the function $\envelope{i_l}{i_r}$ return true if the state-index pair $(i_l,i_r)$ is allowed by the alignment envelope, and false otherwise.
Further, assume that $Z$ is a sparse container data structure, such that
$Z(m)$ always evaluates to zero if $m=\mstate$ is not in the envelope.

\subsubsection{Construction of alignment envelopes}

Let $\recognize(S)$ be defined such that it has only one nonzero-weighted path
\[
X_0 \to W_0 \stackrel{\charat{S}{1}}{\to} M_1 \to W_1 \stackrel{\charat{S}{2}}{\to} M_2 \to \ldots \to W_{L-1} \stackrel{\charat{S}{\seqlen{S}}}{\to} M_{\seqlen{S}} \to W_{\seqlen{S}} \to X_{\seqlen{S}}
\]
so a $\recognize(S)$-state is either the start state ($X_0$), the end state ($X_{\seqlen{S}}$), a wait state ($W_i$) or a match state ($M_i$).
All these states have the form $\phi_i$ where $i$ represents the number of symbols of $S$ that have to be read in order to reach that state,
i.e. a ``co-ordinate'' into $S$.
All $\recognize(S)$-states are labeled with such co-ordinates, as are the states of
any transducer that is a composition involving $\recognize(S)$,
such as $G_n$ or $H_n$.

For example, in a simple case involving a root node (1) with two children (2,3) whose sequences are constrained to be $S_2,S_3$,
the evidence transducer is $G = R \compose G_{\mbox{root}} = R \compose (G_2 \fork G_3) = R \compose (\forkfun{B_2 \compose \recognize(S_2)}{B_3 \compose \recognize(S_3)})$
=
\begin{parsetree}
 ( .$R$. ( .$B_2$. .$\recognize[S_2]$. ) ( .$B_3$. .$\recognize[S_3]$. )  )
\end{parsetree}

All states of $G$ have the form $g=(r,b_2,\phi_2 i_2,b_3,\phi_3 i_3)$ where $\phi_2, \phi_3 \in \{ X, W, M \}$,
so $\phi_2 i_2 \in \{ X_{i_2}, W_{i_2}, M_{i_2} \}$ and similarly for $\phi_3 i_3$.
Thus, each state in $G$ is associated with a co-ordinate pair $(i_2,i_3)$ into $(S_2,S_3)$, as well as a state-type pair $(\phi_2,\phi_3)$.

Let $n$ be a node in the tree,
let ${\cal L}_n$ be the set of indices of leaf nodes descended from $n$,
and let $G_n$ be the phylogenetic transducer for the subtree rooted at $n$,
defined in \secref{EvidenceExpandedModel}.
Let $\States_n$ be the state space of $G_n$.

If $m \in {\cal L}_n$ is a leaf node descended from $n$,
then $G_n$ includes, as a component, the transducer $\recognize(S_m)$.
Any $G_n$-state, $g \in \States_n$, is a tuple, one element of which is a $\recognize(S_m)$-state, $\phi_i$, where $i$ is a co-ordinate (into sequence $S_m$) and $\phi$ is a state-type.
Define $i_m(g)$ to be the co-ordinate and $\phi_m(g)$ to be the corresponding state-type.

Let $A_n:\States_n \to 2^{{\cal L}_n}$ be the function returning the set of {\em absorbing leaf indices} for a state, such that the existence of a finite-weight transition $g' \to g$ implies that $i_m(g) = i_m(g') + 1$ for all $m \in A_n(g)$.

Let $(l,r)$ be two sibling nodes.
The {\em alignment envelope} is the set of sibling state-pairs from $G_l$ and $G_r$ that can be aligned.
The function $E\colon \States_l \times \States_r \to \{ 0,1 \}$ indicates membership of the envelope.
For example, this basic envelope allows only sibling co-ordinates separated by a distance $s$ or less

\[
E_{\mbox{basic}}(f,g) = \max_{m \in A_l(f), n \in A_r(g)} |i_m(f)-i_n(g)| \leq s
\]

An alignment envelope can be based on a {\em guide alignment}. 
For leaf nodes $x,y$ and $1 \leq i \leq \seqlen{S_x}$, let $\mathcal{G}(x,i,y)$ be the number of residues of sequence $S_y$
in the section of the guide alignment from the first column, up to and including the column containing residue $i$ of sequence $S_x$.

This envelope excludes a pair of sibling states if they include a homology between residues which is more than $s$ from the homology of those characters contained in the guide alignment:
\[
E_{\mbox{guide}}(f,g) = \max_{m \in A_l(f), n \in A_r(g)} \max(\ |\mathcal{G}(m,i_m(f),n)-i_n(g)| \ ,  |\mathcal{G}(n,i_n(g),m)-i_m(f)|\ ) \leq s
\]

Let $K(x,i,y,j)$ be the number of match columns (those alignment columns in which both $S_x$ and $S_y$ have a non-gap character) between the column containing residue $i$ of sequence $S_x$ and the column containing residue $j$ of sequence $S_y$.  
This envelope excludes a pair of sibling states if they include a homology between residues which is more than $s$ {\em matches} from the homology of those characters contained in the guide alignment:

\begin{align*}
E_{\mbox{guide}}(f,g) & = & \max_{m \in A_l(f), n \in A_r(g)} \max(\ |\mathcal{G}(m,i_m(f),n)-K(m,i_m(f),n,i_n(g))|, \ \\
& &|\mathcal{G}(n,i_n(g),m)-K(n,i_n(g),m,i_m(f))|\ ) \leq s
\end{align*}

\subsection{Explicit construction of profile $E_n$ from $M_n$, following DP}
\seclabel{Mn2En}

Having sampled a set of paths through $M_n$, profile $E_n$ is constructed by applying a series of transformations.
Refer to the previous sections for definitions pertaining to $M_n$ and $E_n$.

\subsubsection{Transformation $M_n \to M'_n$: sampled paths}

Let $\Pi' \subseteq \Pi_{M'_n}$ be a set of complete paths through $M_n$,
corresponding to $K$ random samples from $P(\pi|M_n)$.

The state space of $M'_n$ is the union of all states used by the paths in  $\Pi_{M'_n}$

\[ 
\statesof{M'_n} = \bigcup_{\pi \in \Pi'} \bigcup_{i=1}^{|\pi|} \pi_i  
\]
where $\pi_i$ is the $i^{th}$ state in a path $\pi \in (\transitionsof{M_n})^\ast$.

The I/O and transition weight functions for $M'_n$ are the same as those of $M_n$:

\begin{eqnarray*}
\transweightfun{M'_n}(m,m') & = & \transweightfun{M_n}(m,m') \\
\emitweightfun{M'_n}(\epsilon,\epsilon,m') & = & \emitweightfun{M_n}(\epsilon,\epsilon,m')
\end{eqnarray*}

\hl{
$M'_n$  can be updated after each path $\pi$ is sampled:}

\[
\statesof{M'_n} = \statesof{M'_n}  \cup \bigcup_{i=1}^{|\pi|} \pi_i  
\]

\hl{
For some arbitrary limiting factor $\profileSizeLimit \geq L$,  if  $\numberofstates{M'_n} \geq \profileSizeLimit$ 
upon addition of states of $\pi$, 
sampling may be terminated.  This allows bounding $\numberofstates{E_n}$ as the 
algorithm traverses up the phylogeny. 
}

\begin{center}
\begin{figure}[h!]
\includegraphics[width=1\textwidth]{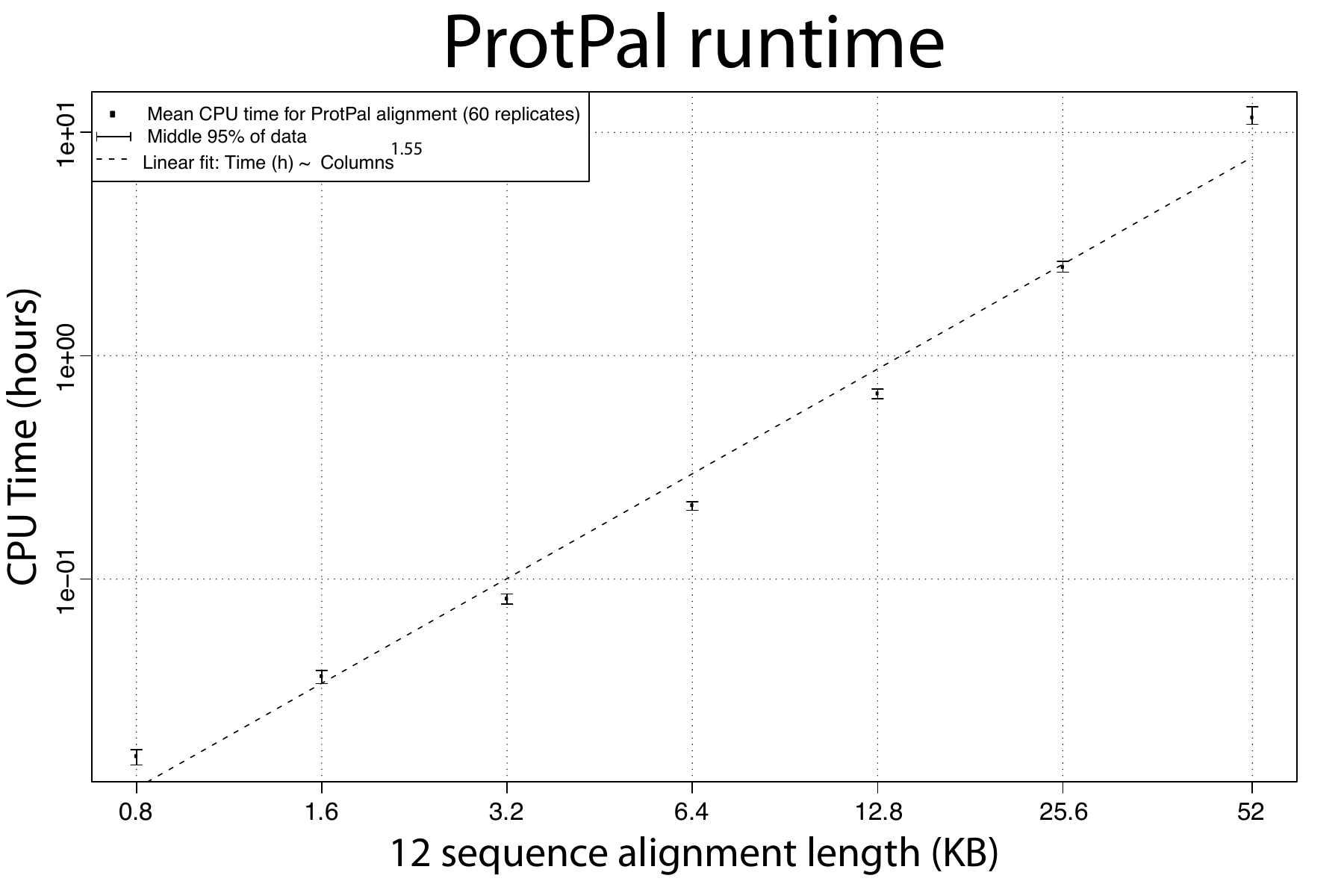}
\caption{When using an alignment envelope, our implementation of the alignment algorithm
scales sub-quadratically with respect to sequence length.  Alignments consisting of 12 sequences
were simulated for lengths ranging from 800bp to 52kb and aligned using our software.
Mean and middle 90\% quantiles (60 replicates) for  CPU time required on a 2GHz, 2GB RAM Linux machine are
shown next to a log-log linear fit (dashed line: CPU Hours $\propto Columns^{1.55}$)).  }
\figlabel{time}
\end{figure}
\end{center}

\subsubsection{Transformation $M'_n \to E''_n$: stripping out the prior}

Let $E''_n$ be a the transducer constructed via removing the $R$ states from the states of $M'_n$.  

\[
\statesof{E''_n} = \{ \hstate : \exists \mstate \in M'_n \}
\]

The I/O and transition weight functions for $E''_n$ are the same as those of $H_n$:

\begin{eqnarray*}
\transweightfun{E''_n}(e,e') & = & \transweightfun{H_n}(e,e') \\
\emitweightfun{E''_n}(\epsilon,\epsilon,e') & = & \emitweightfun{H_n}(\epsilon,\epsilon,e')
\end{eqnarray*}

\subsubsection{Transformation $E''_n \to E'_n$: eliminating null states}
\seclabel{NullStateElim}
Let $E'_n$ be the transducer derived from $E''_n$ by marginalizing its null states.  

The state space of $E'_n$ is the set of non-null states in $E''_n$:

\[
\statesof{E'_n} = \{ e : e \in E''_n, type(e) \in \{S,E,D,W\} \}
\]

The transition weight function is the same as that of $H_n$ (and also $E''_n$) with paths through null states marginalized.
Let 

\begin{eqnarray*}
\pi_{ {E''_n}(e,e')} &  =  & \{  \pi: \pi_1 =e, \pi_{|\pi|}=e', type(\pi_i) = N\ \forall\ i: 2 \leq i < |\pi| \} \\
 \transweightfun{E'_n}(e,e') & = & \sum_{\pi \in \pi_{E''_n(e,e') }} \transweightfun{H_n}(\pi) 
\end{eqnarray*}

The transition weight function resulting from summing over null states in $E''_n$ can be done with a preorder traversal of the state graph, outlined in Algorithm~\ref{removeNullStates}.   The $stack$ data structure has operations $stack.push(e)$, which adds $e$ to the top of the stack, and $stack.pop()$, which removes and returns the top element of the stack.  The $weights$ container maps states in $\statesof{E''_n}$ to real-valued weights.  

\newcommand\newTransHash{t_{\mbox{new}}}
\begin{algorithm}
\KwIn{$\statesof{E''_n}, \transweightfun{E''_n}$}
\KwOut{$\transweightfun{E'_n}$}
Let $\statesof{E'_n} = \{ e : e \in E''_n, type(e) \in \{S,E,D,W\} \}$\\
Initialize $\newTransHash(e,e') = 0$ $ \forall (e,e') \in \statesof{E'_n}$\\
\ForEach { $source\_state \in \statesof{E''_n}$}{
Initialize $stack = [source\_state]$\\
Initialize $weights[source\_state] = 0$\\

\BlankLine

\While{$stack \neq [] $}{

Set $e=stack.pop()$\\
\ForEach {$e' \in \{e' \in \statesof{E''_n}: \transweightfun{E''_n}(e,e') \neq 0\}$ }{

\If{ $type(e') \neq N$}{ 
  $\newTransHash(source, e') += weights[e]\cdot  \transweightfun{E''_n}(e,e')$\\
}
\Else{
$stack.push(e')$\\
$weights[e'] = weight[e] \cdot \transweightfun{E''_n}(e,e')$\\
}

}  
}  
}  
\KwRet{ $\transweightfun{E'_n}(e,e') \equiv \newTransHash(e,e')$ $\forall (e,e') \in \statesof{E'_n}$}
\label{removeNullStates}
\caption{Pseudocode for transforming the transition weight function of $E''_n$ into that of $E'_n$ via summing over null state paths (insertions).  
This is done after stochastic sampling as the first of two steps transforming a sampled $M_n$ transducer into a recognizer $E_n$, described in \secref{Mn2En}.
  Summing over null states ensures that these states cannot align to sibling states in the parent round
of profile-profile alignment.  
An insertion at branch $n$ is, by definition, not homologous to any characters outside the $n$-rooted subtree, and null state elimination is how this is explicitly enforced in our algorithm.} 
\end{algorithm}

Besides the states $\{ \startstate, \laststate, \profileterminate \}$, the remaining states in $E'_n$ are of type $D$, which we now index in ascending topological order: 

\[
\statesof{E'_n} = \{ \startstate, \laststate , \profileterminate \} \cup  \{\profiledelete{n} : 1 \leq n \leq \numStates{n} \}
\]

where $ 1\leq i,j\leq \numStates{n}$,

\begin{eqnarray*}
\newTransName{n}(i,j) & \equiv & \transweightfun{E'_n}(\profiledelete{i},\profiledelete{j}) \\
\newTransName{n}(0,j) & \equiv & \transweightfun{E'_n}(\startstate,\profiledelete{j})  \\
\newTransName{n}(i,\numStates{n}+1) & \equiv & \transweightfun{E'_n}(\profiledelete{i}, \profileterminate)  \\
\end{eqnarray*}

\subsubsection{Transformation $E'_n \to E_n$: adding wait states}

Let $E_n$ be the transducer derived from transforming $E'_n$ into strict normal form.  Since $N$ states were removed in the transformation from $E''_n \rightarrow E'_n$, we need only add $W$ states before each $D$ state:

\[
\statesof{E_n} = \{ \startstate, \laststate , \profileterminate \} \cup  \{\profiledelete{n} : 1 \leq n \leq N \}
\cup  \{\profilewait{n} : 1 \leq n \leq N \}
\]

Note the following correspondences between the previously-defined  $\transweightfun{E_n}(e,e')$ and new notation. 

\begin{eqnarray*}
\transweightfun{E_n}(\profiledelete{i},\profilewait{j}) & = & \newTransName{n}(i,j) \\
\transweightfun{E_n}(\profilewait{j},\profiledelete{j}) & = & 1 \\
\transviawait{E_n}(\profiledelete{i},\profiledelete{j}) & = & \newTransName{n}(i,j) \\
\transtowait{E_n}(\profiledelete{i},\profilewait{j}) & = & \newTransName{n}(i,j) \\
\transtowait{E_n}(\profilewait{i},\profilewait{j}) & = & \delta(i=j) \\
\end{eqnarray*}

\subsection{Message-passing interpretation}
\seclabel{MessagePassing}

In the interpretation of Felsenstein's pruning algorithm \cite{Felsenstein81} and Elston and Stewart's more general peeling algorithm \cite{ElstonStewart71} as message-passing on factor graphs \cite{KschischangEtAl98},
the tip-to-root messages are functions of the form
$P(\outputs_n|\outputn{n}=x)$ where $\outputn{n}$ (a random variable) denotes the sequence at node $n$,
$x$ denotes a particular value for this r.v.,
and
$\outputs_n = \{ \outputn{m} : m \in \leaves_n \}$
denotes the observation of sequences at nodes in $\leaves_n$, the set of leaf nodes that have $n$ as a common ancestor.

These tip-to-root messages are equivalent to our evidence-expanded transducers (\secref{EvidenceExpandedModel}):
\[
P(\outputs_n|\outputn{n}=x) = \wtrans{\weight}{x}{G_n}{\epsilon}
\]

The corresponding root-to-tip messages take the form
$P(\bar{\outputs}_n,\outputn{n}=x)$
where
$\bar{\outputs}_n = \{ \outputn{m} : m \in \leaves, m \notin \leaves_n \}$ 
denotes the observation of sequences at leaf nodes that do {\em not} have $n$ as a common ancestor.
These messages can be combined with the tip-to-root messages to yield posterior probabilities of ancestral sequences
\[
P(\outputn{n}=x|\outputs) = \frac{ P(\bar{\outputs}_n,\outputn{n}=x) P(\outputs_n|\outputn{n}=x)}{P(\outputs)}
\]

We can define a recursion for transducers that model these root-to-tip messages, just as with the tip-to-root messages.

First, define $J_1 = R$.

Next, suppose that $n>1$ is a node with parent $p$ and sibling $s$.
Define
\[
J_n = J_p \compose (B_n \fork (B_s \compose G_s))
=
\begin{parsetree}
 ( .$J_p$. ( .$\idfork$. ( .$B_n$. .. ) ( .$B_s$. .$G_s$. )  ) )
\end{parsetree}
\]

Note that $J_n$ is a generator that outputs only the sequence at node $n$
(because $G_s$ has null output).
Note also that $J_n G_n \transequiv G_0$.

The root-to-tip message is
\[
P(\bar{\outputs}_n,\outputn{n}=x) = \wtrans{\weight}{\epsilon}{J_n}{x}
\]

The equations for $G_n$ and $J_n$ are transducer formulations of the pruning and peeling recursions

\begin{eqnarray*}
P(\outputs_n|\outputn{n}=x) & = & \left( \sum_y P(\outputn{l}=y|\outputn{n}=x) P(\outputs_l|\outputn{l}=y) \right) \left( \sum_z P(\outputn{r}=z|\outputn{n}=x) P(\outputs_r|\outputn{r}=z) \right) \\
P(\bar{\outputs}_n,\outputn{n}=x) & = & \sum_y P(\bar{\outputs}_p,\outputn{p}=y) P(\outputn{n}=x|\outputn{p}=y) \sum_z P(\outputn{s}=z|\outputn{p}=y) P(\outputs_s|\outputn{s}=z)
\end{eqnarray*}
where $(l,r)$ are the left and right children of node $n$.
For comparison,
\begin{eqnarray*}
G_n & = & (B_l \compose G_l) \fork (B_r \compose G_r) \\
J_n & = & J_p \compose (B_n \fork (B_s \compose G_s)) \\
\wtrans{\weight}{x}{B_n}{y} & = & P(\outputn{n}=y|\outputn{p}=x) \\
\wtrans{\weight}{x}{T\compose U}{z} & = & \sum_{y} \wtrans{\weight}{x}{T}{y} \wtrans{\weight}{y}{U}{z} \\
\wtrans{\weight}{x}{T_l}{\outputs_l} \wtrans{\weight}{x}{T_r}{\outputs_r} & = & \wtrans{\weight}{x}{T_l \fork T_r}{\outputs_n}
\end{eqnarray*}

\section{Conclusions}

In this article we have presented an algorithm that may be viewed in two equivalent ways: a form of Felsenstein's pruning algorithm generalized from individual characters to entire sequences, 
or a phylogenetic generalization of progressive alignment.   
Our algorithm  extends the concept of a character substitution matrix (e.g. \cite{JukesCantor69, HasegawaEtAl85}) to finite-state transducers, replacing matrix multiplication with transducer composition.

We described a hierarchical  approximation technique enabling inference 
in \hl{ ${\cal O}(c^2\profileSizeLimit^4N)$} time and memory 
(typical-case ${\cal O}((c\profileSizeLimit)^2 N)$) , as opposed to ${\cal O}(L^N)$ 
for exact, exhaustive inference ($N$ sequences of length $L$, limiting factor $m\geq L$ 
and branch transducer with $c$ states). 
\hl{Empirical tests indicate that } adding additional constraints
 (in the form of an \hl{``alignment envelope'')  brings typical-case} time complexity 
down to \hl{ ${\cal O}(c^2\profileSizeLimit^{1.55} N)$}, making the algorithm practical for typical alignment problems.  

Much of computational biology depends on a multiple sequence alignment as input, yet the uncertainty and bias engendered by an alignment is rarely accounted for.  
Further, most alignment programs account for the phylogeny relating sequences either in a heuristic sense, or not at all.  
Previous studies have indicated that alignment uncertainty and/or accuracy  strongly affects downstream analyses \cite{WongEtAl2008}, particularly if evolutionary inferences are to be made \cite{LoytynojaGoldman2008}.

In this work we have described  the mathematics and algorithms required for an alignment algorithm that is simultaneously explicitly phylogenetic and avoids conditioning on a single multiple alignment.
  In extensive simulations (in separate work, submitted), we find that our implementation of this algorithm recovers significantly more accurate reconstructions of simulated indel histories, 
indicating the need for mathematically rigorous alignment algorithms, particularly for evolutionary applications. 

The source code to this paper, including the graphviz and phylocomposer files used to produce the diagrams,
can be found at \url{https://github.com/ihh/transducer-tutorial}

\section{Methods}

\newcommand{\indelseqgen}{indel-seq-gen}
\newcommand{\protpal}{ProtPal}
\newcommand{\prank}{PRANK}
\newcommand{\clustalw}{CLUSTALW}
\newcommand{\rateRatio}[1]{\frac{\hat{\lambda}^{#1}_{\hat{H}}}{\lambda^#1}}
\paragraph{Accuracy simulation study} Our simulation study is comprised of alignments simulated  using 5 different indel rates $\lambda$ (0.005, 0.01, 0.02, 0.04, and 0.08 indels per unit time), each with 3 different substitution rates (0.5, 1, and 2 expected  substitutions per unit time) and 100 replicates.  
Time is defined such that a sequence evolving for time $t$ with substitution rate $r$ is expected to accumulate $rt$ subsitutions per site.

We employed an independent third-party simulation program, indel-seq-gen,specifically designed to generate realistic protein evolutionary histories \cite{StropeEtAl2009}. indel-seq-gen is capable of modeling an empirically-fitted indel length distribution, rate variation among sites, and a neighbor-aware distribution over inserted sequences allowing for small  local duplications.  Since the indel and substitution model used by indel-seq-gen\ are separate from (and richer than) those used by ProtPal, ProtPal has no unfair advantage in this test.

indel-seq-gen v2.0.6 was run with the following command:\\
{\tt cat guidetree.tree| indel-seq-gen -m JTT -u xia --num\_gamma\_cats 3 -a 0.372 --branch\_scale r/b --outfile simulated\_alignment.fa  --quiet --outfile\_format f -s 10000  --write\_anc}

The above command uses the ``JTT'' substitution model,  the ``xia'' indel fill model (based on neighbor effects, estimated from E coli k-12 proteins \cite{StropeEtAl2009}), and 3 gamma-distributed rate categories with shape 0.372.  Branch lengths are scaled by the substitution rate for simulation rate $r$, normalized by the inverse of indel-seq-gen's underlying substitution rate ($b=1.2$) so as to adhere to the above definition of evolutionary ``time''.   Similarly, indel rates, which are set in the guide tree file {\tt guidetree.tree}, are scaled by $\frac{b}{r}$ so that $t\lambda^*$ insertions/deletions are expected over time $t$ for rate $\lambda^*$.

The  simulations were run on a tree of twelve sequenced {\em Drosophila} genomes \cite{ClarkEtAl2007}.
This  tree was made available to both ProtPal and  PRANK.
To specify ancestral inference, the guide tree, and ``insertions opening forever'', PRANK used the extra options
``{\tt -writeanc -t <treefile> +F}''.
PRANK's {\tt -F} option allows insertions to match characters at alignments closer to the root.  This can be a useful heuristic safeguard when an incorrect tree may produce errors in subtree alignments that cannot be corrected at internal nodes closer to the root.  
Since the true guide tree is provided to PRANK,  it is safe to treat insertions in a strict phylogenetic manner via the {\tt +F} option.  

For computational efficiency,  ProtPal was provided with a  \clustalw\ guide alignment.  Any alignment of the sequences can be used as a guide, and we chose \clustalw\ for its general poor performance in other tests, 
so that ProtPal would gain no unfair advantage by the information contained in the guide alignment. 

Indel rates were estimated by counting indel events in MAP reconstructed histories:
\begin{equation}
\eqnlabel{mapHistoryParams}                      
\hat{\lambda}_{\hat{H}} = argmax_{\lambda'} P(\lambda'|\hat{H},S,T) = 
argmax_{\lambda'} P(\hat{H},S|T,\lambda')
\end{equation}
where the latter step assumes a flat prior, $P(\lambda') = \mbox{const.}$

The root mean squared error (RMSE) for each error distribution was computed as follows:
\begin{equation}
\eqnlabel{RMSE}
RMSE = \sqrt{ \sum_{replicates} (\rateRatio{*}-1)^2 }
\end{equation}

\paragraph{Time simulation study} In order to empirically determine the time complexity of our method,
alignments of varying lengths (800, 1600, 3200, 6400, 12800, 25600, 52000 positions) were simulated
using indel-seq-gen (default parameters) on the 12 {\em Drosophila} species tree 
and aligned with ProtPal using a  guide alignment with restriction
parameter ({\tt -s} option described in \secref{alignmentEnvelopes}) set to 10. 
The analysis time on a 2GHz, 2GB machine was averaged over 60 replicates and plotted next to 
a log-log fit computed using R (www.r-project.org).

\paragraph{Authors' contributions:} OW and IH developed the algorithm and wrote the paper.  BP and GL made substantial intellectual contributions to developing the algorithm.  

\paragraph{Funding:} OW and IH were partially supported by NIH/NHGRI grant R01-GM076705.

\appendix
\section{Additional notation}
This section defines commonplace notation used earlier in the document.
\subsection{Sequences and alignments}
{\em Sequence notation:}
Let $\Omega^\ast$ be the set of sequences over $\Omega$, including the empty sequence $\epsilon$.
Let $x+y$ denote the concatenation of sequences $x$ and $y$, and $\sum_n x_n$ the concatenation of sequences $\{ x_n \}$.

{\em Gapped-pair alignment:}
A gapped pairwise alignment is a sequence of individual columns of the form $(\omega_1,\omega_2)$ where $\omega_1 \in \gappedalphabet{1}$ and $\omega_2 \in \gappedalphabet{2}$.

{\em Map from alignment to sequences:}
The function $S_k:\left(\gappedpair{1}{2}\right)^\ast \to \Omega_k^\ast$ returns the $k$'th row of a pairwise alignment, with gaps removed
\begin{eqnarray*}
S_1(x) & = & \sum_{(\omega_1,\omega_2) \in x} \omega_1 \\
S_2(x) & = & \sum_{(\omega_1,\omega_2) \in x} \omega_2
\end{eqnarray*}

{\em Transition paths:}
A transition path $\pi \in \Pi$ 
is a sequence of transitions of the form $(\phi_1,\omega_I,\omega_O,\phi_2)$
where
$\phi_1,\phi_2 \in \States$,
$\omega_I \in \gappedalphabet{I}$, and
$\omega_O \in \gappedalphabet{O}$.

{\em Input and output sequences:}
Define the input and output sequences
$S_I:\Pi \to \Omega_I^\ast$ and
$S_O:\Pi \to \Omega_O^\ast$
\begin{eqnarray*}
S_I(\pi) & = & \sum_{(\phi_1,\omega_I,\omega_O,\phi_2) \in \pi} \omega_I \\
S_O(\pi) & = & \sum_{(\phi_1,\omega_I,\omega_O,\phi_2) \in \pi} \omega_O
\end{eqnarray*}

\section{Composition and intersection in Mealy transducers}
\applabel{Mealy}

This paper mostly discusses Moore machines (wherein character input and output is associated with states).
For completeness, we give constructions of composition (\secref{Composition}) and intersection (\secref{Fork})
for the more general Mealy machines (wherein I/O is associated with transitions).

\subsection{Mealy normal form}
\applabel{MealyNormal}
We first introduce a normal form for Mealy transducers, analogous to the Moore-normal form.
A letter transducer
 $T = (\Omega_I, \Omega_O, \States, \startstate, \laststate, \Transitions, \weight)$ 
is Mealy-normal if the state space $\States$ can be partitioned into two disjoint subsets, $\States_{\mbox{unready}}$ and $\States_{\mbox{ready}}$,
such that
\begin{eqnarray*}
\weight(\phi_U,\omega_{\mbox{in}},\omega,\phi) & = & 0 \quad \forall \phi_U \in \States_{\mbox{unready}}, \omega_{\mbox{in}} \in \Omega_I \\
\weight(\phi_R,\epsilon,\omega,\phi) & = & 0 \quad \forall \phi_R \in \States_{\mbox{ready}}
\end{eqnarray*}

where
 $\omega \in \gappedalphabet{O}$
and
 $\phi \in \States$.
That is, there are no absorbing transitions from ``unready'' states, and no non-absorbing transitions from ``ready'' states.
Such a Mealy-normal transducer can easily be constructed from any general letter transducer
simply by identifying any states that have both absorbing and non-absorbing outgoing transitions,
splitting each such state into a ``ready'' and an ``unready'' state,
assigning the absorbing transitions to the ``ready'' state and the non-absorbing states to the ``unready'' state,
and adding a dummy transition (of unit weight) from the unready to the ready state.

\subsection{Mealy machine composition}
\applabel{MealyComposition}

{\em Transducer composition:}
Given letter transducers
 $T = (\Omega_X, \Omega_Y, \States, \startstate, \laststate, \Transitions, \weight)$ and
 $U = (\Omega_Y, \Omega_Z, \States', \startstate', \laststate', \Transitions', \weight')$,
in Mealy normal form,
we construct $T \compose U = (\Omega_X, \Omega_Z, \States'' \ldots \weight'')$ as follows.

Let $\States'' \subset \States \times \States'$,
$\startstate''=(\startstate,\startstate')$, $\laststate''=(\laststate,\laststate')$
and
\begin{eqnarray*}
\lefteqn{\weight''((t,u),\omega_x,\omega_z,(t',u')) =} & & \\
& & \left\{ \begin{array}{ll}
\delta(t=t') \delta(\omega_x=\epsilon) \weight(u,\epsilon,\omega_z,u') & \mbox{if $u \in \States_{\mbox{unready}}$} \\
\displaystyle
\delta(u=u') \delta(\omega_z=\epsilon) \weight'(t,\omega_x,\epsilon,t')
+ \sum_{\omega_y \in \Omega_Y} \weight(t,\omega_x,\omega_y,t') \weight'(u,\omega_y,\omega_z,u') & \mbox{if $u \in \States_{\mbox{ready}}$}
\end{array} \right.
\end{eqnarray*}

\subsection{Mealy machine intersection}
\applabel{MealyFork}

{\em Transducer intersection:}
Given letter transducers
 $T = (\Omega_X, \Omega_T, \States, \startstate, \laststate, \Transitions, \weight)$ and
 $U = (\Omega_X, \Omega_U, \States', \startstate', \laststate', \Transitions', \weight')$,
in Mealy normal form,
we construct $T\fork U = (\Omega_X, \Omega_V, \States'' \ldots \weight'')$
where $\Omega_V \subseteq \gappedpair{T}{U}$ as follows.

Let $\States'' \subset \States \times \States'$,
$\startstate''=(\startstate,\startstate')$, $\laststate''=(\laststate,\laststate')$
and
\begin{eqnarray*}
\lefteqn{\weight''((t,u),\omega_x,(\omega_y,\omega_z),(t',u')) =} & & \\
 & & \left\{ \begin{array}{ll}
\delta(t=t') \delta(\omega_x=\omega_y=\epsilon) \weight'(u,\epsilon,\omega_z,u') & \mbox{if $u \in \States'_{\mbox{unready}}$} \\
\delta(u=u') \delta(\omega_x=\omega_z=\epsilon) \weight(t,\epsilon,\omega_y,t') & \mbox{if $t \in \States_{\mbox{unready}}$ and $u \in \States'_{\mbox{ready}}$} \\
\weight(t,\omega_x,\omega_y,t') \weight'(u,\omega_x,\omega_z,u') & \mbox{if $t \in \States_{\mbox{ready}}$ and $u \in \States'_{\mbox{ready}}$}
\end{array} \right.
\end{eqnarray*}

\bibliography{../latex-inputs/alignment,../latex-inputs/ncrna,../latex-inputs/genomics,../latex-inputs/reconstruction}

\end{document}